\documentclass[longauth,bibyear]{aa}
\usepackage{times}
\usepackage{amsmath,mathrsfs,amssymb,amsbsy}
\usepackage{graphicx}
\usepackage{subfigure}
\usepackage{paralist}
\usepackage{color}
\usepackage{natbib}
\usepackage{aas_macros}
\graphicspath{{plots/}}
\newcommand{\alphabib}{../Bibfile/Alpha_Gap}

\def\teff{\rm T_{\rm eff}}
\def\logg{\log g}

\def\feh{\rm[Fe/H]}
\def\afe{\rm [\alpha/Fe]}
\def\meta{\rm [M/H]}

\def\vphi{\rm V_\phi}
\def\vr{\rm V_R}
\def\vz{\rm V_Z}
\def\kms{\,{\rm km\,s^{-1}}}

\def\kpc{\,{\rm kpc}}
\def\pc{\,{\rm pc}}

\def\dex{\,{\rm dex}}
\def\Gyr{\,{\rm Gyr}}
\def\url#1{{\tt#1}}
\def\Trend{\rm [\alpha/Fe]-[Fe/H]}
\def\mgfe{\rm [Mg/Fe]}
\def\K{\,{\rm K}}
\def\nm{\,{\rm nm}}
\def\mas{\,{\rm mas}}
\def\yr{\,{\rm yr^{-1}}}

\begin{document}

%G. Kordopatis, R.F.G. Wyse,  G. Gilmore, A. Recio-Blanco, P. de Laverny,V. Hill, V. Adibekyan,  U. Heiter, I. Minchev, B. Famaey, T. Bensby, S. Feltzing, G. Guiglion, A.J. Korn, S. Mikolaitis, A. Vallenari, A. Bayo,  G. Carraro,  E. Flaccomio, E. Franciosini, A. Hourihane, P. Jofre, S.E. Koposov, C. Lardo, J. Lewis, K. Lind, L. Magrini, L. Morbidelli, E. Pancino, S. Randich, G.G. Sacco, C.C. Worley, S. Zaggia

\newcommand{\edit}[1]{\textcolor{red}{#1}}                    
\newcommand{\change}[1]{\textcolor{magenta}{#1}}                    
\newcommand{\query}[1]{\textcolor{blue}{(#1)}}                    
\newcommand{\ccomment}[1]{\textcolor{cyan}{(#1)}}                         
\author{
	G.~Kordopatis\inst{\ref{aip}} 
	%1st Tier
	\and R.F.G.~Wyse\inst{\ref{jhu}} 
	\and G.~Gilmore\inst{\ref{ioa}} 
	\and A.~Recio-Blanco\inst{\ref{oca}} 
	\and P.~de~Laverny\inst{\ref{oca}} 
	%2nd Tier
	\and V.~Hill\inst{\ref{oca}}
	\and V.~Adibekyan\inst{\ref{Porto}}
	\and U.~Heiter\inst{\ref{Uppsala}} 
	\and I.~Minchev\inst{\ref{aip}}    
	\and B.~Famaey\inst{\ref{strasbourg}}
	\and T.~Bensby\inst{\ref{Lund}}
	\and S.~Feltzing\inst{\ref{Lund}}
	\and G.~Guiglion\inst{\ref{oca}} 
	\and A.J.~Korn\inst{\ref{Uppsala}}	
	\and \v{S}.~Mikolaitis\inst{\ref{oca},\ref{Vilnius}}
	\and M.~Schultheis\inst{\ref{oca}}
	\and A.~Vallenari\inst{\ref{Padova}}
	%Alphabetical
	\and A.~Bayo\inst{\ref{Valparaiso}}
	\and G.~Carraro\inst{\ref{ESO_chile}}
	\and E.~Flaccomio\inst{\ref{Palermo}}
	\and E.~Franciosini\inst{\ref{Arcetri}}
	\and A.~Hourihane\inst{\ref{ioa}}
	\and P.~Jofr\'e\inst{\ref{ioa}}
	\and S.~E. Koposov\inst{\ref{ioa},\ref{Moscow}}
	\and C.~Lardo\inst{\ref{Liverpool}}
	\and J.~Lewis\inst{\ref{ioa}}
	\and K.~Lind\inst{\ref{Uppsala}}
	\and L.~Magrini\inst{\ref{Arcetri}}
	\and L.~Morbidelli\inst{\ref{Arcetri}}
	\and E.~Pancino\inst{\ref{Bologna},\ref{Roma}}
	\and S.~Randich\inst{\ref{Arcetri}}
	\and G.G.~Sacco\inst{\ref{Arcetri}}
	\and C.C.~Worley\inst{\ref{ioa}}
	\and S.~Zaggia\inst{\ref{Padova}}
	}

\institute{
	Leibniz-Institut f\"ur  Astrophysik Potsdam (AIP), An der Sternwarte 16, 14482 Potsdam, Germany \label{aip} 
	\and 	Johns Hopkins University, Homewood Campus, 3400 N Charles Street, Baltimore, MD 21218, USA\label{jhu}
	\and Institute of Astronomy, University of Cambridge, Madingley Road, Cambridge CB3 0HA, UK\label{ioa}
	\and Laboratoire Lagrange (UMR7293), Universit\'e de Nice Sophia Antipolis, CNRS,Observatoire de la C\^ote d'Azur, CS 34229,F-06304 Nice cedex 4, France\label{oca}
	\and 	Instituto de Astrof\'{\i}'sica e Ci\^{e}ncia do Espa\c{c}o, Universidade do Porto, CAUP, Rua das Estrelas, PT4150-762 Porto, Portugal\label{Porto}
	\and{Department of Physics and Astronomy, Uppsala University, Box 516, SE-751 20 Uppsala, Sweden\label{Uppsala}}
	\and{Observatoire Astronomique de Strasbourg, Universit\'e de Strasbourg, CNRS UMR 7550, 11, rue de l'Universit\'e, F-67000 Strasbourg, France\label{strasbourg}}
	\and Lund Observatory, Department of Astronomy and Theoretical Physics, Box 43, SE-221 00 Lund, Sweden\label{Lund}
	\and{Institute of Theoretical Physics and Astronomy, Vilnius University, A. Go\v{s}tauto 12, LT-01108 Vilnius, Lithuania\label{Vilnius}}
	\and{INAF - Padova Observatory, Vicolo dell'Osservatorio 5, 35122 Padova, Italy\label{Padova}}
	\and{Instituto de F\'isica y Astronomi\'ia, Universidad de Valparai\'iso, Chile\label{Valparaiso}}
	\and European Southern Observatory, Alonso de Cordova 3107 Vitacura, Santiago de Chile, Chile\label{ESO_chile}		
	\and INAF - Osservatorio Astronomico di Palermo, Piazza del Parlamento 1, 90134, Palermo, Italy\label{Palermo}
	\and{Moscow MV Lomonosov State University, Sternberg Astronomical Institute, Moscow 119992, Russia\label{Moscow}}
	\and Astrophysics Research Institute, Liverpool John Moores University, 146 Brownlow Hill, Liverpool L3 5RF, United Kingdom\label{Liverpool}
	\and INAF - Osservatorio Astrofisico di Arcetri, Largo E. Fermi 5, 50125, Florence, Italy\label{Arcetri}
	\and INAF - Osservatorio Astronomico di Bologna, via Ranzani 1, 40127, Bologna, Italy\label{Bologna}
	\and ASI Science Data Center, Via del Politecnico SNC, 00133 Roma, Italy\label{Roma}
	}

\title{The Gaia-ESO Survey: Characterisation of the  $\afe$ sequences in the Milky Way discs\thanks{Based on observations collected with the FLAMES spectrograph at the VLT/UT2 telescope (Paranal Observatory, ESO, Chile), for the Gaia-ESO Large Public Survey, programme 188.B-3002.}}

\abstract
%Context
{ 
High-resolution spectroscopic surveys of stars indicate that the Milky Way thin and thick discs follow different paths in the chemical space defined by $\afe$ vs $\feh$, suggesting possibly different formation mechanisms for each of these structures. 
}
%%%%%%%%%%%%%%%%%%%%%%%%%%%%%%%%%%%%%%
%Aims
%%%%%%%%%%%%%%%%%%%%%%%%%%%%%%%%%%%%%%
{We investigate, using the \emph{Gaia-ESO Survey internal Data-Release 2}, the properties of the double sequence of the Milky Way discs (defined chemically as the high$-\alpha$ and low$-\alpha$ populations), and discuss their compatibility with discs defined by  other means such as  metallicity, kinematics or positions.  }
%%%%%%%%%%%%%%%%%%%%%%%%%%%%%%%%%%%%%%
%Methods
%%%%%%%%%%%%%%%%%%%%%%%%%%%%%%%%%%%%%%
{This investigation uses  two different approaches: in velocity space for stars located in the extended Solar neighbourhood, and in chemical space for stars at different ranges of Galactocentric radii and heights from the Galactic mid-plane. The separation we find in  
velocity space allows us to  investigate, in a novel manner, the extent in metallicity of each of the two sequences, identifying them with the two discs, without making any assumption about the shape of their metallicity distribution functions.  Then, using the separation in  chemical space, adopting the magnesium abundance as a tracer of the $\alpha-$elements,   we  characterise the spatial variation of the slopes of the $\afe-\feh$ sequences for the thick and thin discs and the way in which the relative proportions of the two discs change across the Galaxy.}
%%%%%%%%%%%%%%%%%%%%%%%%%%%%%%%%%%%%%%
%Results
%%%%%%%%%%%%%%%%%%%%%%%%%%%%%%%%%%%%%%
{ We find that the thick disc, defined as the stars tracing the high$-\alpha$ sequence,  extends up to super-solar metallicities ($\feh \approx +0.2\dex$) and the thin disc, defined as the stars tracing the low$-\alpha$ sequence, at least down to $\feh\approx -0.8\dex$, with hints pointing towards even lower values. Radial and vertical gradients in $\alpha-$abundances are found for the thin disc, with mild spatial variations in its $\Trend$ paths, whereas for the thick disc we do not detect any such spatial variations, in agreement with  results obtained recently from other high-resolution  spectroscopic surveys.}
%%%%%%%%%%%%%%%%%%%%%%%%%%%%%%%%%%%%%%
%Conclusions
%%%%%%%%%%%%%%%%%%%%%%%%%%%%%%%%%%%%%%
{The small variations in the spatial $\Trend$ paths of the thin disc do not allow us to distinguish between formation models of this structure. On the other hand, the lack of radial gradients and $\Trend$ variations for the thick disc indicate that the mechanism responsible for 
the mixing of the metals in the young Galaxy (e.g. radial stellar migration or turbulent gaseous disc) was more efficient before the (present) thin disc started forming.}

\keywords{Galaxy: abundances, Galaxy: disk, Galaxy: stellar content, Galaxy: evolution, Galaxy: dynamics and kinematics, stars: abundances}

\titlerunning{Characterisation of the  $\afe$ sequences in the Milky Way discs}
\authorrunning{G.~Kordopatis et al.}

\maketitle

%%%%%%%%%%%%%%%%%%%%%%%%%%%%
\section{Introduction}
The decomposition and separation of the Galactic disc into thin and
thick counter-parts, first suggested by \citet{Gilmore83}, has many
consequences in terms of Galaxy formation and evolution.
 Indeed, the differences in chemistry and kinematics between the stars of these two
structures put constraints on both the role of mergers in
the Galaxy in the last $8-12\Gyr$
\citep[e.g.:][]{Wyse01,Abadi03,Villalobos08} and on the
importance of internal evolution mechanisms such as radial migration,
dissipational cooling dependent on metallicity or clumpy turbulent
discs \citep[e.g.:][]{Sellwood02,Schonrich09b,Minchev13,Bournaud09}.

Among the literature, there is no unique way of selecting stars belonging to either of the discs. Often, kinematic criteria are used, but selections based on chemical composition  or spatial position are also employed, introducing potential biases in the distribution functions of the studied structures.  
Indeed, the first disc divisions, mainly based on star counts and density profiles, have found that the properties of the discs have an important overlap in kinematics and chemistry \citep{Gilmore89}.

The advent of multi-object spectroscopy and large spectroscopic surveys such as SEGUE \citep{Yanny09} or RAVE \citep{Steinmetz06},  confirmed these  facts and measured the different kinematics relations of the discs \citep[e.g.][]{Nordstrom04,Kordopatis11b,Kordopatis13a} and metallicities \citep[{[Fe/H]}$_{\rm thin}~\sim -0.1\dex$, {[Fe/H]}$_{\rm thick}~\sim -0.5\dex$, e.g:][]{Bensby07,Reddy08, Ruchti11,Kordopatis13c}. However, despite some specific attempts to characterise the wings of the distributions \citep[e.g.:][]{Norris85, Chiba00,Carollo10, Ruchti11, Kordopatis13c, Kordopatis15} the overlap in kinematics and metallicity for the discs implies that the shape and extent of their distribution functions remain still quite uncertain. The best discriminant parameter seems to be the stellar age,  the thick disc stars being consistently older than $9\Gyr$, \citep[e.g.:][]{Wyse88,Edvardsson93,Bensby05,Bergemann14, Haywood13}. Yet, stellar age is a parameter difficult to obtain, requiring either asteroseismic data, or parallax measurements combined with high-resolution spectroscopy.

In addition to age, the different formation histories of the two discs seems also
 strongly supported from the individual abundances measurements
 obtained for very local volumes of low mass FGK stars from very high-resolution spectra
 \citep[$R \geq 35\,000$, e.g.:][]{Fuhrmann98,
 Fuhrmann08,Fuhrmann11,Reddy06, Adibekyan12, Bensby14}. 
In particular, the study of %\citet{Fuhrmann98, Fuhrmann08,Fuhrmann11} and 
\citet{Bensby14}  used kinematic criteria to define the thin  and thick discs, and found that each of these structures followed different sequences in the parameter space formed by the abundance ratio of $\alpha-$elements over the iron, as a function of iron $(\afe-\feh)$, with potentially a gap between these sequences.
 These results support previous
conclusions that the thick disc stars were formed on a short
time-scale, with relatively little pollution of the interstellar
medium from supernovae of type Ia which are the main source of iron
\citep[e.g:.][and references therein]{Wyse88,Freeman02}. 

All of the
previously cited  high-resolution surveys were, however, limited in volume (typically only probing up to $\sim100\pc$
from the Sun) and often with complex selection functions combining metallicity and kinematics. % \citep[e.g.][]{Bensby14}. 
 Further high-resolution spectroscopic investigation
was therefore needed in order to assess whether this separation was a
global feature in the discs 
or a result of an observational bias. The gap could indeed disappear or become a trough farther away from the Sun, in which case the properties of the stars with intermediate $\alpha-$abundances would shed important light on the way the disc evolved chemically and dynamically.

\citet{Recio-Blanco14} used the first data-release (iDR1) of
the Gaia-ESO Survey \citep{Gilmore12}, analysing faint FGK stellar
populations with high resolution spectra ($R\sim 20\,000$) from
the FLAMES/GIRAFFE spectrograph, to identify these two distinct
sequences out to much larger volumes (up to distances of $5\kpc$ from
the Sun) and made a first attempt to describe their
chemo-dynamical characteristics. This analysis was pushed
further by \citet{Mikolaitis14}, who derived, for  the same
observed dataset, abundances of eight elements (Mg, Al, Si, Ca, Ti,
Fe, Cr, Ni, Y).  \citet{Mikolaitis14}
concluded, consistent with other analyses, that the gap
between the sequences is the most significant when considering
the [Mg/Fe] abundance ratio as a function of [Fe/H], and that
the two discs show no significant  intrinsic
dispersion in their $\alpha-$abundances at fixed [Fe/H].
This lack of scatter in the abundances along the metallicity
for both thin and thick disc indicates
first of all that the initial mass function (IMF) of the stars 
contributing to the $\afe$ ratio at low metallicities and the fraction of binary stars at the origin of supernovae type Ia, was constant
over the formation time of the stars they pre-enrich, 
%\query{also I think since see a decline, not a plateau, that the Type Ia progenitors were the same. Agreed? }.
 and in addition
that the interstellar medium (ISM) should be very well mixed in order
to have the same (small) dispersion at all metallicities.

 Using  only data for Red Clump stars  from the \emph{Apache Point Observatory Galactic Evolution
Experiment} (APOGEE), \citet{Nidever14} measured the $\Trend$ trend
for the thick disc and found that it had only very small 
variations with Galactocentric radius. Comparing with the results of  a simple Galactic chemical
evolution model, the authors concluded that the small amplitude, less than 10 per cent,  of the spatial
variations is possibly indicative of an essentially constant star formation
efficiency, and constant outflow rate balancing the star-formation rate, during the early stages of the Milky Way while the stars that now form the thick disk were formed. Their favoured model has a star-formation efficiency corresponding to a gas consumption timescale of $2\Gyr$, close to that
inferred for clumpy discs observed at redshift 2 \citep[e.g.][]{Tacconi13}.

We here use the FLAMES/GIRAFFE \emph{Gaia-ESO second internal data-release} (iDR2) with its increased statistics (roughly a factor of two on the number of observed targets compared to iDR1), its enhanced analysis (use of a wider wavelength range in the spectral analysis due to an improvement of the parameter determination pipelines) and better quality  (higher signal-to-noise ratio for the spectra, on average). 
We aim to characterise in a robust statistical way the chemical paths followed by the thin and thick discs, in velocity and space,  in order to provide a thorough insight on its possible origins. The understanding of the chemical path origins will eventually further constrain disc formation theories,  by indicating, for example,  whether the thick disc was formed entirely {\it in situ} or by the contribution of extragalactic material (gas and/or stars).

Section~\ref{sect:dataset} describes the dataset and the quality cuts that have been used in this work. Section~\ref{sect:gap} identifies in a novel manner the presence of the two disc populations by showing the way $\alpha-$abundances correlate with velocities. Section~\ref{sect:alpha_histograms} identifies the two populations in terms of star-counts and investigates how their properties change at different Galactic regions. Finally, Sect.~\ref{sect:conclusions} summarises and concludes the present study. 
 Note that in this paper we use the term \emph{metallicity} to designate both the ratio of the global metal abundance to that of hydrogen ($\meta$) and the ratio of  the true iron abundance to that of hydrogen ($\feh$). In general, the abundance pipeline for Gaia-ESO survey provides results such that $\meta \approx \feh$ (Recio-Blanco et al., in prep.), but we make explicit mention of one or the other parameter  when confusion could occur.

\begin{figure}
\centering
\includegraphics[width=0.8\linewidth, angle=0]{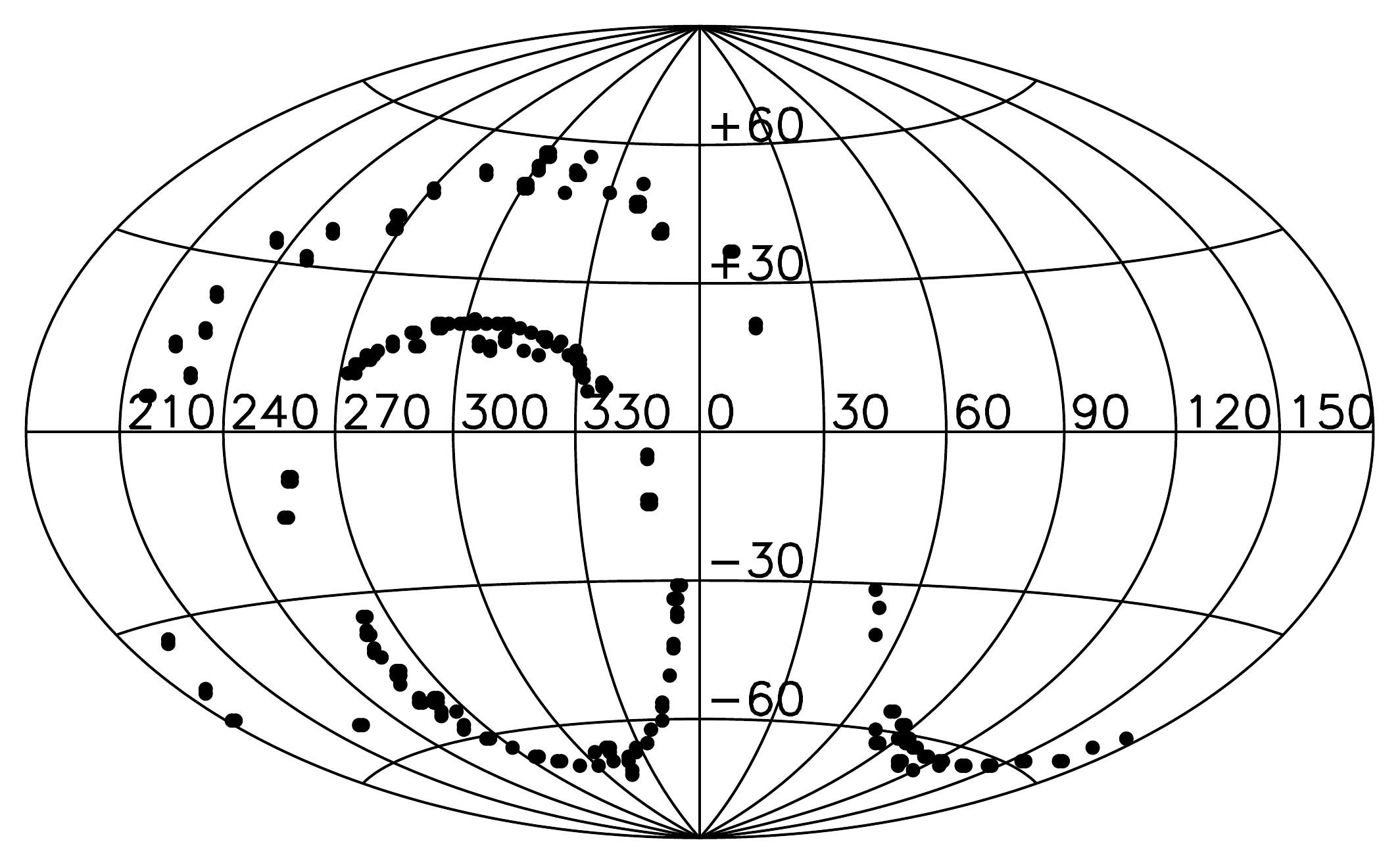}
\caption{Aitoff projection of the Galactic coordinates of the Gaia-ESO iDR2 targets used in this study.}
\label{fig:Aitoff}
\end{figure}

%%%%%%%%%%%%%%%%%%%%%%%%%%%%%%%%%%%%%%%%%%%%%%
\section{Description of the data: Gaia-ESO iDR2}
\label{sect:dataset}

We use throughout this paper the parameters derived from spectra obtained with the FLAMES/GIRAFFE  HR10 and HR21 settings (see Lewis et al., in preparation, for a description of the data reduction procedures). The Galactic coordinates $(\ell, b)$ of the observed targets are illustrated in Fig.~\ref{fig:Aitoff}. 
{
The source catalogue from which these targets are selected is the VISTA Hemisphere Survey (VHS, McMahon et al in prep.),  with the primary selection being: 
\begin{itemize}%{\labelitemii}{$\bullet$}
\item[$\bullet$]
$0.0 < (J-K) < 0.45$ for $14.0 <J<17.5$
\item[$\bullet$]
$0.4 < (J-K) < 0.70$ for $12.5 < J < 15$. 
\end{itemize}

Extinction was taken into account by shifting the colour-boxes by $0.5\times E(B-V)$, where $E(B-V)$  is the median \citet{Schlegel98} extinction in that line-of-sight.
If not enough targets were available within the colour cuts, additional
targets were  assigned by relaxing the red-edge of the colour-cut \citep[see][and Gilmore et al. in prep. for further details]{Gilmore12}.
}

We  adopted the recommended parameter values from the Gaia-ESO  analysis  for the effective temperature ($\teff$), surface gravity ($\logg$) and elemental abundances and the iron abundance we use is derived from  individual Fe~I lines.
The determination of the atmospheric parameters for the Gaia-ESO survey is described in Recio-Blanco et al. For the sake of completeness we give here a brief description of how these were obtained and refer the interested reader to the original reference for further details.  We note that throughout the paper, the signal-to-noise ratio (S/N) refers to the value achieved in the HR10 setup (centred at $548.8\nm$). For a given star, the S/N in the HR21 setting (centred at $875.7\nm$) is approximately twice  as large.

%%%%%%%%%%%%%%%%%%%%%%%%%%%%%%%%%%%%%%%
 \begin{figure*}
\centering
$\begin{array}{ccc}
\includegraphics[width=0.33\linewidth, angle=0]{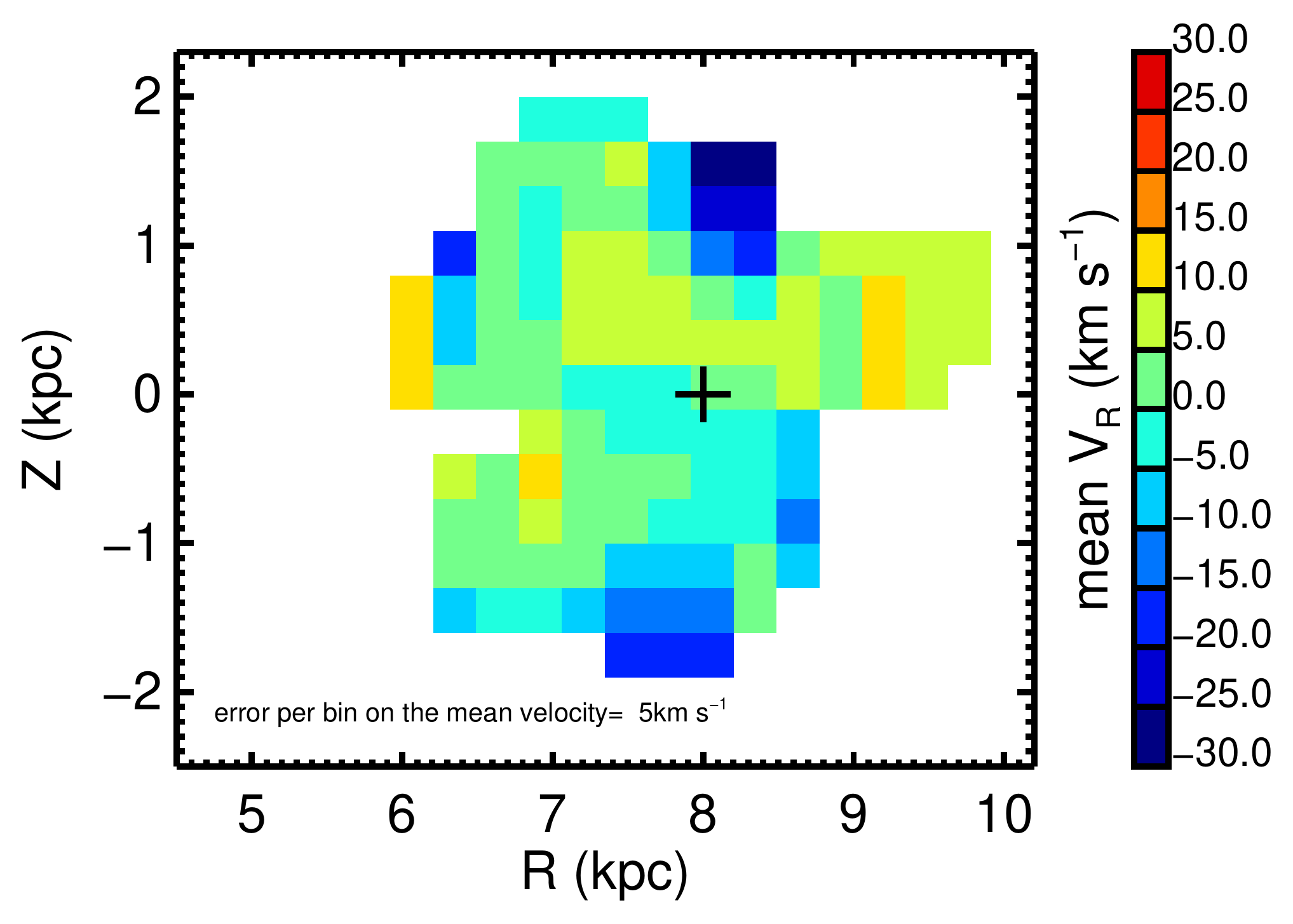}& 
\includegraphics[width=0.33\linewidth, angle=0]{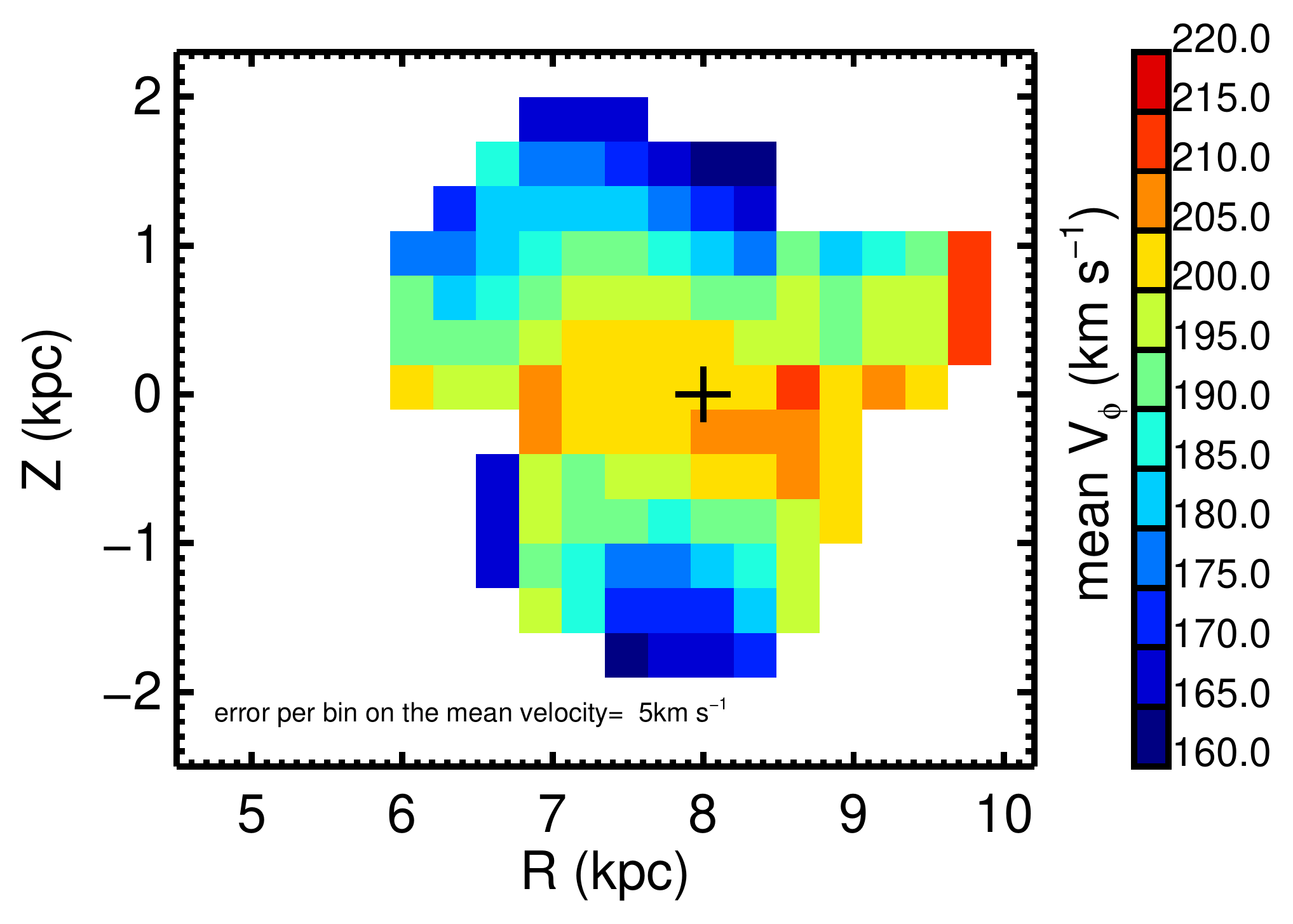} &
\includegraphics[width=0.32\linewidth, angle=0]{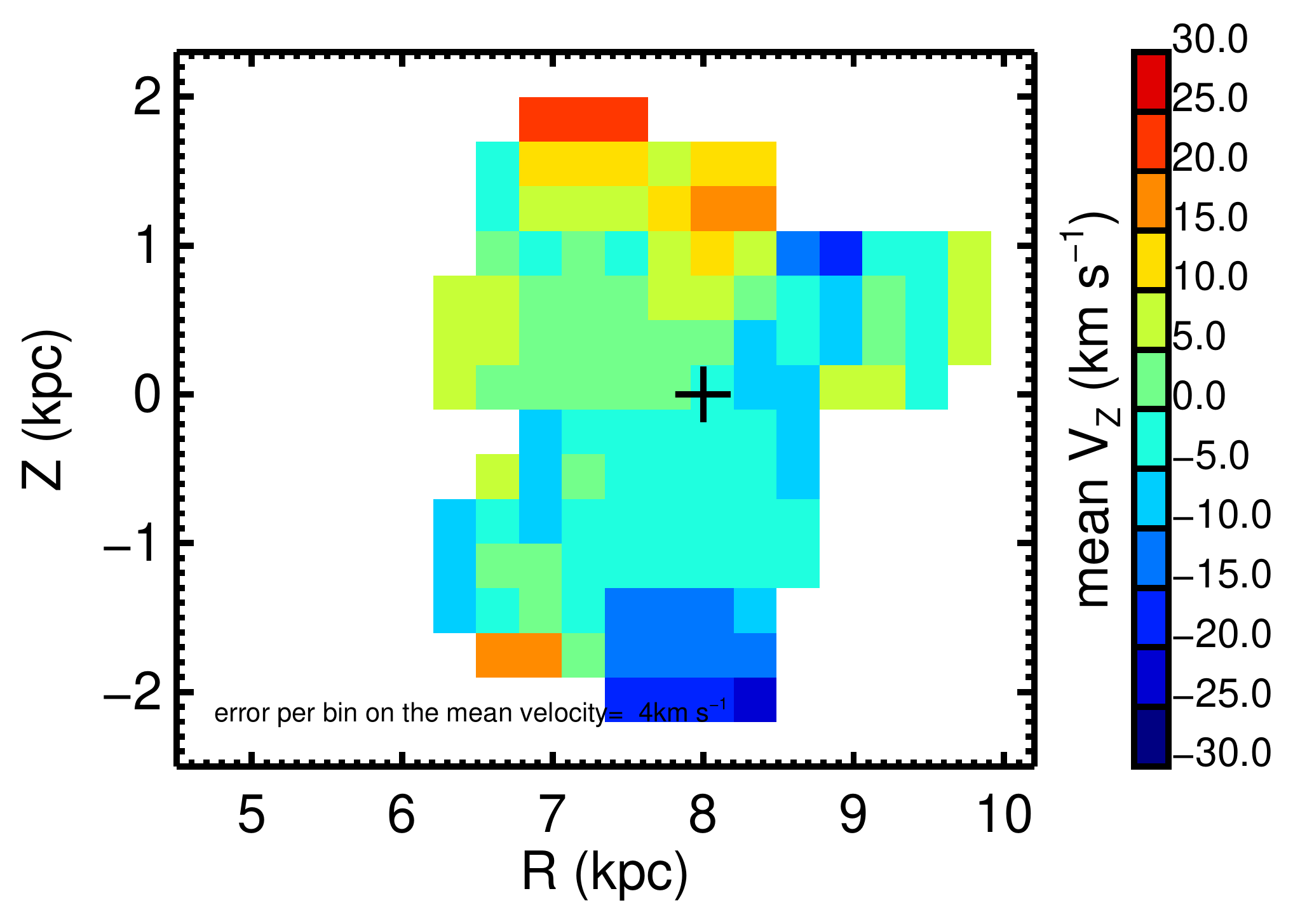}
\end{array}$
\caption{Mean radial (left), azimuthal (middle) and vertical (right) Galactocentric velocities of the FLAMES/GIRAFFE stars in the Gaia-ESO internal Data-Release\,2  (iDR2) as a function of their position in the Galaxy (in a cylindrical reference frame).  For each panel, only the stars having 
errors less than $65\kms$ in that particular component of space velocity are selected, resulting to 3162, 3667, 3243 targets, for $\vr$, $\vphi$ and $\vz$  respectively.
Mean values have been computed in bins of $0.4\kpc$ size,  provided each position-bin contained at least 15 stars, resulting to an error on the mean velocity of $5\kms$ per bin.}
\label{fig:velocity_map}
\end{figure*}

%%%%%%%%%%%%%%%%%%%%%%%%%%%%%%%%%%%%%%%
\subsection{Determination of the atmospheric parameters and abundances}
\label{sect:Param_determination}

{

The stellar atmospheric parameters  and abundances were determined by the members of the Gaia-ESO 
work package in charge of the GIRAFFE spectrum analysis for FGK-type stars. 
The individual spectra were  analysed by three different procedures:
\emph{MATISSE} \citep{Recio-Blanco06}, \emph{FERRE} \citep[][and further developments]{Allende-Prieto06}
and \emph{Spectroscopy Made Easy} \citep[SME,][and further developments]{Valenti96}.

Each of these procedures performed well with respect to APOGEE results and well-studied  open and globular clusters (i.e. the comparisons resulted in low-amplitude dispersions and  simple  biases).  
The random errors on the parameter values were reduced by putting  the results from each procedure on the same scale, then combining them. This was achieved by first choosing  the results of a given procedure as  the reference, and then  multi-linear transformations as a function of $\teff$ and $\logg$ were used to put the results from the two other procedures on the same scale as the reference one. 
The adopted transformations were of the form:
\begin{equation}
\Delta \theta_{p} = \theta_{p}- \theta_{r} = c_0 + c_1\cdot \mathrm{T}_{\mathrm{eff},p} + c_2\cdot \logg_p
\end{equation}
where $\theta$ is one of  $\teff$, $\logg$, $\meta$ or $\afe$, the subscript $r$ indicates the value from the  reference procedure, and the subscript $p$ indicates the value from either of the other  procedures  (see Recio-Blanco et al., in prep., for more details).

The resulting random errors for each method were fairly similar ($\sim 60\K$ for $\teff$, $0.14$ for $\logg$, $0.06$ for $\meta$ and $0.05$ for $\afe$) and,
as a consequence, the results for each star were combined by a simple average with the same weight given to each method. 
The averaged results were then calibrated with respect to 19 benchmarks stars \citep{Jofre14}; this corrected for systematic biases of $+83$\,K for $\teff$ and $+0.23\dex$ for $\meta$.

Once the final estimates of the values of the atmospheric parameters were obtained, the individual abundances, including the magnesium (Mg) 
and iron (Fe) abundances used in this work, were derived from three independent methods: 
an automated spectral synthesis method \citep{Mikolaitis14}, a Gauss-Newton method using 
a pre-computed synthetic spectra grid \citep{Guiglion13} and SME \citep{Valenti96}. In line with the
Gaia-ESO consortium analysis, all the abundance analysis methods used MARCS \citep{Gustafsson08} 
model atmospheres. In addition, the input atomic and molecular data were provided by the linelist group of Gaia-ESO,
who collated the most recent and complete experimental and theoretical data sources.
The comparison of the results of each method showed that a simple shift was enough
to put them all on the same scale. This shift was independent of both the atmospheric
parameters and  abundances,  and corrected for small  offsets with
respect to the Solar abundance (of order $\pm 0.05\dex$).
Finally, for each star and each element, the results
of the three methods were averaged with equal weight, and the elemental abundances relative to the Sun (denoted [X/Fe] for element X) 
were obtained using the \citet{Grevesse10} Solar elemental  abundances. 

The final mean (median) errors, corresponding to the parameter dispersion within the analysis nodes that have determined the atmospheric parameters and abundances  are: $70\K$ ($36\K$) for $\teff$, $0.14\dex$ ($0.08\dex$) for $\logg$, $0.08\dex$ ($0.06\dex$) for $\feh$ and $0.06\dex$ ($0.05\dex$) for $\mgfe$. 
}

%%%%%%%%%%%%%%%%%%%%%%%%%%%%%%%%%%%%%%%%%%%%
\subsection{Determination of  Galactocentric positions and velocities}
The absolute magnitudes of the stars have been obtained by projecting the atmospheric parameters ($\teff$, $\logg$, $\meta$) and  
VISTA ${\rm (J-K_s)}$ colour of the stars  
on the Yonsei-Yale isochrones \citep{Demarque04}, as described in \cite{Kordopatis11b,Kordopatis13a}, with the updates made in \cite{Recio-Blanco14} concerning the treatment of the reddening.
 The derived distances have a mean uncertainty of 15 per cent.  
They  were combined with the Galactic coordinates, line-of-sight velocities of the stars  and  the proper motions from the PPMXL catalogue \citep{Roeser10} in order to compute the Galactocentric positions and velocities in a cylindrical frame, with $\vr, \vphi$ and $\vz$ defined as positive with increasing $R,\phi$ and $Z$ (with the last towards the north Galactic pole). 
 Typical errors of the proper motions are $4-10\mas\yr$ \citep{Roeser10}, resulting to transverse velocity errors at $2\kpc$ of $37-94\kms$.
 %mu (arcsec /year) = Vt (km/s) / 4.74*d (pc)
 The adopted Solar motion  with respect to the Local Standard of Rest (LSR) is that of \citet{Schonrich10}, 
namely $\rm U_\odot=11.1, V_\odot=12.24, W_\odot=7.25\kms$. Finally, the LSR is presumed to be on a circular orbit with azimuthal velocity $\rm V_c=220\kms$, and we take the Sun to be located at $(\rm R_0, Z_0)=(8,0)\kpc$.
The associated errors on the Galactocentric positions and velocities have been obtained by running 5000 Monte-Carlo realisations on the distances, proper motions and radial velocities and by measuring the dispersion of the derived parameters.

The mean stellar velocities (after the quality selections, see
below) as a function of Galactocentric position are illustrated in
Fig.~\ref{fig:velocity_map}, with a binning of $0.4\kpc$. The maps
show consistent results with those already published  in the literature. 
 In particular,  we detect a mild radial gradient in the velocity component along the radial direction $\vr$, consistent with that found by  \citet{Siebert11}. This is visible  at negative Z,  going from $\sim 2 \kms$ at $R\sim 6.5\kpc$ to $\sim -5\kms$ at $9\kpc$  ($\vr$, left panel of Fig.~\ref{fig:velocity_map}). We also detect a compression/rarefaction pattern in the
vertical velocity component, seen as the difference in sign of $<\vz>$ at $Z=\pm1\kpc$ and $R=6,9\kpc$ (right panel, 
Fig.~\ref{fig:velocity_map}), consistent with the patterns identified in SEGUE G-dwarf data by \citet{Widrow12} and in RAVE data by \citet{Williams13} and \citet{Kordopatis13c}.  These motions may be induced by external perturbations such as a satellite galaxy \citep[e.g.][]{Widrow14} or internal perturbations such as spiral arms \citep[e.g.][]{Faure14}.

\begin{figure}
\centering
\includegraphics[width=0.9\linewidth, angle=0]{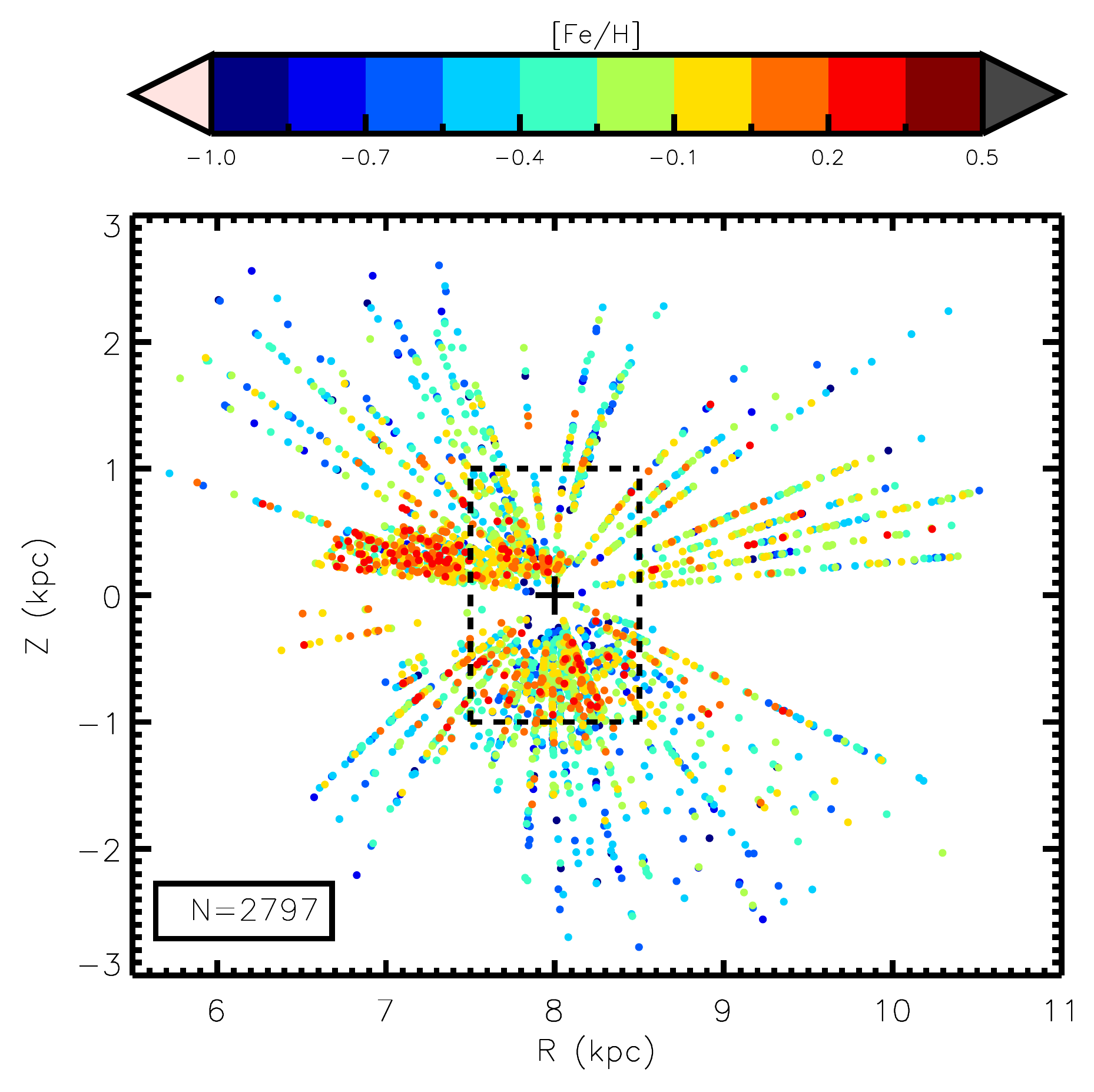}
\caption{Scatter plot of the iron abundances of the stars having each of the components of  their Galactocentric space velocity known better than $65\kms$ as a function of their measured Galactocentric radius and distance from the plane. The position of the Sun, at $(R_0,Z_0)=(8, 0)\kpc$, is indicated with a black `$+$' symbol. The dashed box represent the selection made for the ``local'' sample around the Sun (see Sect.~\ref{sect:gap} and Sect.~\ref{sect:Solar_cylinder}). Typical error on positions is 8 per cent.}
\label{fig:R_Z_plane}
\end{figure}

%%%%%%%%%%%%%%%%%%%%%%%%%%%%%%%%%%%%%
\subsection{Selection of subsamples based on quality cuts}
\label{sect:quality_sample}
The Gaia-ESO Survey observes three classes of targets in the
Galaxy \citep[field stars, members of open clusters and
calibration standards, see][]{Gilmore12}.  Our analysis in
this paper focuses on the Milky Way discs and therefore we
kept only the field stars, {\it i.e.} those stars  that are neither 
members of open clusters nor part of the calibration catalogue. 

Furthermore, in order to have a dataset with
reliable parameters, we excluded stars with spectra having a
signal-to-noise ratio (S/N) less than 10 in the HR10 setup (higher S/N
cuts have also been tested, without changing the overall conclusions
of this paper) or having the highest uncertainties in the values of the derived parameters (i.e. being at the tails of the error distributions). As a compromise  between accuracy and sample size, we set  the cuts for the errors in effective temperature  at $\sigma_{\teff} \geq 250\K$ (4\,per cent of the stars), in surface gravity at $\sigma_{\logg} \geq 0.5$ (3\,per cent of the stars), in iron abundance at $\sigma_{\feh} \geq 0.3\dex$ (0.5\,per cent of the stars), and in line-of-sight distance at $\sigma_D/D \geq 0.5$ (4\,per cent of the stars).

In addition, for each velocity component studied, we excluded those stars having errors in the given velocity greater than $65\kms$. 
 The errors in three-dimensional space velocity are dominated by the errors in 
the tangential velocities, so these rejection criteria lead to selection functions for each  velocity component that are dependent  on the stellar distances and the particular lines-of-sight (see Fig.~\ref{fig:R_Z_plane}). 
Indeed, stars close to the Galactic plane tend to have large errors on $\vz$, and stars towards the Galactic poles have poorly constrained $\vr$. 
The removal of  stars that had a derived error in a   given velocity component that was greater than $65\kms$, left us with   3162, 3667 and 3243 stars, for $\vr, \vphi$ and $\vz$, respectively.
The spatial distribution of the 2797 stars that have errors in \emph{all three} of  their velocity components below  $65\kms$  is illustrated on Fig.~\ref{fig:R_Z_plane}, colour coded according to iron abundance. One can see from this plot that the metal-rich population ($\feh>-0.2$~dex) is mostly found close to the plane, below $1\kpc$, as expected for a thin-disc dominated population.

\begin{figure}
\centering
\includegraphics[width=0.99\linewidth, angle=0]{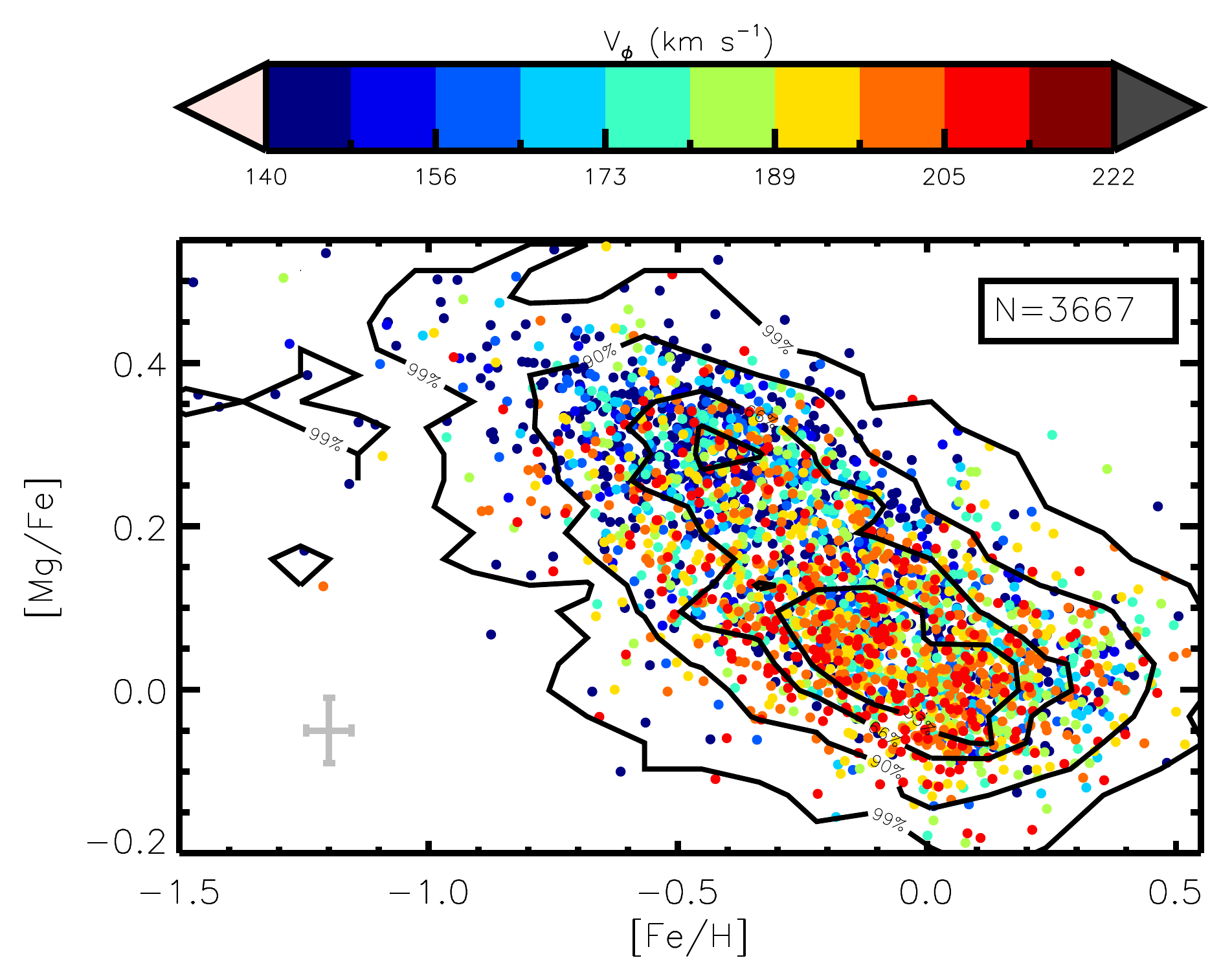}
\caption{$\mgfe$ as a function of $\feh$ for the Gaia-ESO iDR2 stars with S/N$>10$. The mean error bar is represented in grey at the bottom left corner. The range of the plot has been truncated to $-1.5\leq\feh\leq0.55\dex$. The contour lines  are drawn for 33 per cent, 66 per cent, 90 per cent and 99 per cent  of the sample.}
\label{fig:Mg_Feh}
\end{figure}

\begin{figure}
\centering
\includegraphics[width=0.99\linewidth, angle=0]{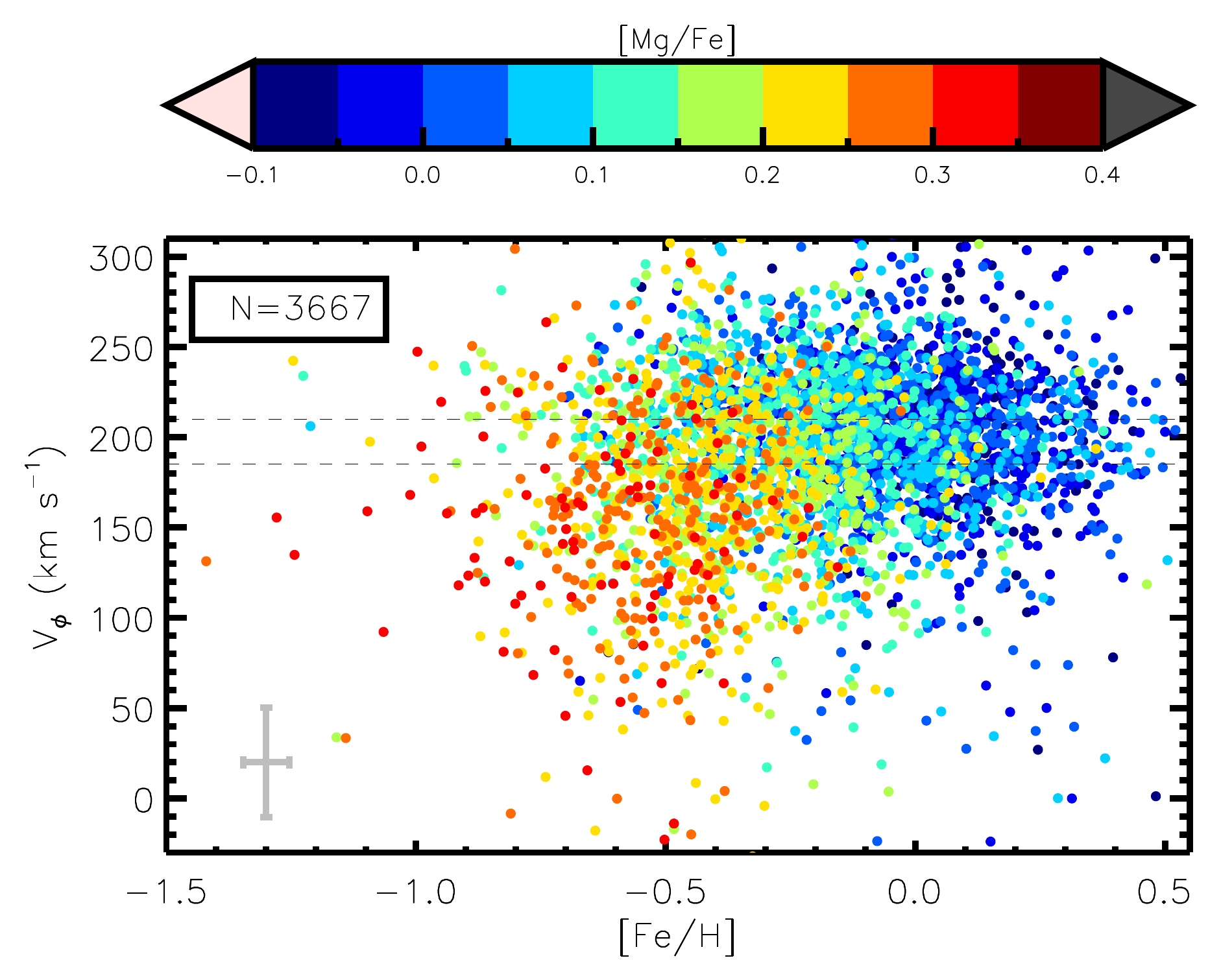}
\caption{Azimuthal velocity as a function of $\feh$, colour-coded according to the $\mgfe$ abundance (the velocity and metallicity ranges have been truncated for visualisation purposes). Dashed horizontal lines are plotted at $\vphi=210\kms$ and $\vphi=185\kms$. Typical errors on $\vphi$ and $\feh$ are represented at the bottom left corner, in grey. }
\label{fig:Mg_vphi}
\end{figure}

Figure~\ref{fig:Mg_Feh} shows the scatter plot of the azimuthal velocities of the 3667 stars that passed the error cut in this velocity component, in the chemical space defined by magnesium and  iron. Following \citet{Mikolaitis14}, magnesium is used as the $\alpha-$element tracer in the remainder of this paper. 
The gap suggested from the  analyses of \citet[][using the HARPS spectrograph]{Adibekyan12}, \citet[][using mainly the MIKE spectrograph]{Bensby14} and \citet[][using the high S/N sample of Gaia-ESO iDR1]{Recio-Blanco14}  is seen in Fig.\,\ref{fig:Mg_Feh} as a trough, due to the low S/N threshold  (S/N$> 10$) we imposed for the analysis at this stage (median S/N$=25$; see Sect.~\ref{sect:mock_tests} for an analysis of mock data illustrating how a real  gap can get populated by noisy data and Sect.~\ref{sect:Solar_cylinder} for an analysis using higher S/N cuts on the Gaia-ESO iDR2 data). 
Indeed, the two populations  -   one  low$-\alpha$ ($\mgfe\lesssim 0.2\dex$) on  orbits with higher azimuthal velocities (mean $\overline{\vphi}\sim210\kms$), and one  high$-\alpha$ ($\mgfe \gtrsim 0.1\dex$) on orbits with lower azimuthal velocities ($\overline{\vphi}\sim185\kms$) -  can be identified with the colour-code. These two populations are more clearly seen in Fig.~\ref{fig:Mg_vphi}, representing $\vphi$ as a function of $\feh$, colour-coded according to the $\mgfe$ abundance. The fact that the high$-\alpha$ population has a larger velocity dispersion is also evident.

Given their different global kinematic properties, throughout this paper the high$-\alpha$ and low$-\alpha$ populations will be referred to as the thick and thin discs, respectively.  
In the next section, we explore the metallicity ranges up to which these two populations extend, by investigating their kinematics, but without decomposing the stellar populations.

%%%%%%%%%%%%%%%%%%%%%%%%%%%%%%%%%%%%%%%%%%%
\section{The local disc dichotomy in terms of Galactocentric velocities}
\label{sect:gap}
In this section we aim to investigate, for the stars in the
  greater Solar neighbourhood ($7.5 \leq R \leq 8.5\kpc$ , $|Z|<1\kpc$) - the Solar suburbs -   the
 metallicity range over which the disc(s) can be
 decomposed into at least two different populations,
 defined by their correlation between kinematics and chemical
 abundances.  This we achieve with no {\it a priori} assumption
 about the shapes of their metallicity distribution
 functions (DF) or their kinematic DFs.  Kinematics provide the 
 means to discriminate among models for the formation and
 evolution of the discs, and the way in which metallicity and
 chemical abundances correlate with kinematics carries important and
 fundamental information on the nature of the stars in our sample \citep*[e.g.][]{Eggen62,Freeman02, Rix13}.  For  the
 remainder of this paper, we define a stellar population by demanding that its member stars show the same
 correlations between chemistry and kinematics over the whole
 range of metallicity that the population spans.  The approach
 presented below is complementary  to, and independent of, that of
 Sect.~\ref{sect:alpha_histograms}, where we fit the $\alpha-$abundances 
 in different metallicity bins in order to trace the evolution
 of the trends for both discs, defined purely by chemistry.

  \subsection{The greater Solar neighbourhood sample}

The disc stars in the greater Solar neighbourhood are selected by requiring  distances close 
to the Galactic plane ($|Z|<1\kpc$), Galactocentric radii between $7.5$ and $8.5\kpc$ ({\it i.e.} distances from the Sun closer than $1.12\kpc$) and azimuthal velocities greater than $0\kms$. 
These selections ensure that:\\

 {\it (i)} The radial and vertical gradients in the discs have minimal effects on  the interpretation of the results. \\

 {\it (ii)} The sample under consideration is as homogeneous as possible towards all the lines-of-sight (see Fig.~\ref{fig:R_Z_plane}). Excluding stars farther than $1\kpc$ from the plane should not particularly bias the kinematics of the selected populations (other than the bias coming from possible  vertical gradients in $\vphi$), provided that the populations are quasi-isothermal in vertical velocity dispersion, which appears to be a reasonable approximation 
 \citep[e.g.:][]{Flynn06,Binney12b,Binney14b}.\\

 {\it (iii) } The contamination by inner halo stars is reduced, in order to focus on the thin and thick disc chemical dichotomy. The adopted azimuthal velocity threshold of  $0\kms$ allows the retention of  metal-weak thick disc stars, given an expected azimuthal velocity distribution for the thick disc that extends  down to at least $60-80\kms$
 \citep[{\it i.e.} $\sim ~ 0\kms$ with a $65\kms$ error, see ][]{Kordopatis13c,Kordopatis13a}. This threshold  simultaneously excludes counter-rotating stars likely to be members of  the halo \citep[in any case, the contribution from the halo should be small,  given the metallicity range and distances above the plane investigated; see for example,][]{Carollo10}.  Note that in this work we treat the metal weak thick disc as part of  the canonical thick disc; whether or not the chemo-dynamical characteristics of the metal-weak thick disc are distinct will be investigated in a future paper. \\

% however, that because it is difficult to disentangle the metal-weak thick disc from the inner halo stars, the chemo-dynamical characteristics of the metal weak thick disc are not investigated specifically in this paper and that they will be studied in a follow-up work.} \\

  \subsection{Description of the method}
  We assess the presence of two  distinct populations in chemical space by measuring the mean velocities ($\mathrm{\overline{V_i}}$) and their dispersions ($\sigma_{V_i}$) for different $\feh$ bins, in addition to  the correlation(s) between the means of each of the velocity components  and the stellar $\alpha-$enhancements (noted simply $\rm \partial V_i / \partial \mgfe$).
While the evolution of the mean velocities as a function of metallicity has long been used to identify the transition between the thin and thick discs \citep[e.g.:][]{Gilmore89,Freeman02,Reddy06}, here the novel approach is to see whether there is a difference in the mean velocities at a given $\feh$, as a function of $\mgfe$. 
Non-zero values for $\rm \partial V_i / \partial \mgfe$  can arise due to the following:

\begin{enumerate}
\item
The estimated values of the atmospheric parameters, distances and velocities have correlated errors and therefore biases.
\item
At least one population has a velocity gradient and/or chemical gradient as a function of Galactocentric radius and/or vertical distance, resulting in different kinematic properties in a given metallicity bin. 
\item
Only one population has an intrinsic correlation between chemistry and kinematics, as  has previously been proposed for the thick disc in previous studies \citep[$\partial \vphi / \partial \feh \approx 50 \kms\dex^{-1}$, e.g:][]{Kordopatis11b, Kordopatis13c}.
\item
 There are at least two different populations with distinct kinematics and chemistry. 
\item
A combination of  some or all of the above. 

\end{enumerate}

 Possibility (1) is not likely. Indeed, as noted earlier  (Sect.~\ref{sect:Param_determination}), the calibration using benchmark stars removed the known biases in the stellar parameters, while  Recio-Blanco et al. (in prep.) found no  significant  
 degeneracies between the parameter determinations that could have led  to artificial correlations between the parameters.
 Nevertheless,  hidden biases may still exist. 
 For example, a bias in metallicity,  effective temperature or surface gravity could affect the final distance estimation and thus the final Galactocentric space velocity. Biases of less than $0.1\dex$ in $\meta$ or less than $100\K$ in $\teff$ (the maximum possible values for this  analysis), could change the distances up to 10 per cent \citep[see][their Table~1, for further details]{Schultheis15}.  At $1\kpc$, this would translate to changes in velocities of $\sim 5\kms$(assuming proper motions of $10\mas\yr$),
  which is too small to have a significant effect on  our analysis.

  %%%%%%%%%%
 %mu (arcsec /year) = Vt (km/s) *  / (4.74 * d (pc) )
% 4.74*2=9.48 kms
 %4.74*2.2=10.428

Concerning possibility (2), the amplitude of chemical gradients in the thin disc has been thoroughly discussed in the literature, with evidence from a variety of stellar tracers for mild radial and vertical metallicity gradients, of the order of $\partial \feh / \partial R \approx -0.04$ to  $-0.07\dex\kpc^{-1}$ and $\partial \feh / \partial Z \approx -0.05 $ to $-0.1 \dex\kpc^{-1}$ \citep[e.g.][]{Gazzano13,Boeche13,Recio-Blanco14, Anders14}. The thick disc shows no evidence for a radial metallicity gradient \citep[e.g.:][]{Cheng12, Bergemann14,Boeche14, Mikolaitis14} and at most a mild vertical metallicity gradient \citep{Ruchti10,Kordopatis11b,Kordopatis13a,Mikolaitis14}. The $\afe$ gradients in both discs are also small ($\lesssim \pm 0.02\dex\kpc^{-1}$),  with the actual amplitude  and even the sign being still a matter of debate \citep[see Sect.\,5 of][]{Mikolaitis14}. 
As far as the kinematic trends are concerned, neither the radial gradient in $\overline{\vr}$ \citep[$\rm \partial \vr / \partial R \approx 3\kms\kpc^{-1}$,][]{Siebert11} nor the ``compression-rarefaction'' patterns reported in $\overline{\vz}$ in both $\rm R$ and $\rm Z$ \citep{Williams13,Kordopatis13b}, are  strong enough to contribute significantly within the  volume of the greater Solar neighbourhood.
 A similar conclusion holds for $\overline{\vphi}$, given the studies of \citet{Lee11} and \citet{Binney14b} based, respectively, on the SDSS-DR8 \citep{Aihara11} and RAVE-DR4 \citep{Kordopatis13b} catalogues for stars located roughly in the same volume as the present study.

The difference in the chemo-dynamical signature between that due to a single
population having an intrinsic correlation between metallicity and
kinematics,  possibility (3) above, or that due to the superposition of several populations each with distinct correlations between 
metallicity and kinematics (possibility (4)) is more subtle. It can however be seen in the
variation of the gradient of kinematics with chemistry, such as $\rm \partial V_i / \partial \mgfe$  with
metallicity. Indeed, a single population, as defined at the
beginning of this section, should exhibit no trend in $\rm
\partial V_i / \partial \mgfe$ as a function of  metallicity.  
A combination of more than one population would produce variations of this quantity if  the 
relative proportions of the populations changes with $\feh$.  Our 
investigation of the extent of the superposition of the stellar
populations using this approach therefore differs from
those previously published in the literature, since here we make no assumptions about the number of these populations that may be present, nor the shape of their
distribution functions \citep[see for example][for a decomposition of
the disc into multiple mono-abundance populations]{Bovy12b}.

\begin{figure*}
\centering
\includegraphics[width=0.75\linewidth, angle=0]{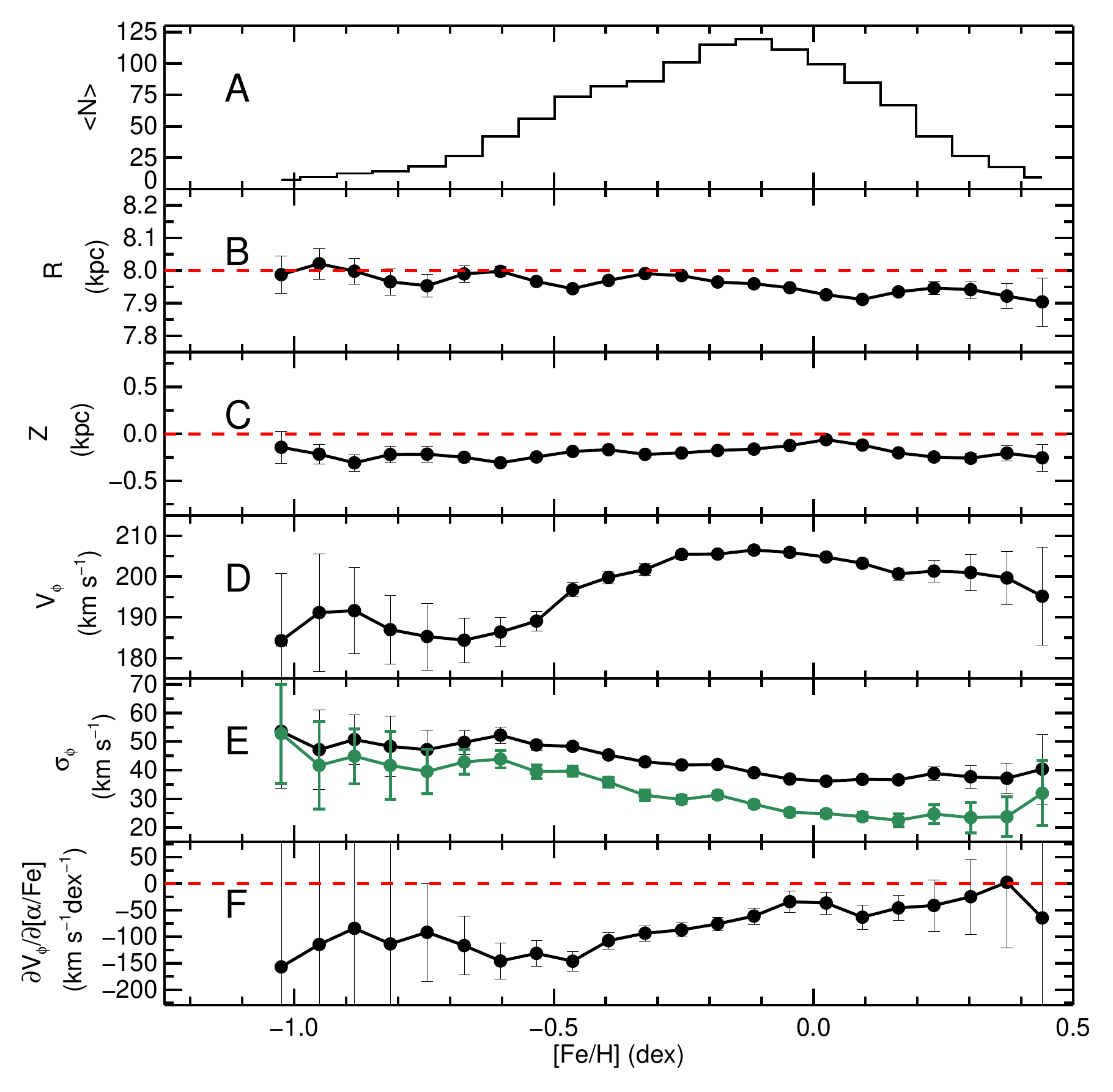}
\caption{Mean number of stars (A), Galactocentric radii (B), distances from the Galactic plane (C), azimuthal velocities (D), azimuthal velocity dispersions (E) as measured from the data (in black) and corrected by the error measurements (in green),  at different metallicity bins, for stars located within the greater  Solar neighbourhood. Panel F shows the correlations between the mean azimuthal velocity and the $\alpha-$abundance ratios for different metallicity bins. The error bars represent the $3\sigma$ Poisson noise averaged over the Monte-Carlo realisations. The bins are $0.07\dex$ wide, chosen as a compromise between the number of stars inside each bin and an adequate coverage of the metallicity range.}
\label{fig:vphi_correlations}
\end{figure*}

\begin{figure*}
\centering
\includegraphics[width=0.75\linewidth, angle=0]{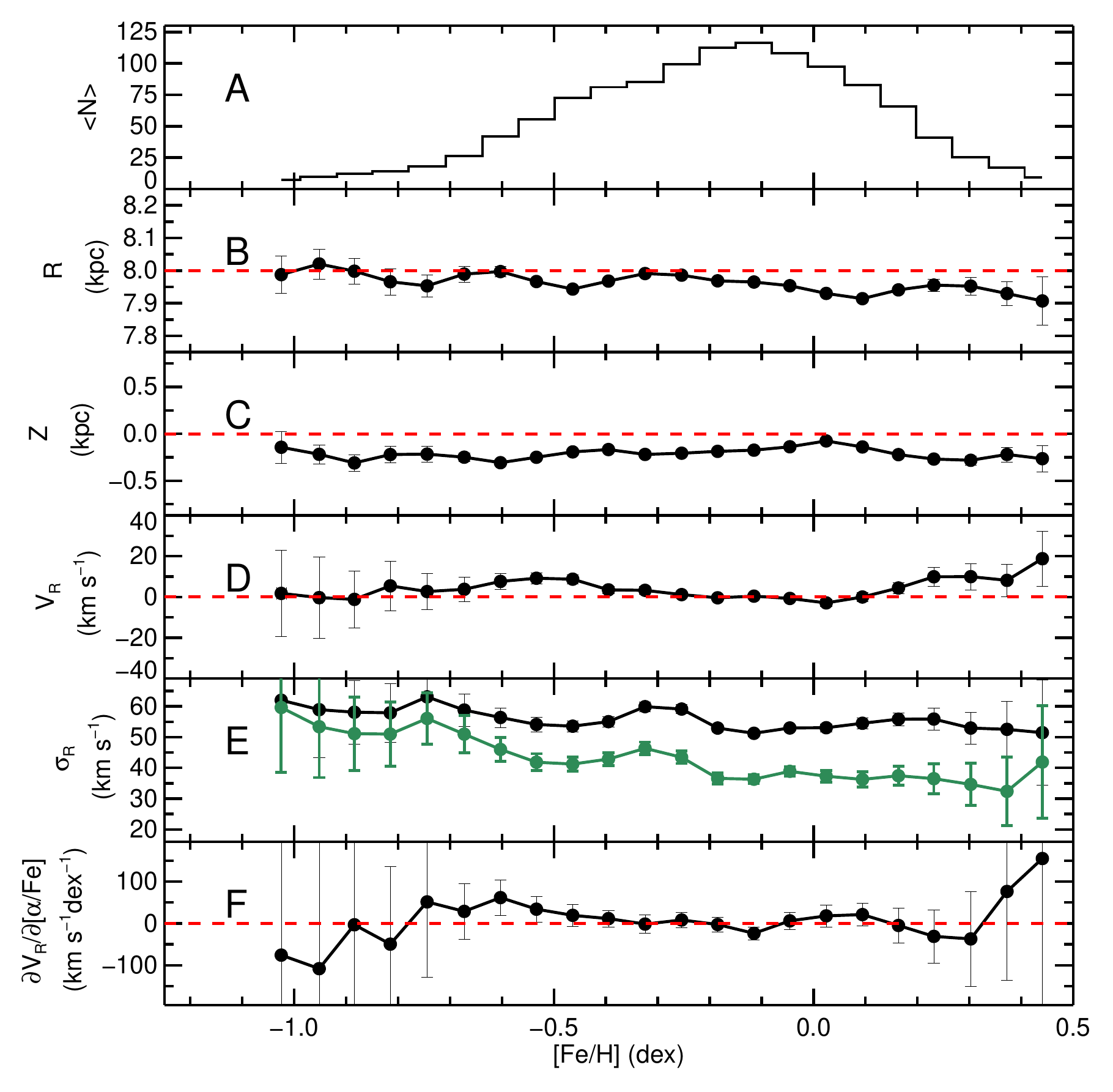}
\caption{Same as Fig.~\ref{fig:vphi_correlations} but for the radial velocity component, $\vr$.}
\label{fig:vr_correlations}
\end{figure*}

\begin{figure*}
\centering
\includegraphics[width=0.75\linewidth, angle=0]{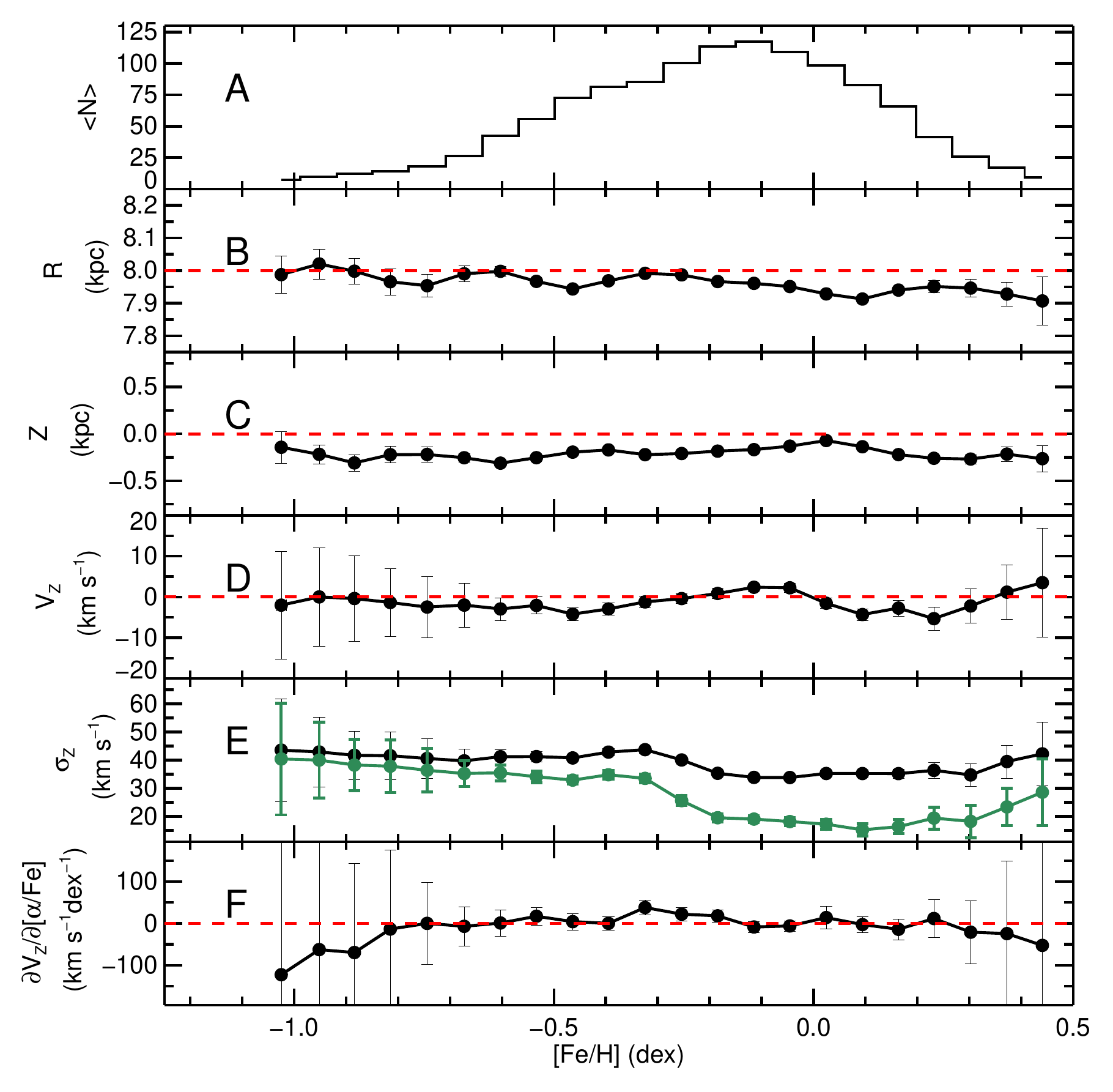}
\caption{Same as Fig.~\ref{fig:vphi_correlations} but for the vertical velocity component, $V_Z$.}
\label{fig:vz_correlations}
\end{figure*}

 Figures~\ref{fig:vphi_correlations}, \ref{fig:vr_correlations}, \ref{fig:vz_correlations} and \ref{fig:Allv_correlations} show the results obtained for the azimuthal ($\vphi$), radial ($\vr$) and vertical ($\vz$) velocity components of the stars. 
Each of these figures represent the mean number of stars per iron abundance bin\footnote{Figures~\ref{fig:vphi_correlations}, \ref{fig:vr_correlations}, \ref{fig:vz_correlations} each contain different number of targets, according to the selection criteria performed based on the velocity errors. The distributions (panels A) are not corrected for the selection function.}  
  (panels A), their mean radial distance from the Galactic centre (panels B), their distance from the plane (panels C), their mean velocity and dispersion (panels D and E), and finally the correlation between the mean velocity and the $\alpha-$abundances (panels F  and Fig.\,\ref{fig:Allv_correlations}).  
They have been obtained with 100 Monte-Carlo realisations on positions, velocities and iron abundance, and by keeping the iron bin boundaries at fixed values.
 Stars can therefore move between bins from one  realisation to the next, as parameter values change with each realisation.

After each Monte-Carlo realisation, the mean velocities and $1\sigma$ velocity dispersions of the stars within each bin were measured, provided that bin contained at least 15 stars. 
 As far as the $\rm \partial V_i / \partial \mgfe$ values are concerned, they have been obtained by fitting a line in the $\rm V_i$ vs $\mgfe$ parameter space and by measuring the slope of that line.
The final %velocity 
values inside each bin were then obtained by computing the average over the realisations. The associated error bars were obtained by dividing the standard deviation of the means by the square root of the mean number of stars inside each bin. 
Finally, the corrected velocity dispersions (green lines in Figs.~\ref{fig:vphi_correlations}, \ref{fig:vr_correlations} and \ref{fig:vz_correlations}) were obtained by removing quadratically two times the mean velocity error of the stars inside each bin,  the factor of two arising since the errors are taken into account twice: first, intrinsic to the measurement, and then again during the Monte-Carlo realisations \citep[see also ][]{Kordopatis13c}. We therefore used 
the following relation to correct the velocity dispersions for observational uncertainties:
\begin{equation}
\label{eqn:dispersion_errors}
\sigma_{V_i}=\sqrt{\sigma_{V_i}^{*2 }- 2 \times mean(error_{V_i})^2}
\end{equation} 
where  $\sigma_{V_i}$ represents  the corrected velocity dispersion of $\vr,\vphi$ or $\vz$, and $\sigma^*_{V_i}$ is the corresponding measured velocity dispersion (averaged over the Monte-Carlo realisations). We refer the reader to Guiglion et al. (submitted) for a thorough analysis of the velocity dispersion of the disc stars as a function of the chemical composition.

 %%%%%%%%%%%%%%%%%%%%%%%%%%%%%%%%%%%%%%%%%%
 \subsection{Azimuthal velocity component}
 \label{sect:vphi_analysis}
 
 The identification of the thin and thick disc through the analysis of the stellar kinematics is most cleanly achieved using the azimuthal velocities of the stars, because of the asymmetric drift in the mean kinematics. 
Indeed, from the trends of the mean  $\vphi$ velocity (Fig.~\ref{fig:vphi_correlations}, panel D), one can identify the thin disc component, with $\overline{\vphi} \sim 205\kms$, dominating the star counts for metallicities above $-0.4\dex$. The thick disc, virtually unpolluted by the halo (due to the imposed cuts: $|Z|<1\kpc$ and $V_\phi>0\kms$), is distinguishable below metallicities of $-0.5\dex$,  where there is a decrease in the mean azimuthal velocities.
The velocity dispersion (Fig.~\ref{fig:vphi_correlations}, panel E) at  metallicities above $-0.2\dex$  has the value  $\sigma_\phi \sim 25\kms$.  Below $-0.5\dex$, this increases to $\sigma_\phi \sim 45\kms$. These values  agree with the estimates of the velocity dispersions of the old thin and thick discs as measured far from the Solar neighbourhood \citep{Kordopatis11b},  those measured  for stars in the RAVE survey within roughly the same volume \citep{Binney14b},  and other values available from  the literature \citep[e.g.:][]{Soubiran03,Lee11}.

\begin{figure}[th]
\centering
\includegraphics[width=0.9\linewidth, angle=0]{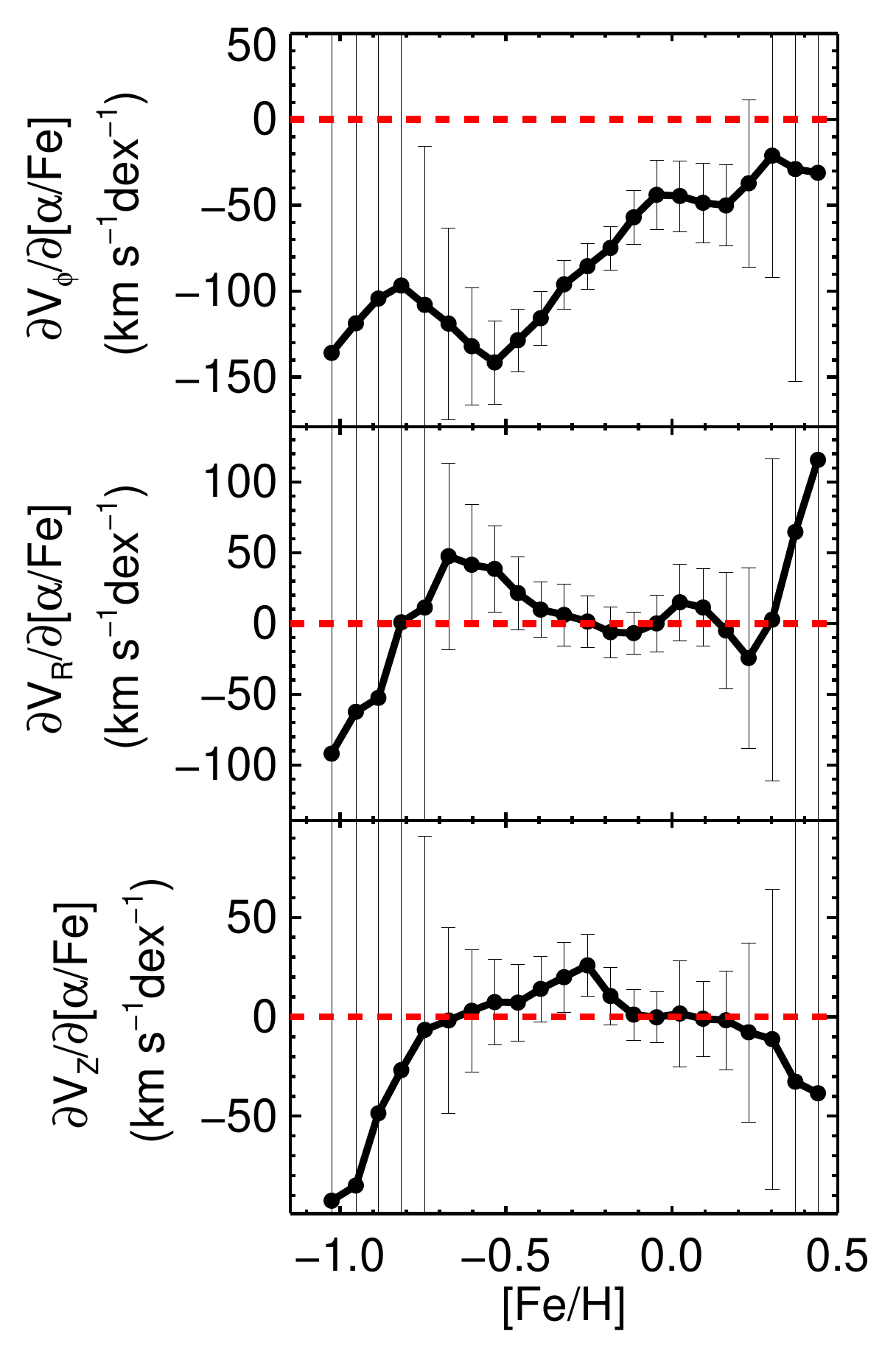}
\caption{ Same as panels F of Figs.~\ref{fig:vphi_correlations} to  \ref{fig:vz_correlations}, but put side by side to facilitate comparison of the trends. The plots have been smoothed  with a boxcar average taking into account the closest neighbour.}
\label{fig:Allv_correlations}
\end{figure}

As far as  the correlations between $\vphi$ and $\afe$ are concerned, (Fig.~\ref{fig:vphi_correlations}, panel F and Fig.~\ref{fig:Allv_correlations}, top plot), the behaviour of $\partial \vphi / \partial \mgfe$ with increasing metallicity shows three regimes:
\begin{itemize}
 \item a value close to zero at super-solar metallicities,
 \item a non-zero value of $\partial \vphi / \partial \mgfe \approx -120\kms\dex^{-1}$ below $-0.5\dex$,  compatible  with little variation within the errors,
 \item an intermediate regime  between metallicities $[-0.5,+0.2]\dex$,  where the amplitude of  $\partial \vphi / \partial \mgfe$ monotonically increases with metallicity. We interpret this variation as indicative of a varying mixture of populations.
  \end{itemize}
  
  We note that these trends are not depending on the adopted binning in metallicity, since they are also recovered when adopting, for example,  overlapping  bins $0.07\dex$ wide, spaced by $0.05\dex$.
  Following from  the discussion at the beginning of
 Sect.~\ref{sect:gap}, these three regimes suggest that the relative
 proportions of the different stellar populations present in our
 sample do not vary significantly  above $+0.2\dex$, since we find no trend of $\partial \vphi / \partial
 \mgfe$ with metallicity in this range. 
 Similar conclusions can be made below $-0.5\dex$, although  the larger error bars in this regime could hide weak trends. 
  In terms of separate thin and thick discs\footnote{ Note that  the metal-weak thick disc is not treated separately from the canonical thick disc, and that its chemo-dynamical properties will be thoroughly investigated in a future paper.}, 
 the
 trends we find  imply that the metallicity distribution of the thick disc extends at
 least up to Solar values, and that of the thin disc goes down to at least  $-0.5\dex$. This overlap in metallicity  was already suggested from the
 $\alpha-$abundance trends derived from very-high resolution studies of
 stars in the immediate Solar neighbourhood  \citep[$d <100\pc$, e.g.][]{Reddy06, Adibekyan13, Bensby14,Nissen15}, and by \citet{Nidever14} for stars in 
 a wide range of Galactic radii ($0\lesssim R \lesssim15\kpc$).

We can further assess the presence of thin disc stars  below $\feh=-0.5\dex$ by making a \emph{rough} separation of the stars within a given $\feh$ bin 
into either high$-\alpha$ ($\mgfe >0.3\dex$) or low$-\alpha$ ($\mgfe < 0.2\dex$,  see also Sect.\,\ref{sect:MWThinDisc}) candidates.
This results in the low$-\alpha$ population having  a median $\vphi \sim 212 \kms$, whereas the median $\vphi$ of the high$-\alpha$ population is $\sim 165\kms$. 
The low$-\alpha$ population has a lag compared to the LSR  that is typical of the thin disc, and the lag for the high$-\alpha$ population is  consistent with estimates for the  canonical and metal weak thick disc, provided there is a correlation between $\vphi$ and $\feh$ of $50\kms\dex^{-1}$, as found by \citet{Kordopatis13b}.  We therefore conclude that the thin disc metallicity distribution extends down to at least $\feh=-0.8\dex$. Below that metallicity, the number of stars is too small to allow us to derive a  reliable estimate of the mean velocity.
The relative proportion of thin disc stars at low metallicities will be discussed further in Sect.\,\ref{sect:alpha_histograms}. 

A natural outcome of the existence of thin disc stars below $\feh < -0.5\dex$  is that the non-zero  $\partial \vphi / \partial \mgfe$ value that we measure  is at least partly due to the mixture of thin and thick disc.
However, within the errors, Fig.~\ref{fig:Allv_correlations} shows a flat behaviour below  $\feh \lesssim -0.5\dex$, suggesting a constant fraction of thick/thin disc.  This result will also be found when separating the stars according to their $\alpha-$abundances. 
Indeed,  while we cannot eliminate the possibility that there exists an intrinsic correlation between $\vphi$ and $\afe$ for the thick disc stars,  as suggested by \citet{Recio-Blanco14}, we note that the differences that we find in the median $\vphi$ values for $\mgfe \sim 0.35\dex$  and $\mgfe \sim 0.1\dex$ ($165\kms$ and $212\kms$, respectively), could be sufficient to explain the measured value of $\partial \vphi / \partial \mgfe \approx -120 \kms\dex^{-1}$, provided a fixed fraction of thick disc stars over thin disc stars of approximately 60 per cent. 
This ratio is in good agreement with the local thin disc / thick disc ratio of 3:2 for the metallicity bin $[-0.8, -0.6]$ derived by \citet{Wyse95} through the combination of a volume-complete local sample ($d\lesssim 30\pc$) and an {\it in situ} sample at $Z \sim 1\kpc$ (after scaling with appropriate density laws and kinematics).

%%%%%%%%%%%%%%%%%%%%%%%%%%%
\subsection{Radial and vertical velocity components}
Due to the fact that both the radial and vertical  velocity distribution functions are centred at zero  for both discs\footnote{The  compression/rarefaction patterns in the kinematics of the discs lead to  only  small deviations from zero means, see \citet{Siebert11a,Williams13, Kordopatis13c}.}, the thin/thick disc dichotomy is harder to identify with these velocity components.  
The investigation of the mean velocities as a function of metallicity confirms indeed  this fact 
(see panel D of Fig.~\ref{fig:vr_correlations} and Fig.~\ref{fig:vz_correlations}), where we find mean velocities centred around $0\kms$.

The velocity dispersions (panels E) of the most metal-poor stars are
found to be $(\sigma_R, \sigma_Z)\approx (50,40)\kms$, typical
of thick disc values, while  the most metal-rich stars have dispersions more typical of the thin disc, being
$(\sigma_R,\sigma_Z)\approx (30,20)\kms$ 
 \citep[e.g.:][]{Soubiran03, Lee11,Bovy12,Sharma14}. In
particular, the abrupt change in the slope $\partial \sigma_{Z} /
\partial \feh$ for metallicities $\gtrsim -0.4\dex$ suggests that
this is the metallicity threshold above which the thin disc becomes
the dominant population. This point will be confirmed in
Sect.~\ref{sect:Solar_cylinder}, where the discs are separated in
chemical space only.

As far as the $\rm \partial V_{R,Z} / \partial \mgfe$ trends are concerned, we find them being consistent with zero at least within the metallicity range $[-0.8,+0.2]\dex$. 
 The results using $\vphi$ showed that the relative proportion of the populations is changing across this metallicity range. Therefore, the  invariant, null value for $\rm \partial V_{R,Z} / \partial \mgfe$ implies that all populations present in our sample share common radial and vertical velocity  distributions, at least in the mean.

An investigation of the detailed trends of the velocity dispersions is outside the scope of this paper, but we did investigate the specific case of  the behaviour of $\partial \sigma_{Z} / \partial \mgfe$ as a function of metallicity, as this may be compared with the independent analysis of Guiglion et al. (2015, submitted).  We find a non-zero value for the metal-poor stars (though still marginally compatible with zero at the 3$\sigma$ level). This value, suggests that in a given metallicity range the thick disc stars have a larger velocity dispersion than the thin disc ones ({\it i.e.} the low$-\alpha$ ones), as expected (see Guiglion et al. for details).

%%%%%%%%%%%%%%%%%%%%%%%%%%%%%%%%%%%%%%%%%
\subsection{Discussion }
Our analysis of the mean azimuthal velocities of the stars shows that
for metallicities between $-0.8$ and $+0.2\dex$, there is a
mixture of  at least two  populations in our sample.  Further,  their
relative proportions vary with metallicity. This is a well-established result for the thin and thick disks defined by kinematics \citep[e.g.][]{Carney89, Wyse95,Bensby05, Bensby07}; the new result here refers to the discs when defined by distinct chemical abundance patterns.

Our results show that the thin disc, defined as the
low$-\alpha$ population, extends in metallicity at least as low as $-0.8\dex$, whereas
the thick disc, defined as the high$-\alpha$ population, extends at
least up to Solar metallicities.  
 Outside the $[-0.8,+0.2]\dex$ range, the large error bars that we derive in our analysis cannot allow to draw any robust conclusions.

%We do not exclude the presence of these populations outside this range, but note that their relative proportion should either be low and/or remain constant, since they are not altering significantly the $\rm \partial V_i / \partial \mgfe$ trends. 

The metallicity ranges over which  we have found the discs to extend  
have to be considered in light of the results of  
\citet{Haywood13} and \citet{Snaith14b}.  These authors concluded that 
chemical evolution was traced by  the thick disk up to high metallicities, then after dilution by metal-poor gas, proceeded following the thin disk sequence from low metallicity to high metallicity.  
% \query{Amusingly WG95 discussed prolems with merger models that the overlap caused also, p2985, and concluded that need large scatter in age-metallicity at all time; we assumed no metal-poor gas inflow.} 
A  thick disc extending up to super-Solar metallicities plus  a  thin disc
extending down to $-0.8\dex$, challenge this view since such a large 
metallicity overlap - over $1\dex$ -  would require  accretion
of a significant mass of metal-poor gas to lower the ISM's metallicity before the  onset of star
formation in the thin disc. We note, however, that  we have not corrected for the 
selection function  in our analysis. Further, we have very limited information concerning stellar ages. 

Future data from Gaia will provide better age estimates from improved distances and absolute magnitudes, in addition to   proper motions that are ten times more accurate than  those we adopted here, which will allow a more refined analysis.

%%%%%%%%%%%%%%%%%%%%%%

\section{Identification and characterisation of the $\mgfe$ sequences in terms of star-counts}
\label{sect:alpha_histograms}

In the previous section we identified and confirmed the
existence of two populations in our Solar suburb
sample, for metallicities between $[-0.8,+0.2]\dex$. In
that analysis we made no assumptions about the underlying metallicity 
distribution functions.  In this section, we  extend  the analysis to an investigation of the $\alpha-$abundance sequences 
across the entire metallicity range of the data. We  aim to characterise the possible  double structure in each  metallicity bin, and
obtain estimates of the mean $\alpha-$enhancement value, and the dispersion, of each component, for different regions of the Galaxy  (i.e. both within  and beyond  the extended Solar neighbourhood). 
As  in the previous sections, we use the magnesium abundance as a
tracer of the $\alpha-$abundance of the stars, and we make the reasonable assumption that at a fixed $\feh$ the Gaia-ESO survey is not biased with respect to the value of $\mgfe$.

\subsection{Model fitting of the data}

The model-fitting is done using a truncated-Newton method
\citep{Dembo83} on the un-binned $\mgfe$ data of a given metallicity range, by
maximising the log-likelihood of the following relation:

\begin{equation}
\mathcal{L}(\mu_1,\sigma_1,\rho_1,\mu_2,\sigma_2,\rho_2)=\prod_{t=1}^N \sum_{i=1}^2 \frac{\rho_i}{\sqrt{2\pi\sigma_i^2}} \cdot e^{-\frac{(\afe_t - \mu_i)^2}{2\sigma_i^2}}
\label{eq:likelihood_2pop}
\end{equation}
where $N$ is the total number of stars considered in a given metallicity range, $\mu_1,\mu_2,$  are the mean $\afe$\ values of the thin and thick disc, $\sigma_1,\sigma_2$ are the dispersions on the $\afe$\ abundances, $\rho_1$ and $\rho_2$ are the relative proportions of each population, and $\afe_t$ is the $\afe$ abundance ratio of a given star measured from its spectrum.  The fit is performed for every Monte-Carlo realisation separately.

Motivated by the results of \citet{Adibekyan13,Bensby14,Recio-Blanco14}, we assume  that the mean $\afe$ abundance of each population  with  $\feh \geq -0.9\dex$  is a function of metallicity, and that the thin disc's mean $\afe$ at any metallicity is always lower than (or equal to) the mean $\afe$ abundance of the thick disc. We note that the thick disc exhibits an $\afe$ plateau at lower metallicities \citep[e.g.][]{Ruchti10}, this plateau ending with the so-called knee of the $\afe-\feh$ distribution.  The exact position of the knee is a matter of debate; its location depends on how quickly star formation and chemical enrichment happened in the thick disc. The quality of our data at the lowest metallicities does not, however, allow us to investigate this issue.

To perform our fitting, the initial assumed values  for the mean $\alpha-$enhancements of the discs are:
\begin{eqnarray}
\mu_{\rm thick}&=&-0.30 \times {\rm [Fe/H]} + 0.2 \label{eqn:thick_enhancement_model}\label{eqn:prior_thick}\\
\mu_{\rm thin}&=&-0.10  \times {\rm [Fe/H]}. \label{eqn:thin_enhancement_model}\label{eqn:prior_thin}
\end{eqnarray}
We allow the fitting procedure to vary these means freely by up to $\pm 0.15\dex$. As far as the dispersions of the distributions are concerned, we allow them to vary in the interval between $[ 0.05, 0.15]\dex$, and start with an initial value  of $0.08\dex$,   larger than the typical errors in $\afe$ (see Sect.~\ref{sect:Param_determination}). 

This set of assumptions  allows the value of  $\mgfe$ at the low metallicity end of the thick disc distribution to be between $0.35$ and
$0.65\dex$ 
\citep[consistent with  the results of][obtained with the Gaia-ESO DR1
dataset]{Mikolaitis14}.  Furthermore,  the priors also ensure that the thin disc sequence has enhancements that are 
close to Solar values at Solar metallicities and that the $\mgfe$ prior for the thick
disc is always higher than that for the thin disc (the starting relations are equal  at ten times the solar iron abundance).

Note that our procedure always fits two components to the data, even  in regions where only one sequence could exist.  
 Since a  model with two components will always fit the data better, we investigate the significance of this fit by deriving the $p-$value of the log-likelihood ratio between the null hypothesis  of having only one  thick/thin disc and the alternative model of having both components.  We therefore define the following test statistic of three degrees of freedom:  
\begin{equation}
D=2\cdot \left(\log \mathcal{L}_2 - \log \mathcal{L}_1 \right)
\label{eq:chi-test}
\end{equation}
where $\mathcal{L}_2$ is the likelihood associated to the model with two components (5 degrees of freedom, see Eq.~\ref{eq:likelihood_2pop}) and  $\mathcal{L}_1$ is the likelihood associated to the null model with only one component (2 degrees of freedom).  
We set that for a given Monte-Carlo realisation the two components model is the more likely if  $D \geq 7.82$, which corresponds to a $p-$value $\leq 0.05$. We then derive the mean fitted populations over all the Monte-Carlo realisations, and conclude that two populations are needed only if the mean is greater than 1.5 ({\it i.e.} the two component model is preferred for more than half of the realisations with 0.95 confidence).

This procedure has been applied to six subsamples, selected by location within  the Galaxy: three ranges in Galactocentric radius (inner Galactic ring: $6.5-7.5\kpc$, Solar suburbs:  $7.5-8.5\kpc$,  outer Galactic ring: $8.5-10\kpc$), and two distances from the plane ($|Z|<1\kpc$ and $1<|Z|<2\kpc$). 
 The results,  obtained  after averaging 100 Monte-Carlo realisations on distances, metallicities and $\mgfe$ abundances, are discussed in the following sections.

\begin{figure*}
\centering
\includegraphics[width=0.75\linewidth, angle=0]{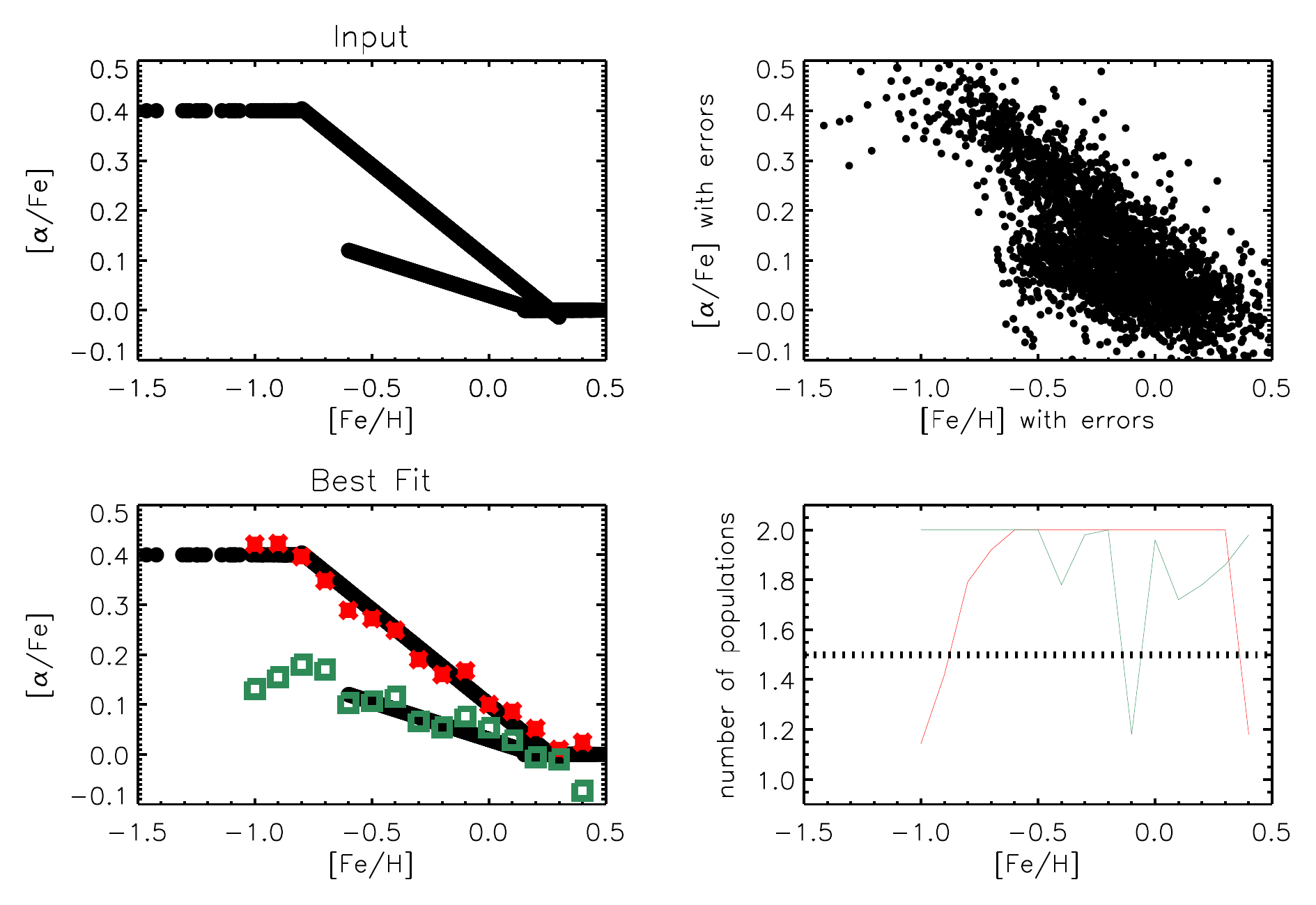}
\caption{{\bf Top:} adopted trends for the mock data used to test our fitting procedure (left) and effect on the mock data of introducing errors in $\feh$ and $\afe$ as the observations (right). {\bf Bottom:} Output of our fitting procedure for the high$-\alpha$ (red crosses) and low$-\alpha$ (green squares) populations performed on the noisy mock data. Over-plotted in black are the error-free points used to create the mock catalogue. The bottom right plot shows the average of the number of populations needed to better fit the data based on the log-likelihood ratio tests of Eq.\,\ref{eq:chi-test}. When the red (green) line is below $1.5$, this indicates that fitting the data with only a thick (thin) disc rather than two components simultaneously is more significant for more than half of the Monte-Carlo realisations.} 
\label{fig:mock}
\end{figure*}

%%%%%%%%%%%%%%%%%%%%%%%%%%%%%%%%%%%%%%%%
\subsection{Tests on mock data }
\label{sect:mock_tests}

We tested our fitting procedure on mock data to verify that it is able
to find the ``correct'' $\alpha-$enhancements of two populations with
chemical trends similar to what is expected in the Milky Way. For that
purpose, we took the $\feh$ values of the stars in our sample, and
assigned for the range $-0.8 < \feh <+0.3$ the trends seen on the top
left plot of Fig.~\ref{fig:mock}: the position of the thick disc knee is located at $-0.8\dex$, the thin disc's lowest metallicity is $-0.6$, and the two sequences merge at $\feh \geq 0.3$. 
These values are selected in order to roughly represent the actual data, while still differing from the priors of Eqs.\,\ref{eqn:prior_thick} and  \ref{eqn:prior_thin}; this should allow us  to verify that our procedure converges towards the correct results.
  The `stars' were randomly assigned to 
either the high$-\alpha$ or low$-\alpha$ population. Then, for each
`star' we assigned the uncertainties of the true Gaia-ESO iDR2
catalogue, to reproduce realistic error bars. The blurring due to the
uncertainties on the iron and $\alpha-$abundances  can be seen in the top-right  panel.  One can notice that the underlying gap in the
chemical paths is now barely visible and that Fig.~\ref{fig:mock}
is qualitatively  similar to Fig.~\ref{fig:Mg_Feh} where the actual
Gaia-ESO data are plotted.

%\ccomment{Fig 10, bottom right panel (number of pops vs [Fe/H]) is barely legible; suggest make the lines thicker.} 

We applied our procedure, starting from equations (\ref{eqn:thick_enhancement_model}) and (\ref{eqn:thin_enhancement_model}), to  100 Monte-Carlo realisations of noisy mock data and derived the average trends
plotted in red crosses and green square symbols on the bottom left panel of
Fig.~\ref{fig:mock}. One can see that the
``true'' trends are nicely recovered, up to Solar metallicities,
validating  our procedure. In particular, the procedure performs well for a gap-width varying with metallicity, even when the separation is very small. 
Our procedure also correctly finds that at the low metallicities a single high$-\alpha$ population is enough to fit the data. 
 Indeed, despite fitting two populations even at the lowest metallicities (green squares below $\feh=-0.6$ in the lower left panel of Fig.~\ref{fig:mock}), the relative weight associated to these points, combined with the test statistic of Eq.~\ref{eq:chi-test}, indicate  that these points are not likely to represent correctly the data (the red line in the bottom right plot of Fig.~\ref{fig:mock} is consistently below 1.5).

The difficulty of fitting the data towards the highest metallicities, where the the separation between the two sequences is small, can be seen on the noisiness of the average log-likelihood ratio, comparing  the two-component model with a model having only the thin disc (green line). We note, however, that the two component model is always the preferred fit, even at this regime, as it should be, based on the input mock data.

In the following sub-sections, we
therefore apply our procedure with confidence to the Gaia-ESO iDR2 data.

%%%%%%%%%%%%%%%%%%%%%%%%%%%%%%%%%%%%%%%%
\subsection{Robustness to signal-to-noise cuts}
\label{sect:SNR_cuts}
\begin{figure*}
\centering
\includegraphics[width=0.6\linewidth, angle=0]{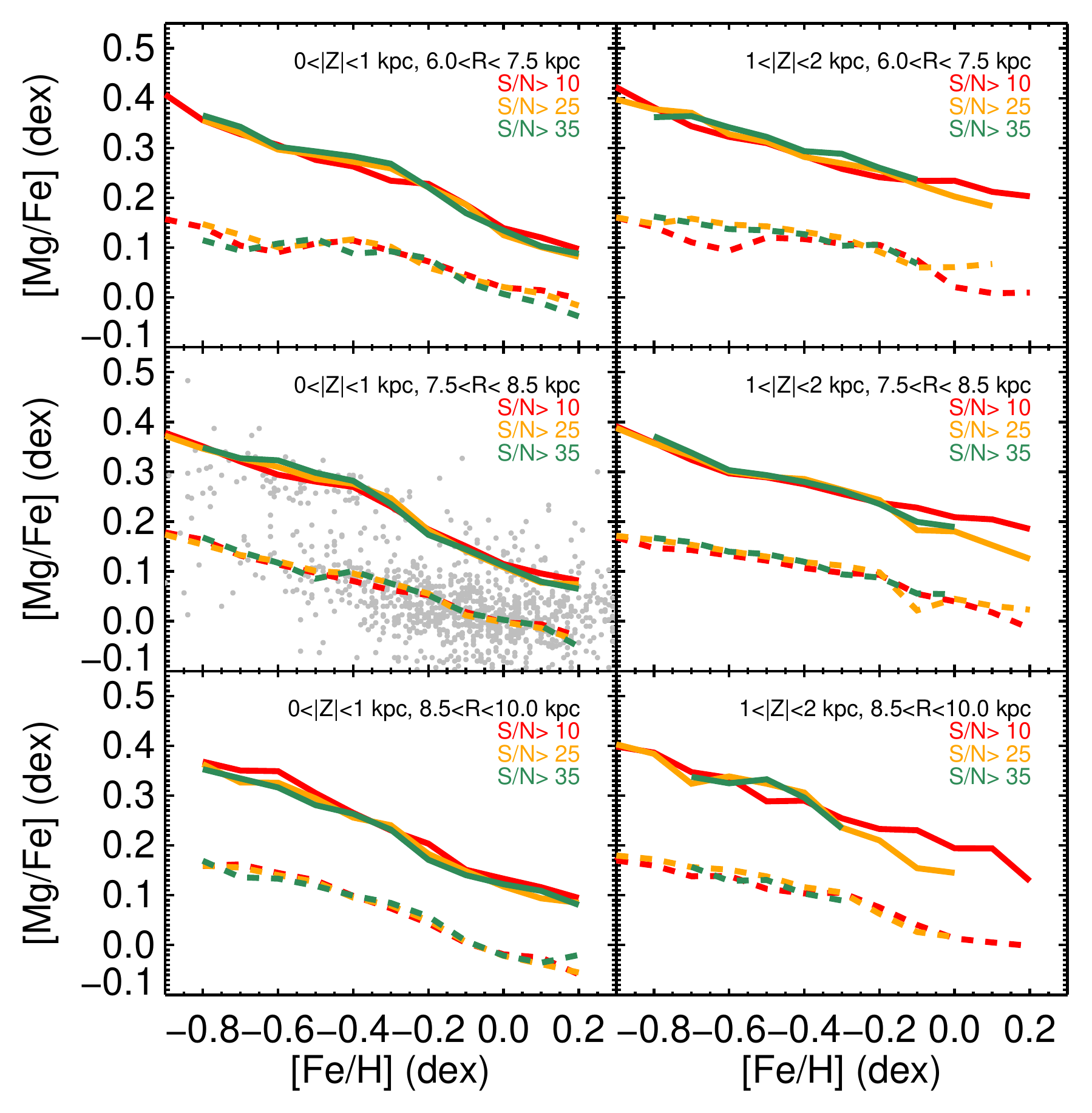}
\caption{Trends of the $\mgfe-\feh$ sequences for the thick disc (plain line) and thin disc (dashed line), for different signal-to-noise (S/N) threshold limits close to the plane ($0<|Z|<1\kpc$, left column) and far from it ($1<|Z|<2\kpc$, right column). The 
top, middle and bottom rows show the Gaia-ESO iDR2 results for the inner Galaxy, Solar suburb and outer Galaxy, respectively. The over-plotted grey points in the panel $7.5 < R < 8.5\kpc$ and $0<|Z|<1\kpc$ are the $\alpha-$abundance measurements of \citet{Adibekyan13}, obtained for a very local sample ($d\lesssim150\pc$). No significant differences in the derived trends are found as a function of S/N, which is indicative of the robustness of our fitting method. }
\label{fig:trends_SNReffect}
\end{figure*}

Figure~\ref{fig:trends_SNReffect} shows the trends of the  high$-\alpha$ and low$-\alpha$ populations at different Galactocentric radii and distances from the plane, colour-coded according to different cuts in S/N of the actual Gaia-ESO spectra. One can see that our results are robust to these cuts, with the overall trends being similar,  independent of the S/N threshold. Lowering the threshold  increases the number of analysed stars and populates  regions of chemical space which statistically have fewer stars, {\it i.e.} at lower and higher metallicities. 
A lower S/N threshold therefore  achieves a better description of the overall behaviour of the trends, but with larger uncertainties on individual estimates, particularly  at the extremes of the metallicities.
In the remainder of this analysis, when not stated otherwise, we adopt S/N$>25$ as providing a compromise between the number of analysed stars and the errors on the trends. All the quality cuts on distance and atmospheric parameter errors remain similar to the ones of Sect.~\ref{sect:quality_sample}.

%%%%%%%%%%%%%%%%%%%%%%%%%%%%%%%%%%%%%%%%

\begin{figure*}
\centering
$\begin{array}{ccc}
\includegraphics[width=0.49\linewidth, angle=0]{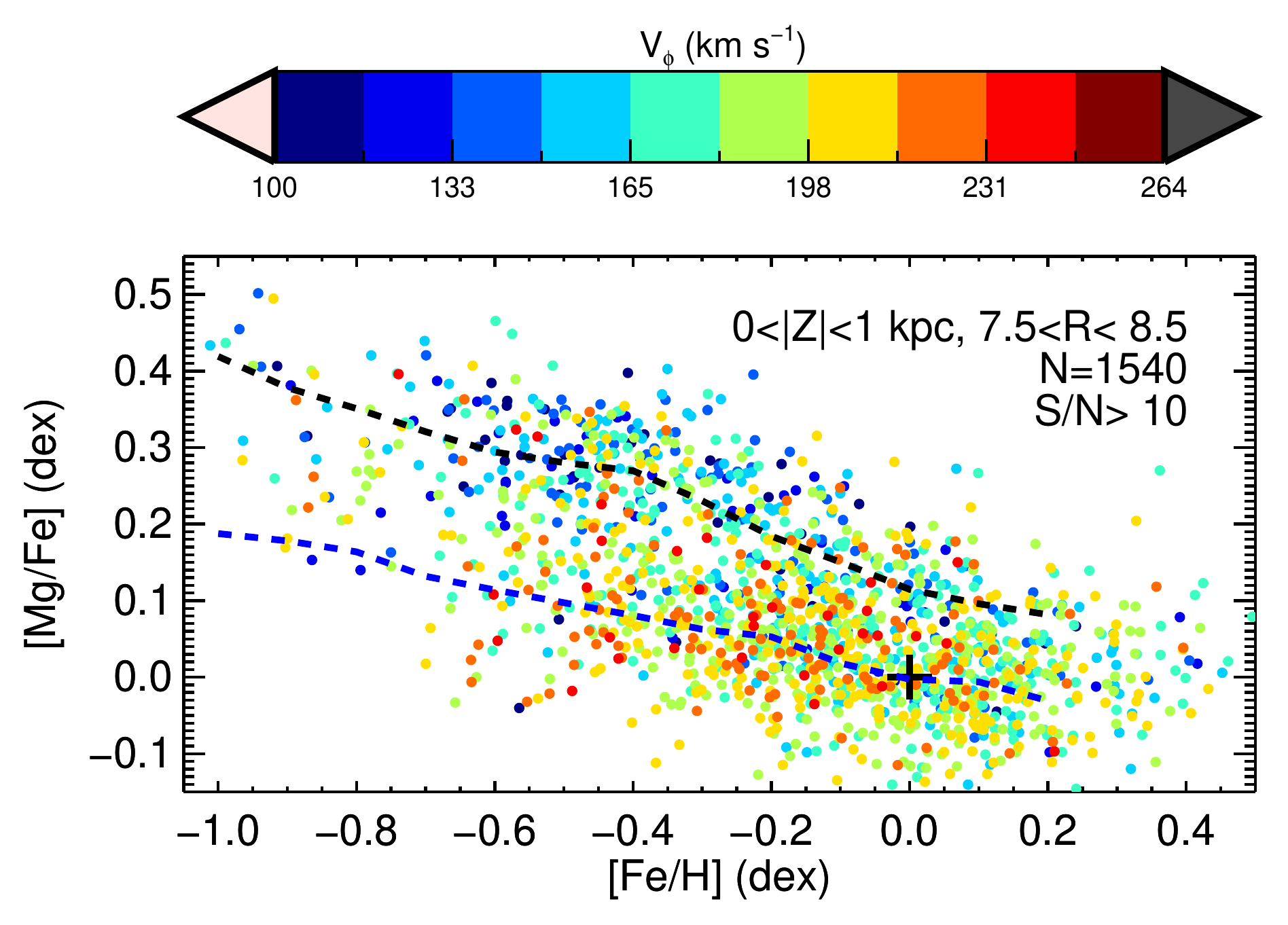} &
\includegraphics[width=0.49\linewidth, angle=0]{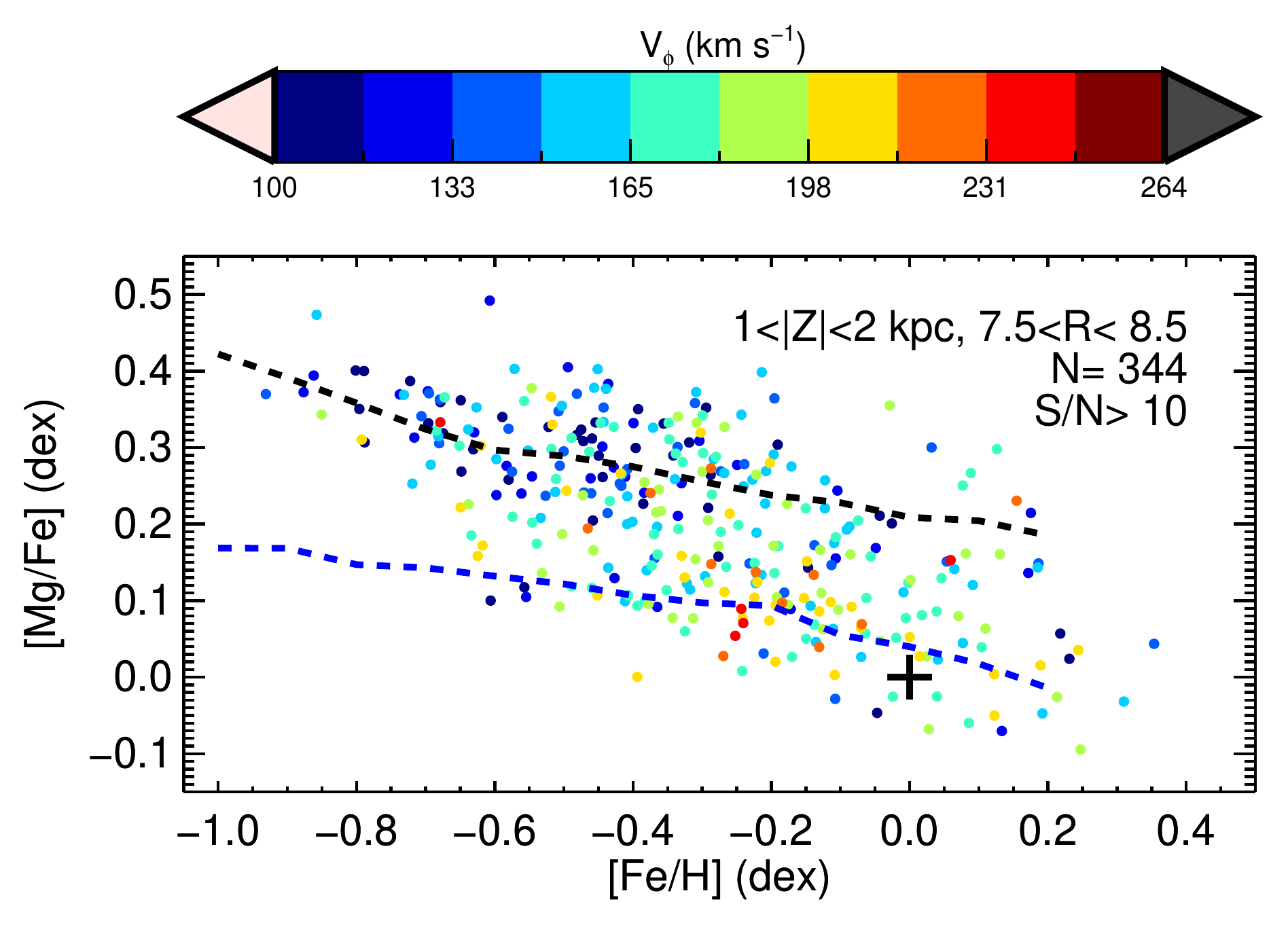}\\ 
\includegraphics[width=0.49\linewidth, angle=0]{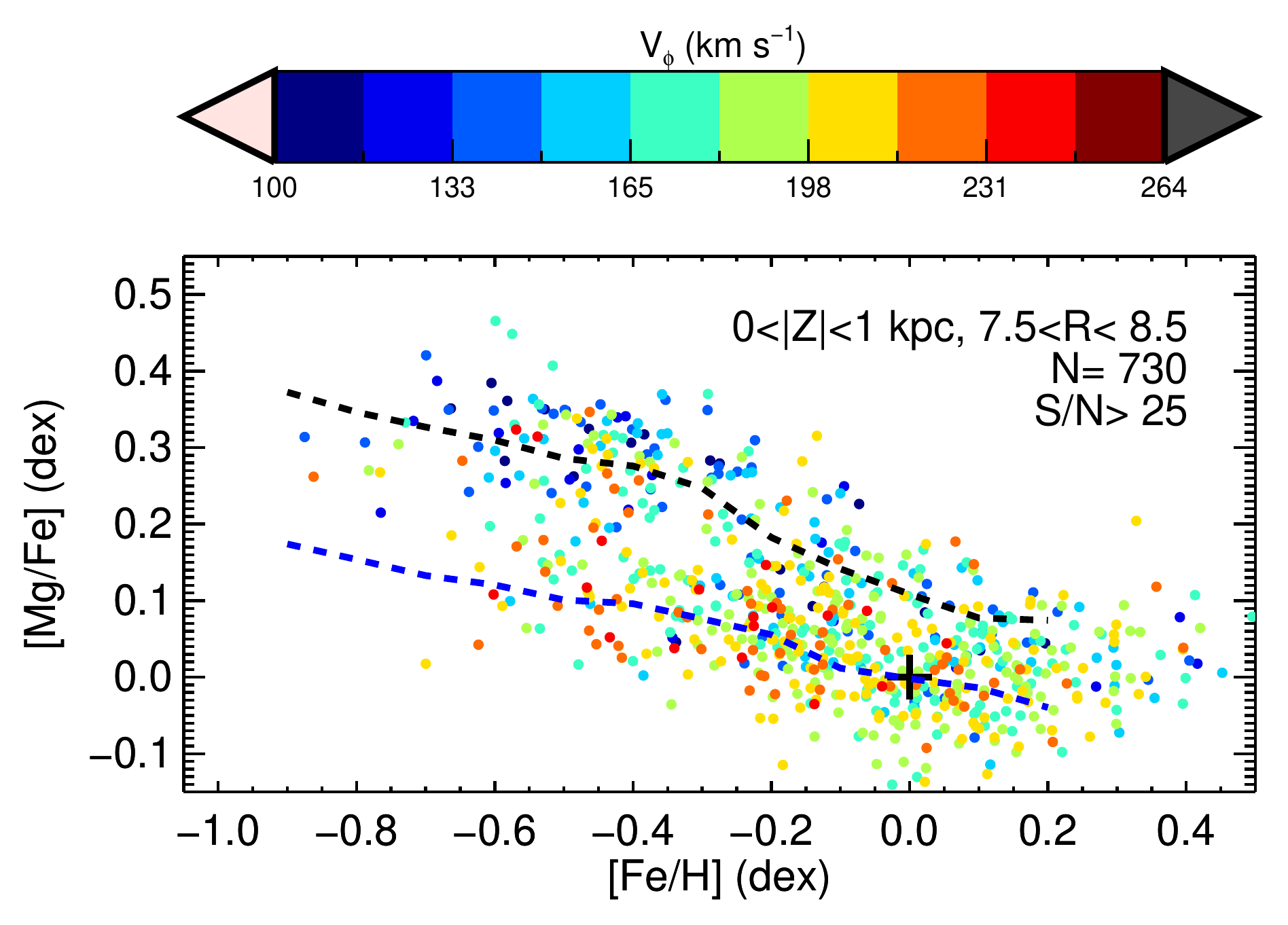} & 
\includegraphics[width=0.49\linewidth, angle=0]{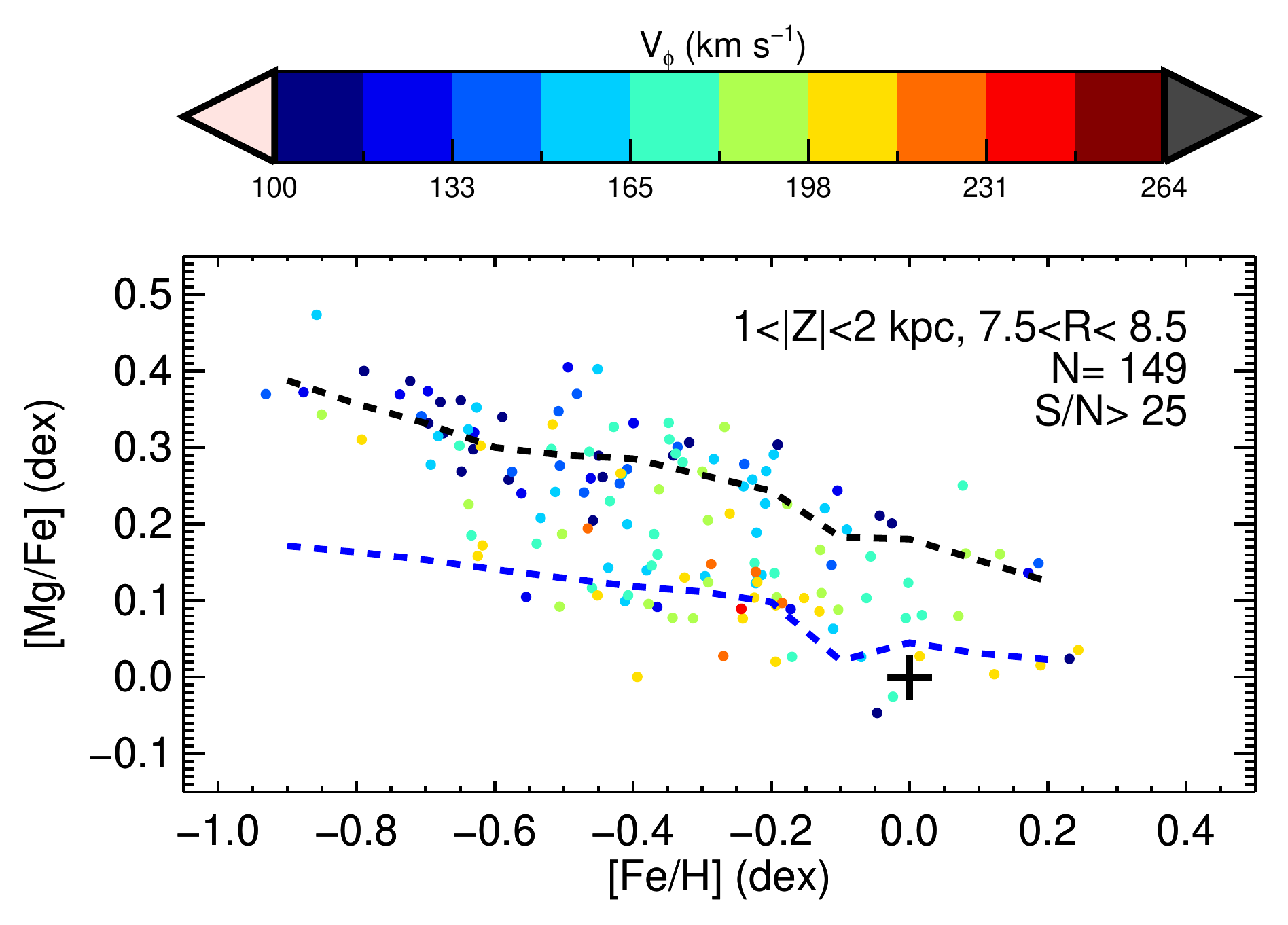}\\
\includegraphics[width=0.49\linewidth, angle=0]{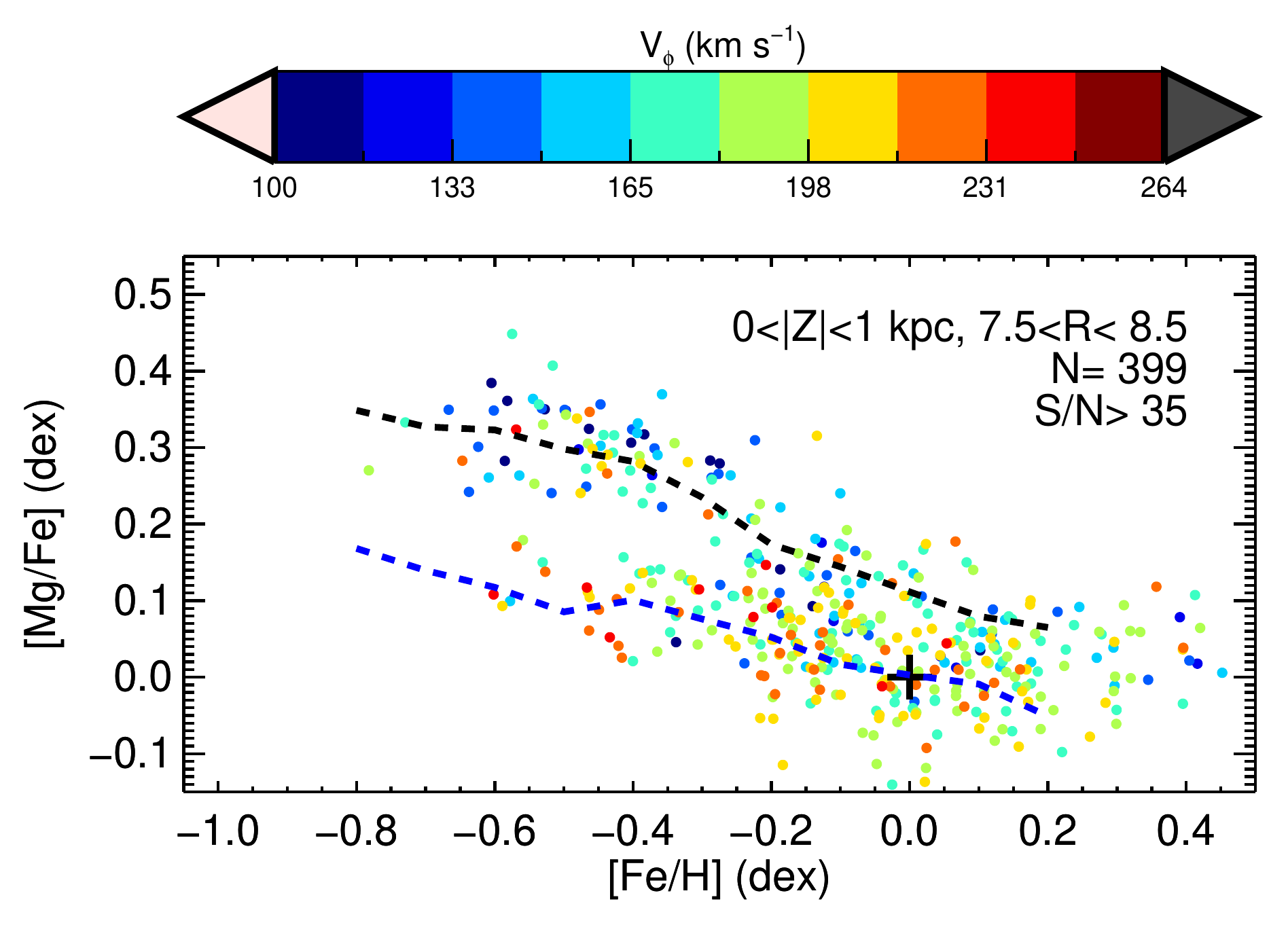} &
\includegraphics[width=0.49\linewidth, angle=0]{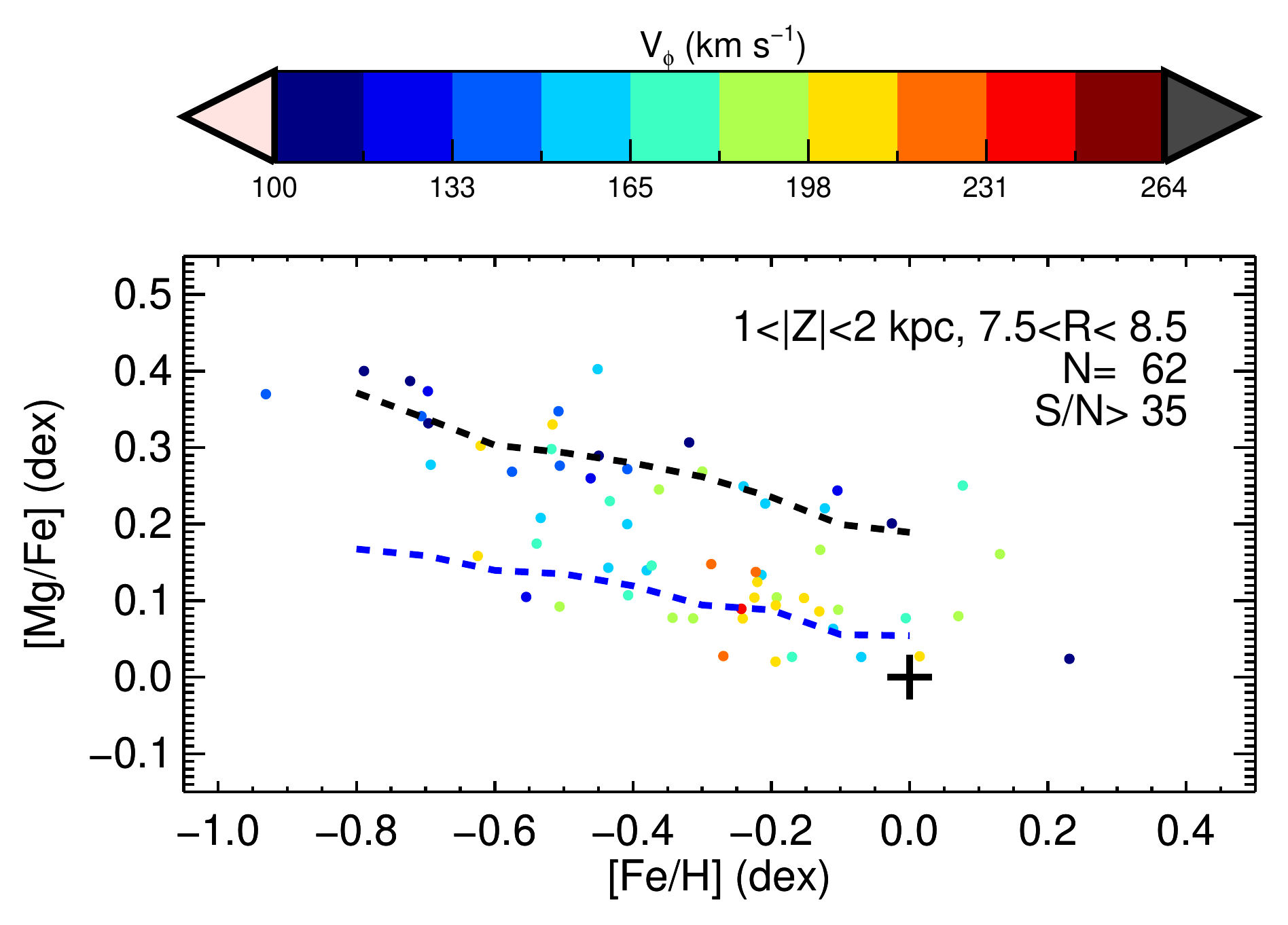} 
\end{array}$
\caption{$\mgfe$ as a function of $\feh$ for the stars in the Solar neighbourhood, closer than $1\kpc$ from the Galactic plane (left panels) and between $1\kpc$ and $2\kpc$ from the plane (right panels). Different selections, according to the signal-to-noise ratio of the spectra, are shown at the top (S/N$>10$) and the bottom (S/N $>35$). The colour-code corresponds to the azimuthal velocity of the stars. The dashed blue (black) line corresponds to the best fit model of the thick (thin) disc. The black `+' sign is located at $\mgfe=0$ and $\feh=0\dex$. }
\label{fig:scatter_and_trends}
\end{figure*}

%%%%%%%%%%%%%%%%%%%%%%%%%%%%
\subsection{Properties within $7.5 < R<8.5\kpc$}
\label{sect:Solar_cylinder} 

\begin{figure}
\centering
$\begin{array}{cc}
\includegraphics[width=0.5\linewidth, angle=0]{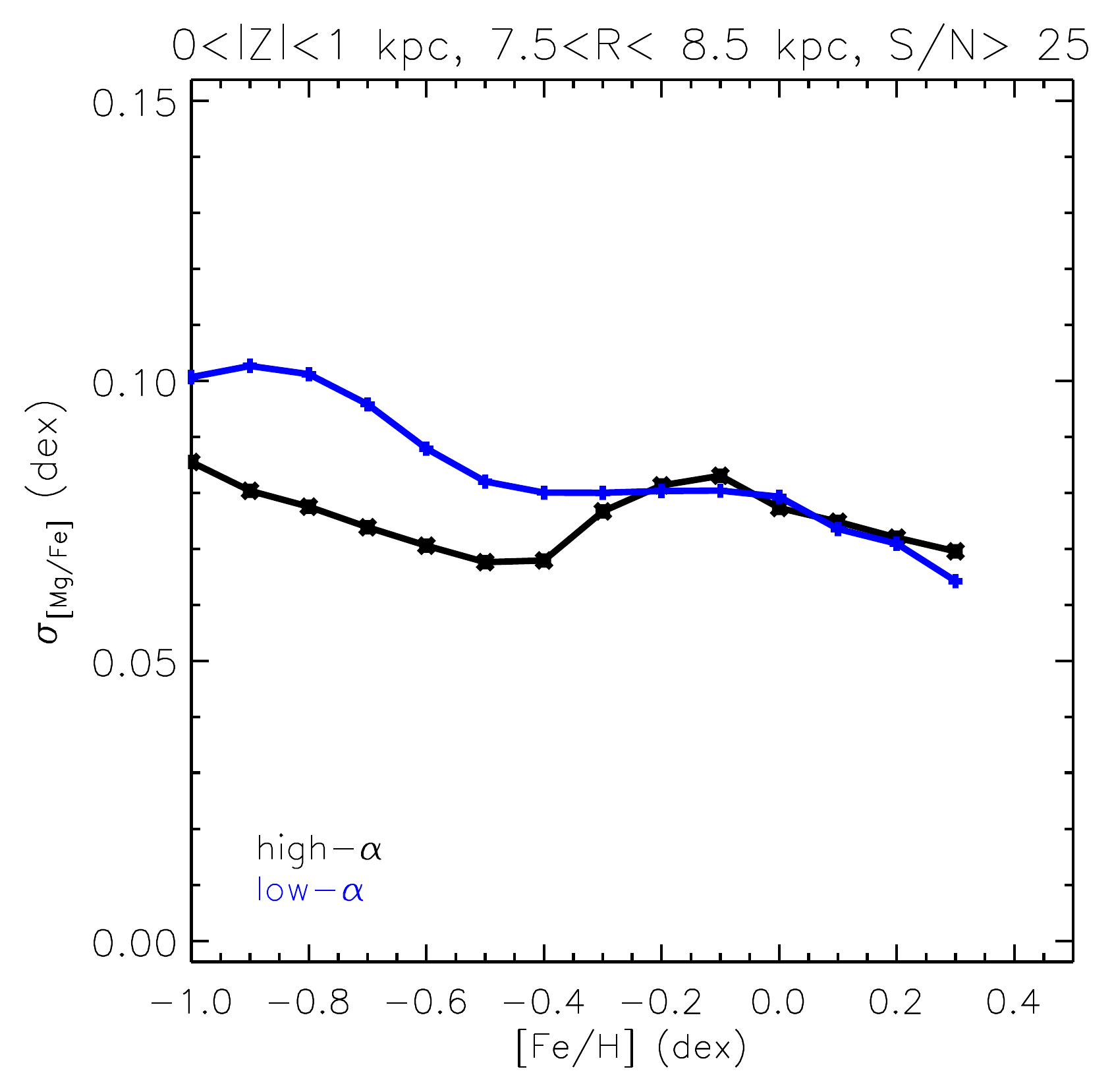}& 
\includegraphics[width=0.5\linewidth, angle=0]{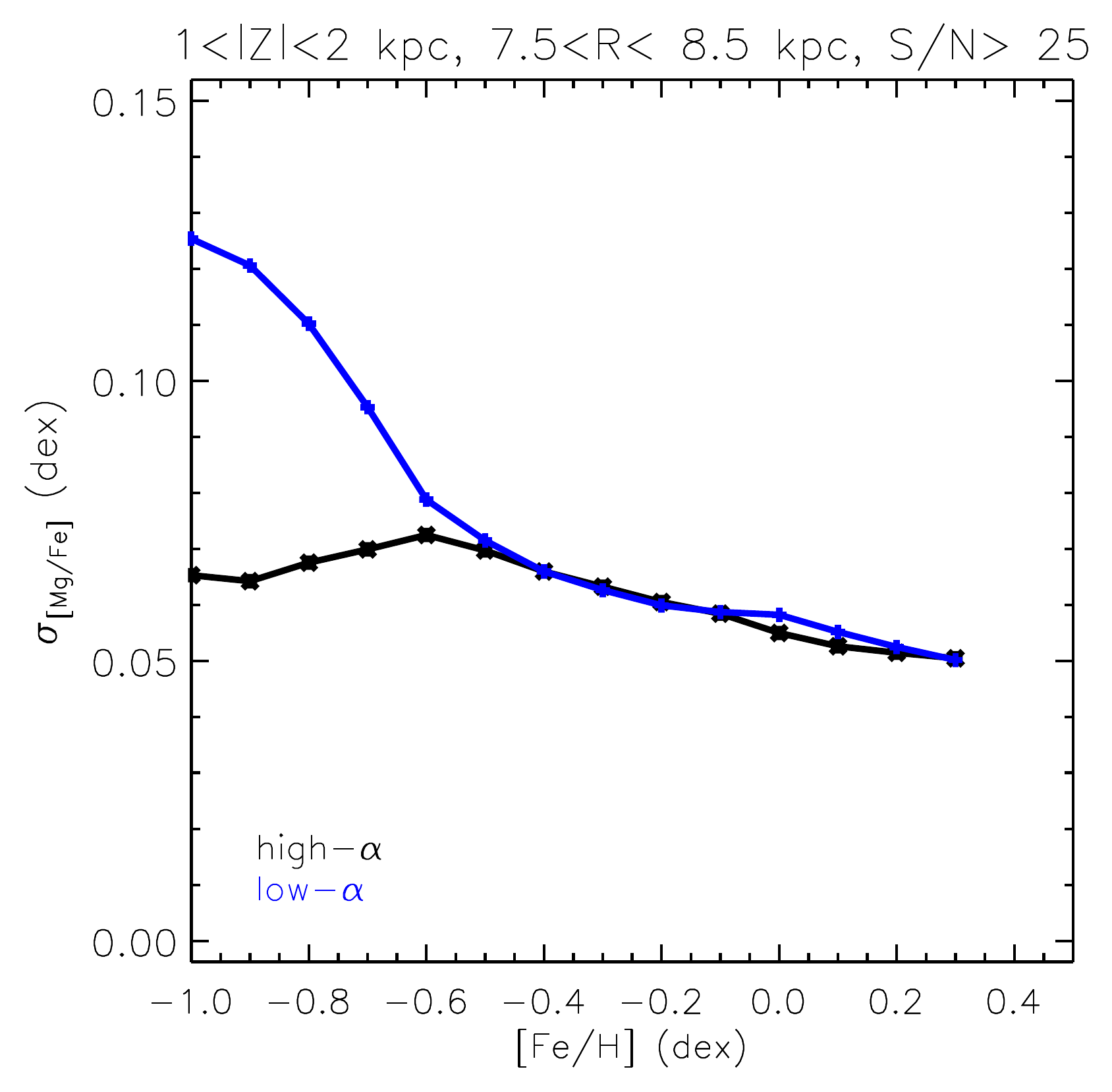}
\end{array}$
\caption{$\mgfe$ dispersions as a function of $\feh$, for the thick disc stars (in black) and the thin disc stars (in blue), close to the Galactic plane (left) and far from it (right). }
\label{fig:dispersions_and_KS}
\end{figure}

We start by describing our results in  the Solar suburbs ($7.5 \leq R \leq 8.5\kpc$) at two different height-bins from the Galactic plane: $|Z|<1\kpc$ and $1<|Z|<2\kpc$. 
Figure~\ref{fig:scatter_and_trends} shows the matching of the recovered trends with the data-points at different S/N cuts, Fig.~\ref{fig:dispersions_and_KS} represents the $\mgfe$ dispersions as a function of $\feh$, and finally middle plots of Fig.~\ref{fig:proportions_MW} and Fig.~\ref{fig:chi2_significance} show the derived relative numbers of the two sequences close and far from the plane, and the mean required populations to fit the data.
For both heights, the two populations are recovered nicely, with no particular indication of a merging of the sequences up to $\feh \approx +0.2$.  Indeed, the mean number of populations above $\feh>0$ (second column of Fig.~\ref{fig:chi2_significance}) is most of the time greater than or close to 1.5, however with a noisy and decreasing trend indicating the difficulty of fitting robustly our data.

As mentioned in the previous Section, the data do not
allow us to recover reliably the position of  a   `knee' in the $\afe$ trend.
However, we do recover the slopes of the decrease of the $\mgfe-\feh$ for
each of the low$-\alpha$ and high$-\alpha$ sequences. These slopes  depend on the histories of star formation and gas accretion/outflow
\citep[e.g.:][see also
Sect.\,\ref{sect:interpretation}]{Snaith14b,Nidever14}. The 
measured slopes are reported in Table\,\ref{tab:slopes_disc} and are discussed in the following
subsections.

%%%%%%%%%%%%%%
\subsubsection{Close to the plane}
\label{sect:close_to_the_plane_Solar_suburb}
For the stars close to the plane,  in the metallicity interval $[-0.9,0.0]\dex$ we find (see Table~\ref{tab:slopes_disc}):
\begin{eqnarray}
\mu_{\rm thin}&=&(-0.20 \pm 0.04) \times {\rm [Fe/H]} + (0.00 \pm 0.02)\\
\mu_{\rm thick}&=&(-0.31 \pm 0.05) \times {\rm [Fe/H]} + (0.12 \pm 0.02)
\end{eqnarray}

%% FOR SNR=20
%MEDIAN

\begin{table*}
\caption{Slopes of the $\mgfe-\feh$ relations for thin and thick disc stars obtained for S/N$\geq 25$.}
\begin{center}
\begin{tabular}{lccccc}\hline \hline
& & \multicolumn{2}{c}{$|Z| (\kpc) <1$} & \multicolumn{2}{c}{$1<|Z| (\kpc) <2$} \\
&&$a$ &$b$ & $a$ &$b$ \\ \hline 
		&$R (\kpc) \leq 7.5$ 		&	$-0.14 \pm 0.07$ 	&$ 0.03 \pm 0.04$ & $-0.16 \pm 0.04$& $ 0.06 \pm 0.02$ \\
Thin disc &7.5$<R (\kpc)  \leq8.5$ 	&  	$-0.20 \pm 0.04$	& $0.00 \pm 0.02$ &$-0.17 \pm 0.05$& $ 0.04 \pm 0.02$ \\
		& $R (\kpc)>8.5$ 		& 	$-0.23 \pm 0.03$      & $0.00 \pm 0.02$ & $-0.14 \pm 0.08$& $  0.03 \pm 0.04$  	\\ \hline
		
		& $R (\kpc) \leq 7.5$ 	&	$-0.27 \pm 0.05$ 	&$ 0.14 \pm 0.02$ & $-0.25 \pm 0.05$& $ 0.19 \pm 0.02$ \\
Thick disc &7.5$<R (\kpc)  \leq8.5$ &  	$-0.31 \pm 0.05$	& $0.12 \pm 0.02$ &$-0.21 \pm 0.05$& $ 0.19 \pm 0.02$ \\
		& $R (\kpc)>8.5$ 		& 	$-0.28 \pm 0.07$     & $0.13 \pm 0.03$ & $-0.19 \pm 0.15$& $ 0.14 \pm 0.07$  	\\ \hline
\end{tabular}
\tablefoot{
The $a$ and $b$ coefficients are defined for $-0.9\leq \feh (dex)\leq 0$, as: $\afe$$=a\times \feh$$ + b$.
}
\end{center}
\label{tab:slopes_disc}
\end{table*}%

These slopes are robust to different S/N cuts (see
Fig.\,\ref{fig:trends_SNReffect}; note we adopted S/N$>25$),
and are found to fit well the trends  of \citet{Adibekyan13} (grey points in Fig.\,\ref{fig:trends_SNReffect}, in the panel representing the Solar suburb). In addition, we found that increasing the S/N threshold
emphasises the gap between the two chemical sequences by
depopulating the region between the chemical sequences (see
Fig.~\ref{fig:scatter_and_trends}). The prominence of this gap is therefore compatible with the results of \citet{Bensby14, Recio-Blanco14}, we note, however, that in order to establish whether this gap is true or just  a trough, depopulated due to low number statistics, additional data will be needed (e.g. the future data-releases of Gaia-ESO). 

 Our results show a steeper slope for the $\mgfe$ sequence of the thick disc compared to the thin disc.  As we will discuss in the following sections,  a steeper slope implies higher ratio of past to present SFR (provided there is a fixed stellar IMF and Type Ia supernova delay distribution function for both discs). Above Solar metallicities (where the determination of the slope is not performed), we notice a flattening of the sequence for the thin disc (see Fig.\,\ref{fig:scatter_and_trends}) that could be compared to the ``banana shape'' quoted in \citet[][their Fig.\,10]{Nidever14}, also seen in \citet[][their Fig.\,15]{Bensby14}.

We find that the $\alpha-$enhancement dispersion,  {\it i.e.} the thickness of the chemical sequence followed by either one of the discs,  is of the order of $\sigma_{\rm [Mg/Fe]_{thin}}\sim 0.08\dex$ and $\sigma_{\rm [Mg/Fe]_{thick}}\sim 0.07\dex$, in agreement with \citet{Mikolaitis14}, obtained using the Gaia-ESO iDR1.
We note, however,  that below $\feh \lesssim -0.7\dex$, the thin disc dispersion increases to $0.1\dex$; this is due to low number statistics, as our test on the significance of the fit indicates that below this metallicity, one component model is on average a better fit to the data.

\begin{figure*}
\centering
$\begin{array}{ccc}

\includegraphics[width=0.33\linewidth, angle=0]{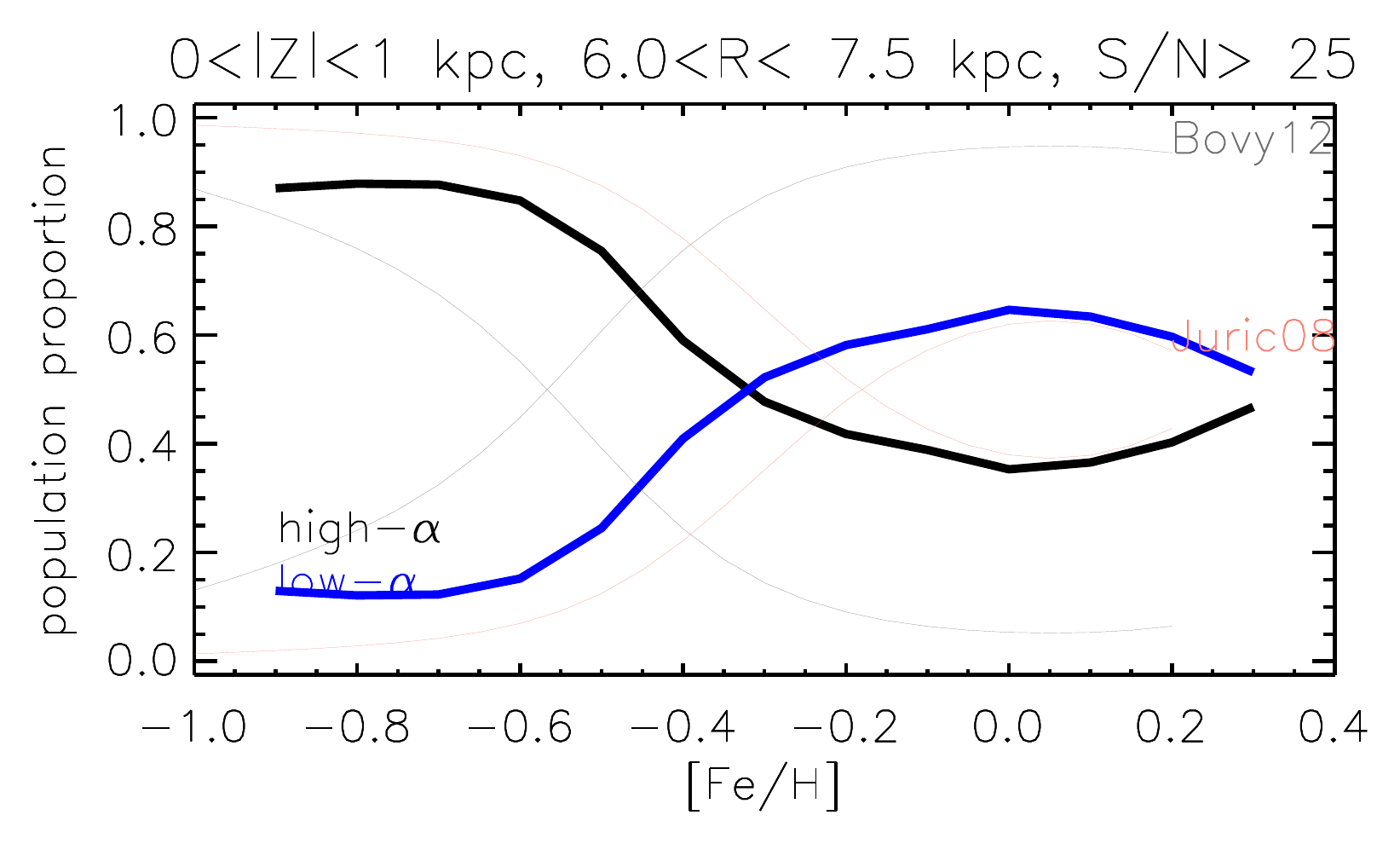}& 
\includegraphics[width=0.33\linewidth, angle=0]{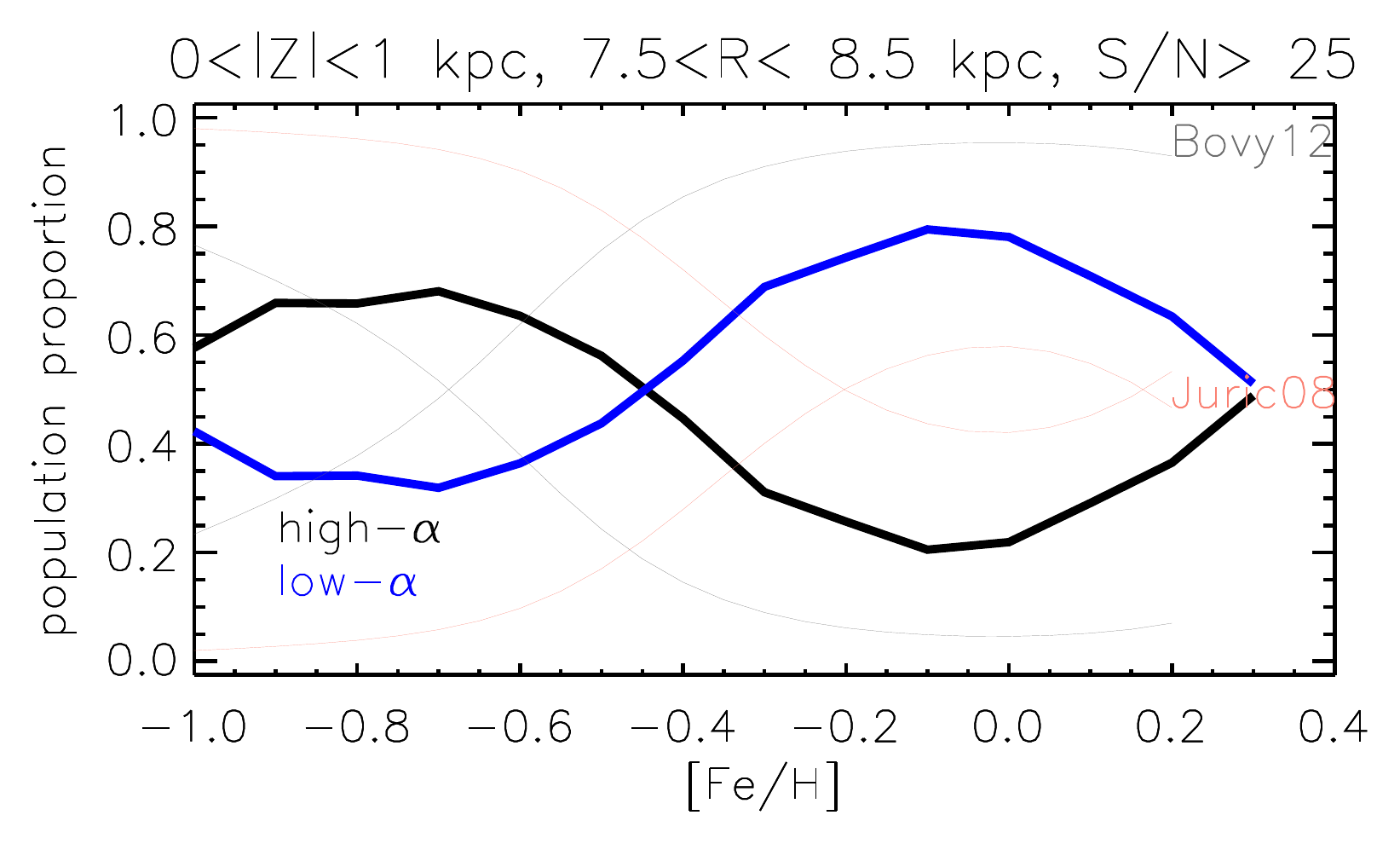} &
\includegraphics[width=0.33\linewidth, angle=0]{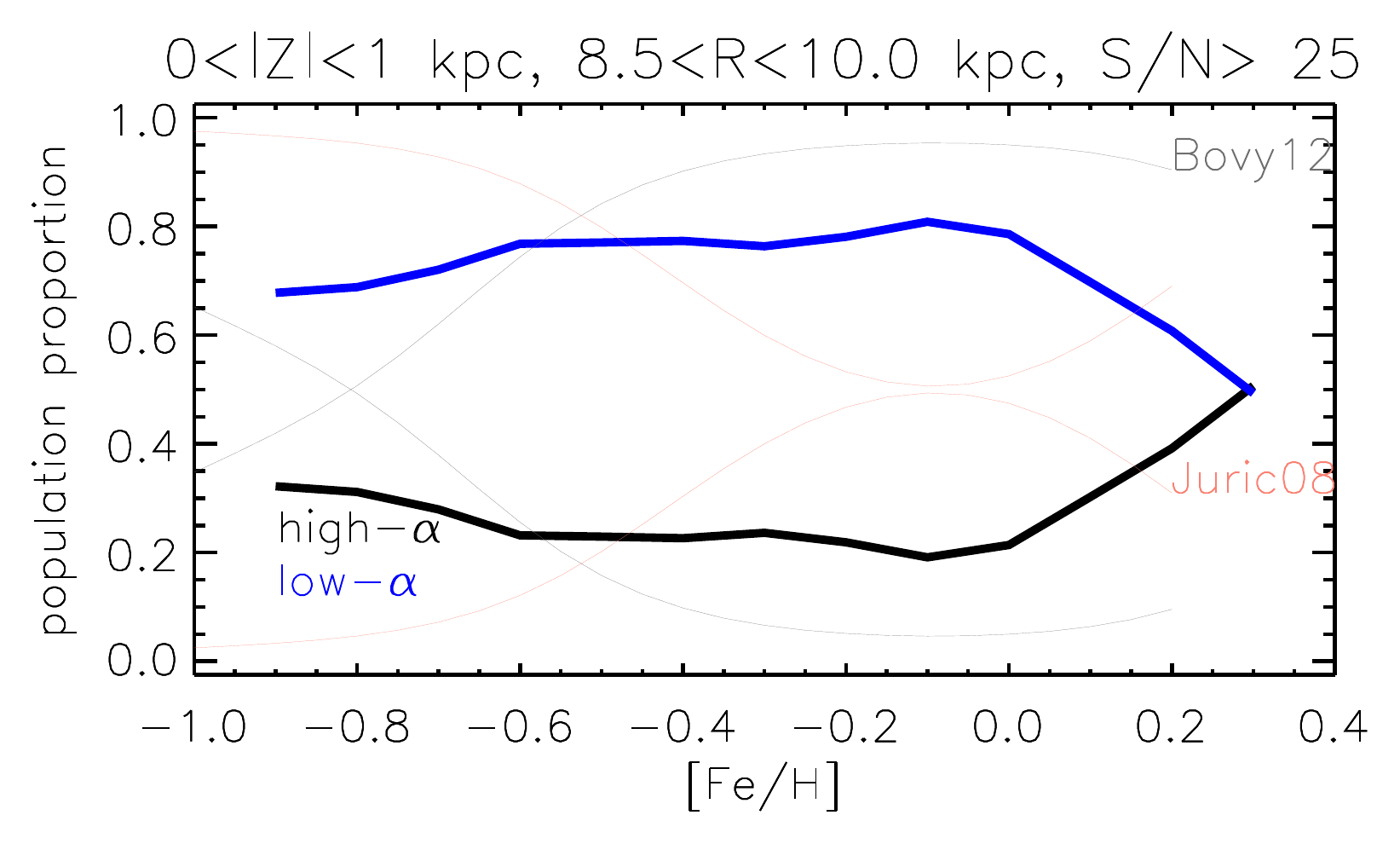} \\
\includegraphics[width=0.33\linewidth, angle=0]{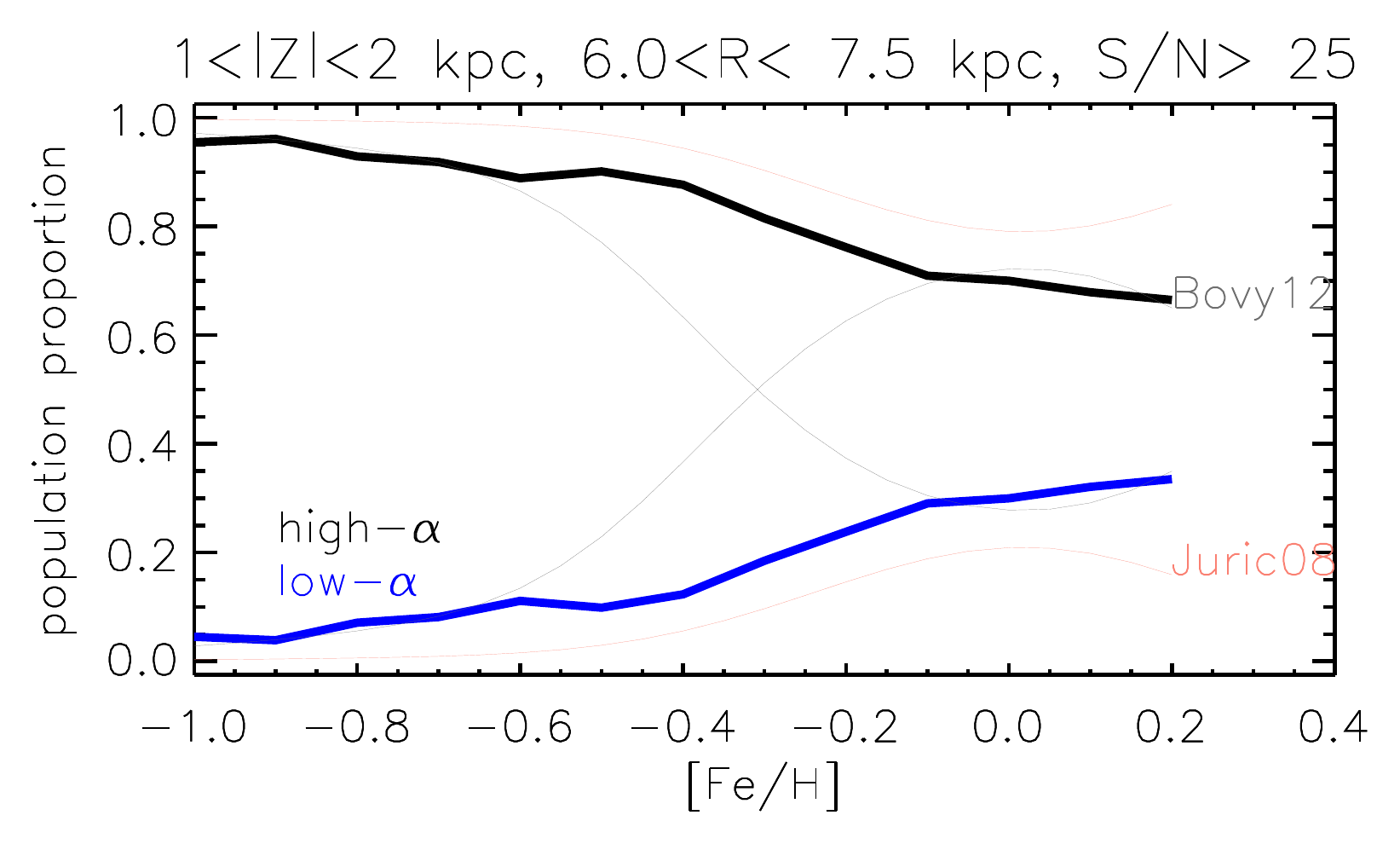}& 
\includegraphics[width=0.33\linewidth, angle=0]{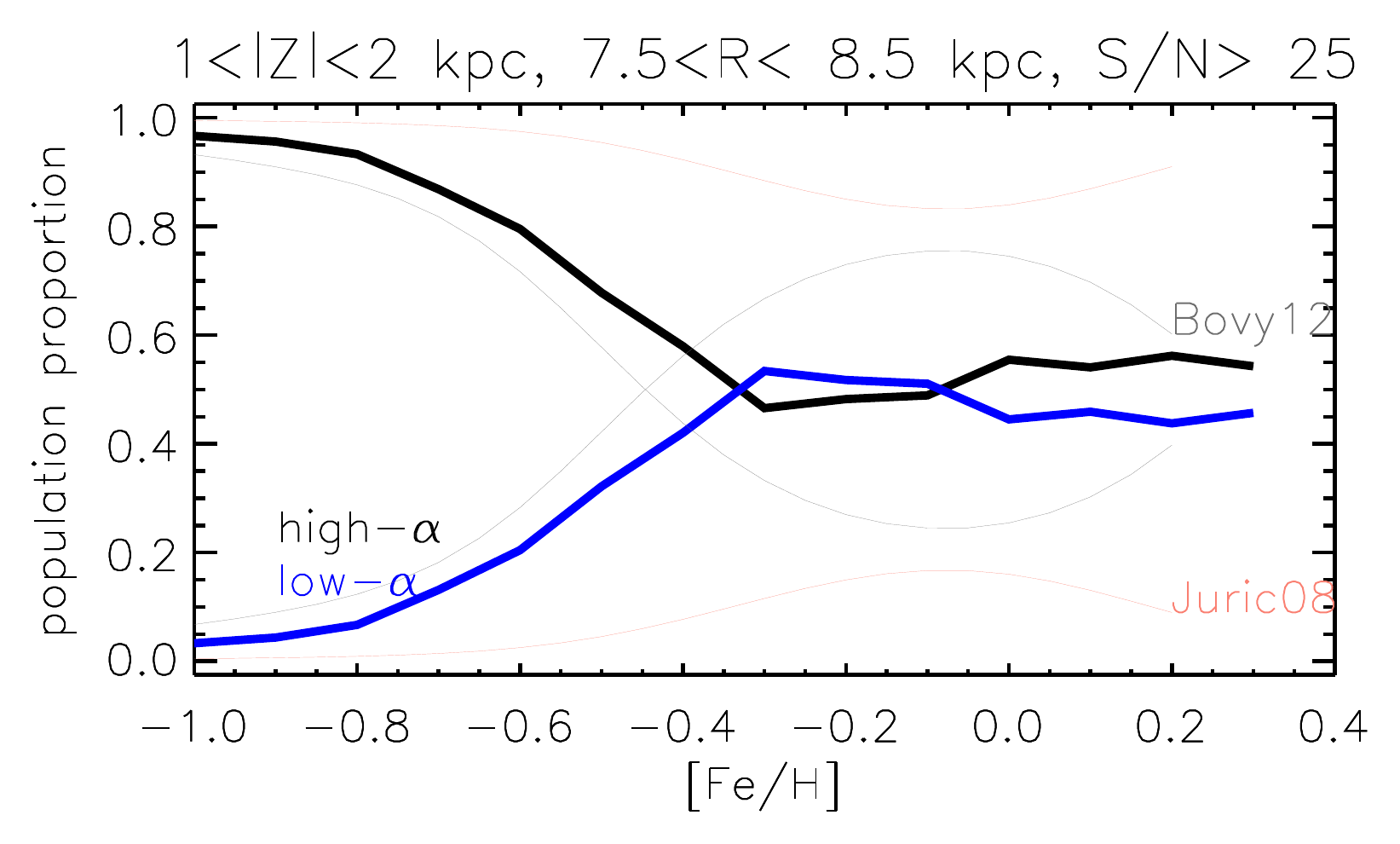} &
\includegraphics[width=0.33\linewidth, angle=0]{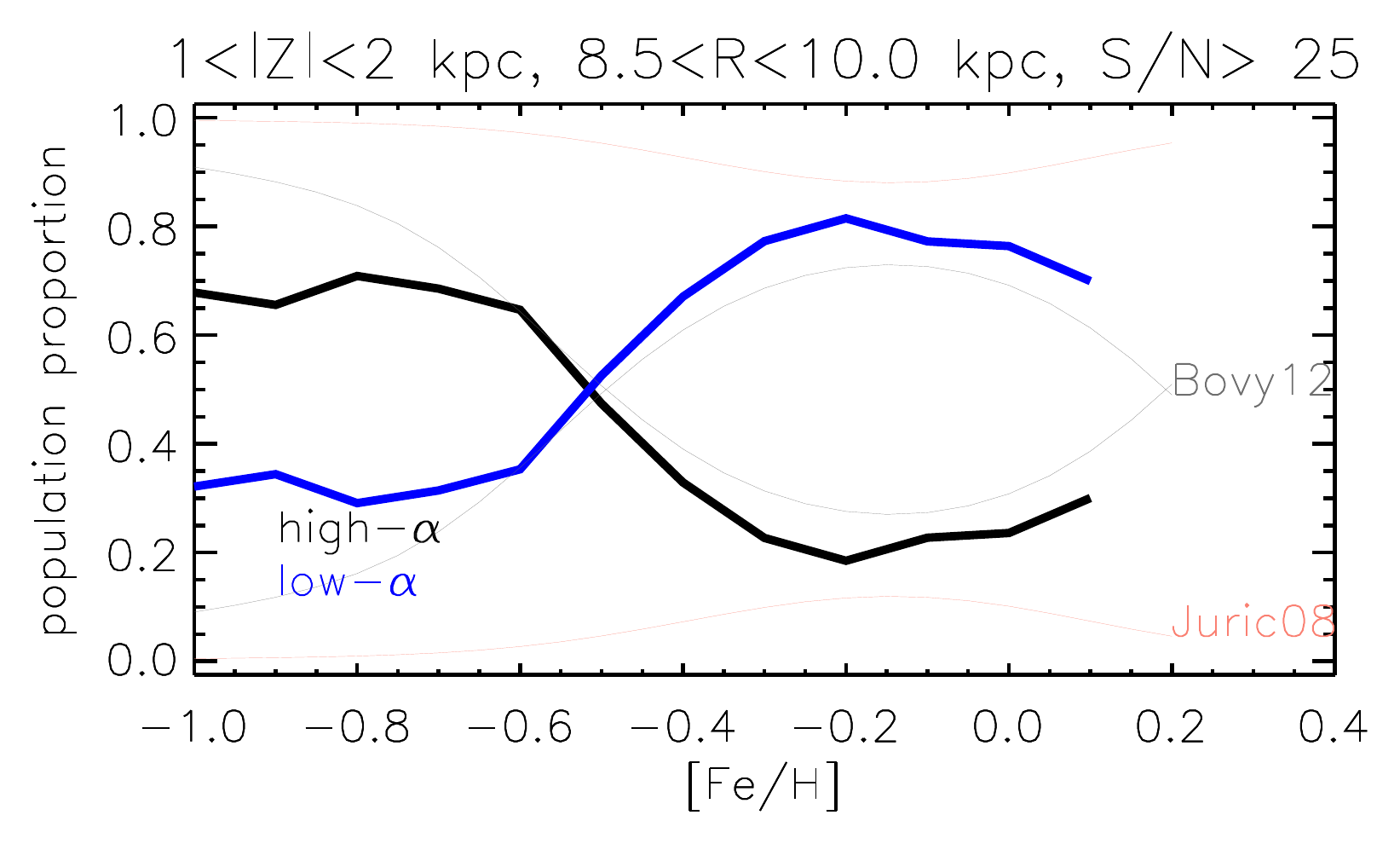}
\end{array}$
\caption{Relative proportion of the high$-\alpha$ (in black) and low$-\alpha$ (in blue) populations as a function of $\feh$, for increasing Galactocentric radii (from left to right) and distances from the Galactic plane (upper panels: close to the plane, lower panels: far from the plane). Plain thick  lines represent measurements derived from observations, smoothed with a boxcar average taking into account the closest neighbour, whereas thin red and grey  lines are derived from the toy model described in Table~\ref{tab:MDF_discs}, with the scale-heights and scale-lengths of \citet{Juric08} and \citet{Bovy12b}, respectively. }
\label{fig:proportions_MW}
\end{figure*}

\begin{figure*}
\centering
$\begin{array}{ccc}
\includegraphics[width=0.33\linewidth, angle=0]{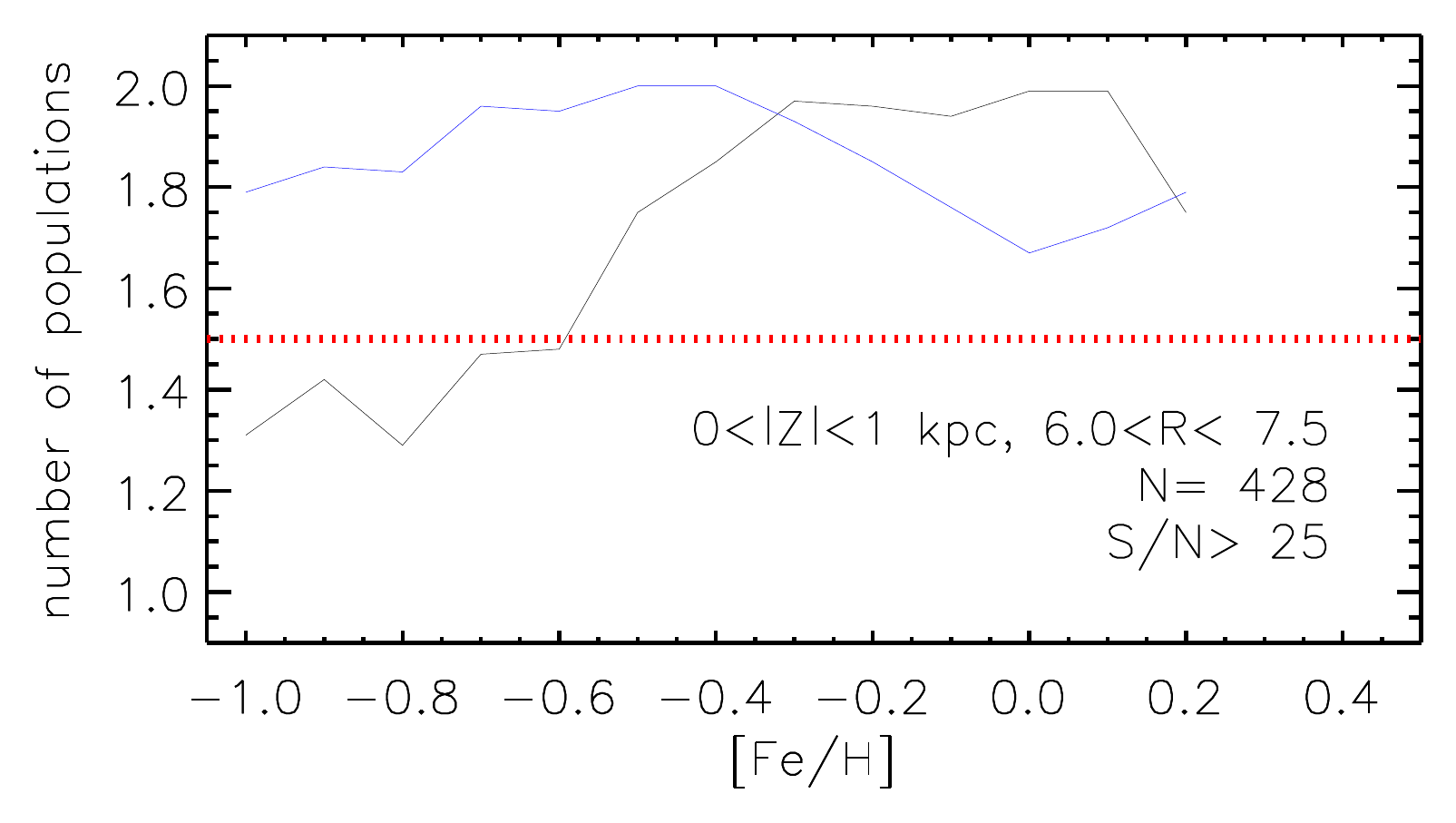}& 
\includegraphics[width=0.33\linewidth, angle=0]{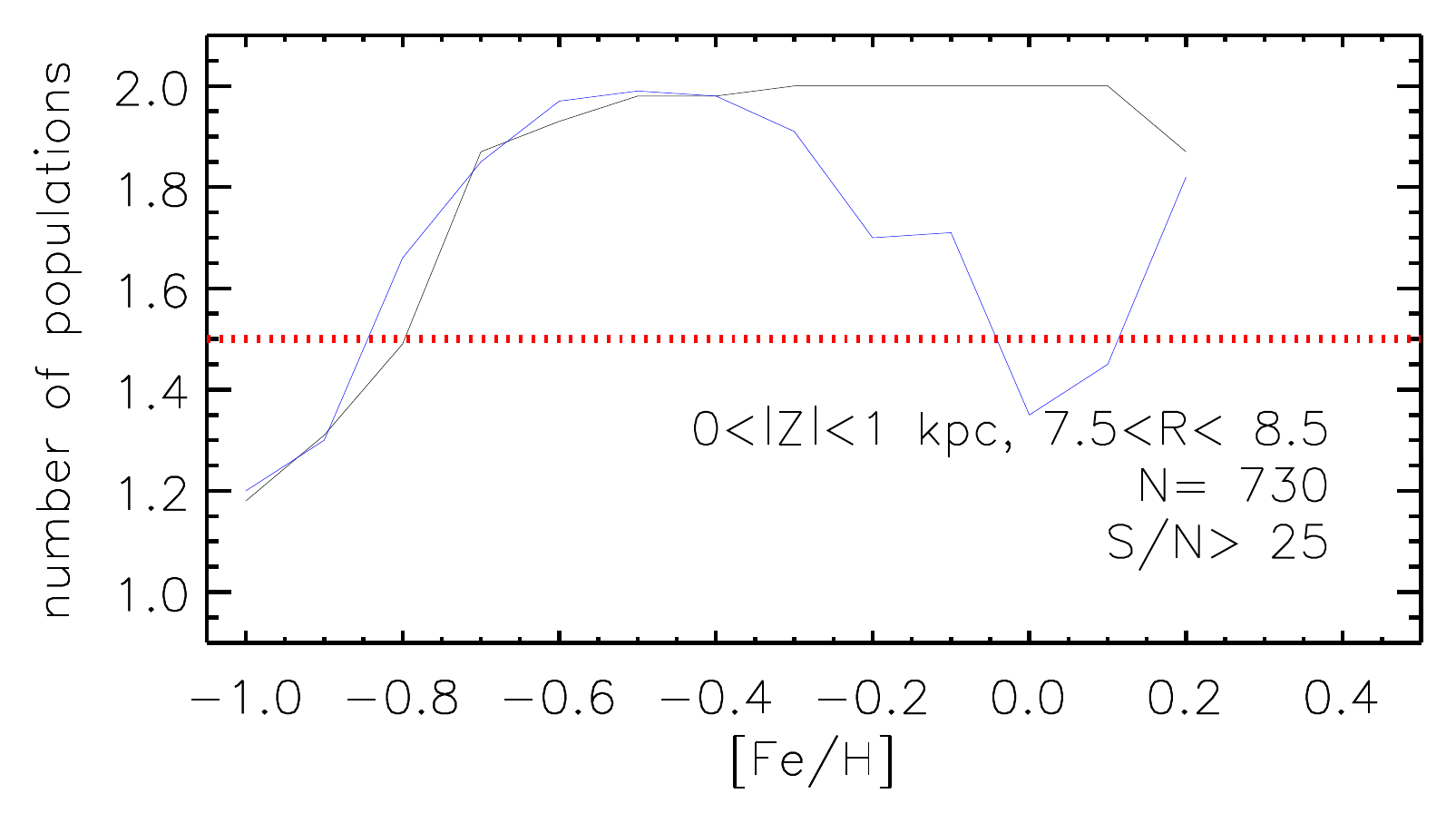} &
\includegraphics[width=0.33\linewidth, angle=0]{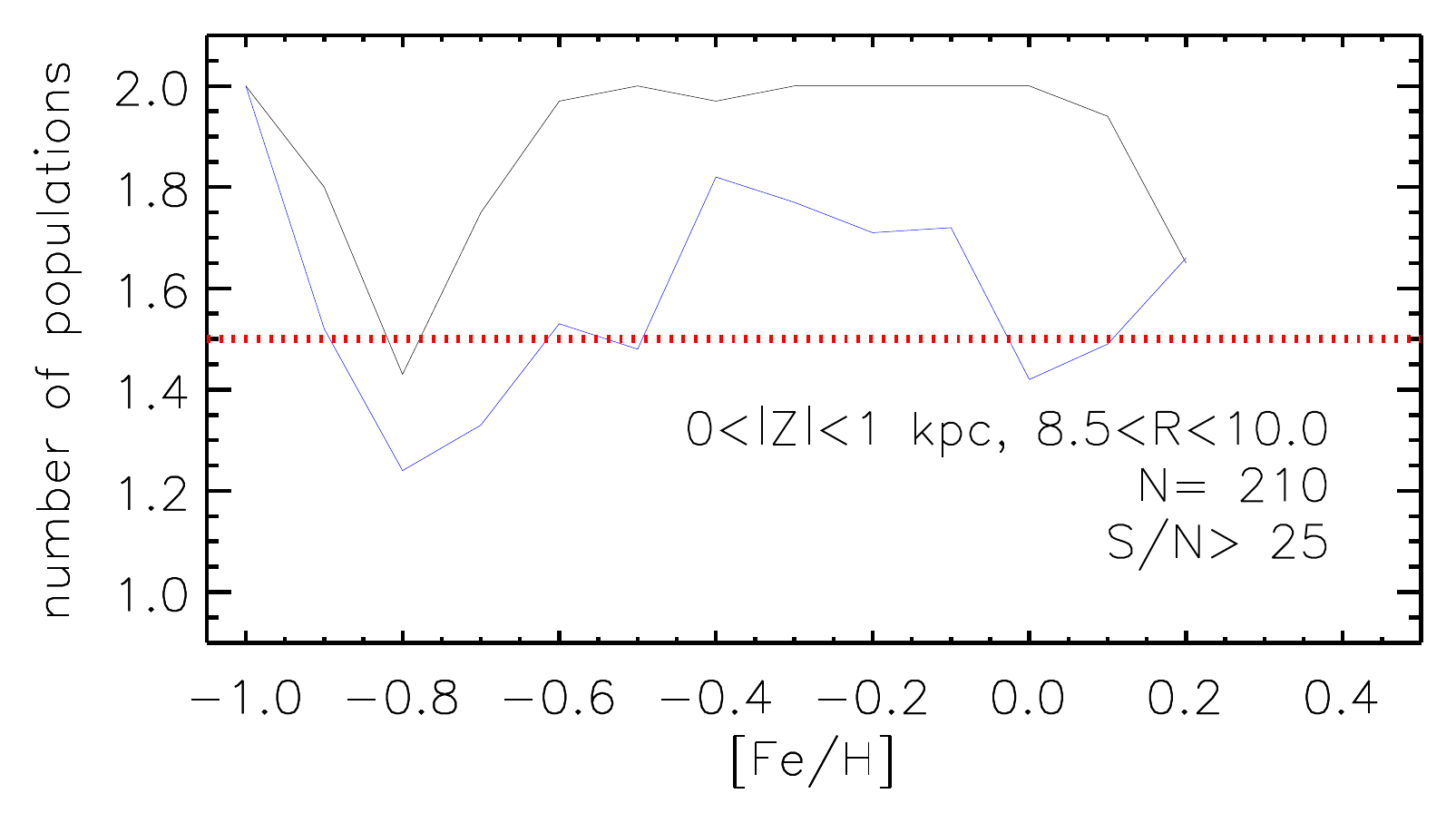} \\
\includegraphics[width=0.33\linewidth, angle=0]{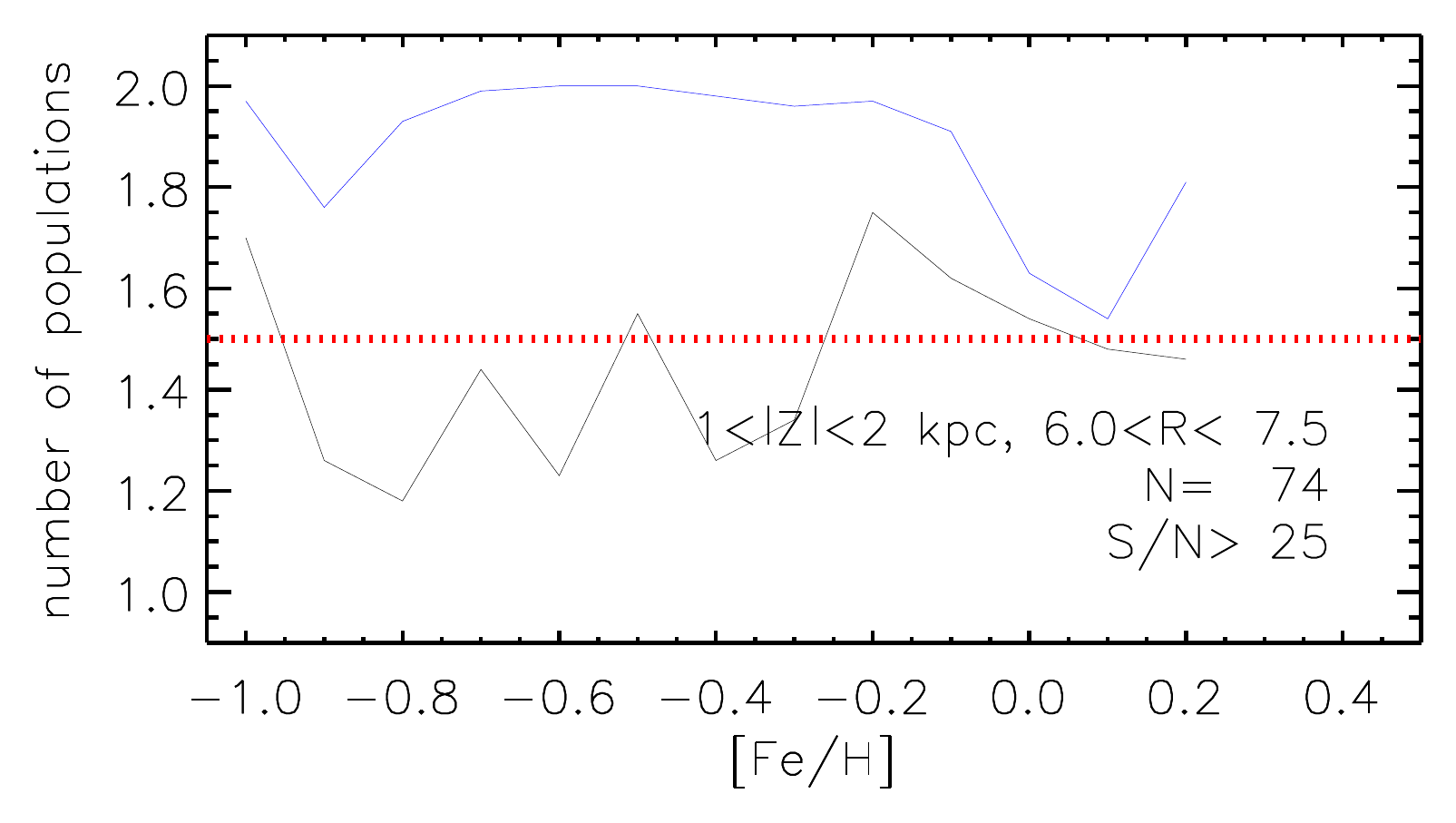}& 
\includegraphics[width=0.33\linewidth, angle=0]{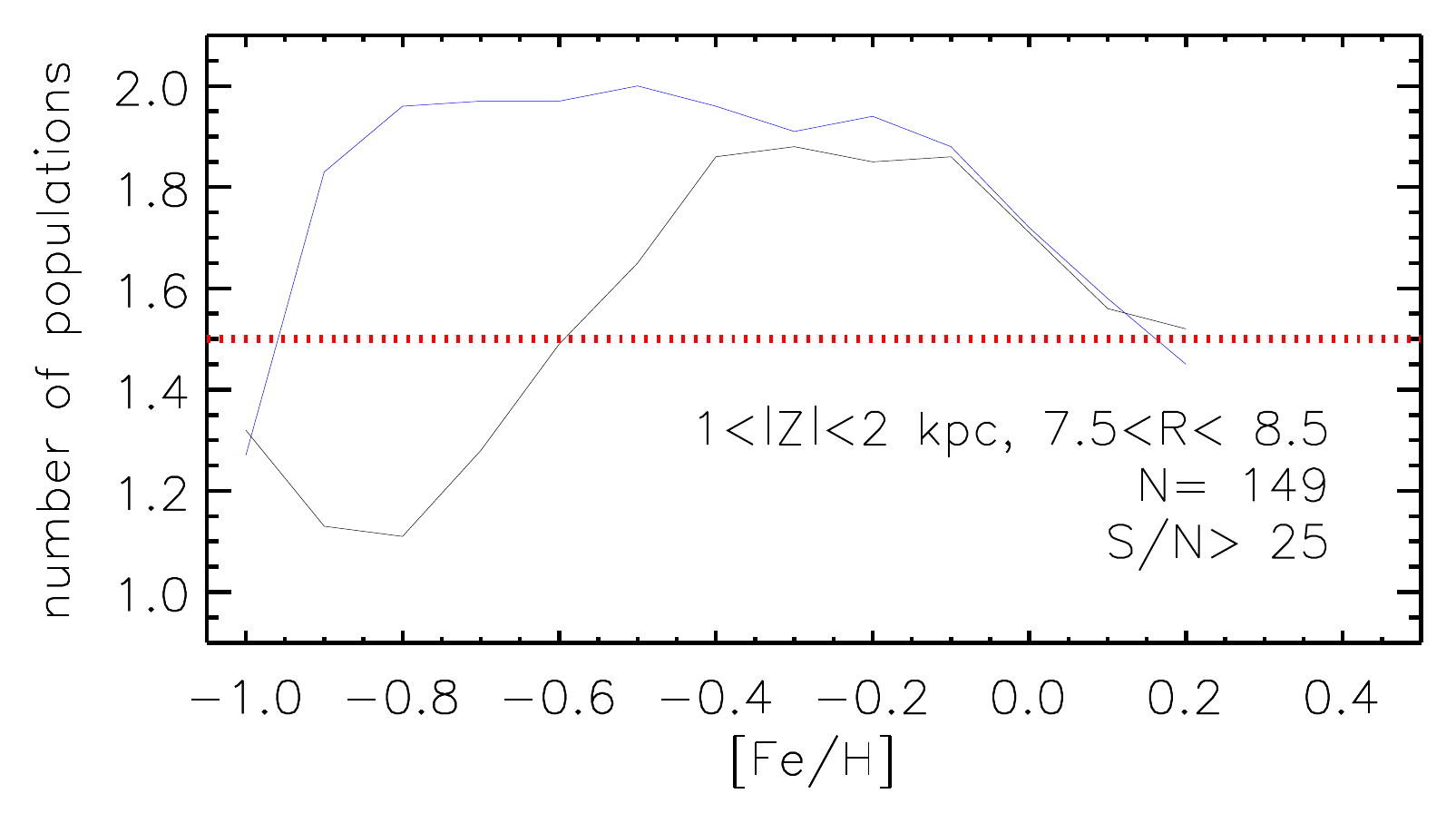} &
\includegraphics[width=0.33\linewidth, angle=0]{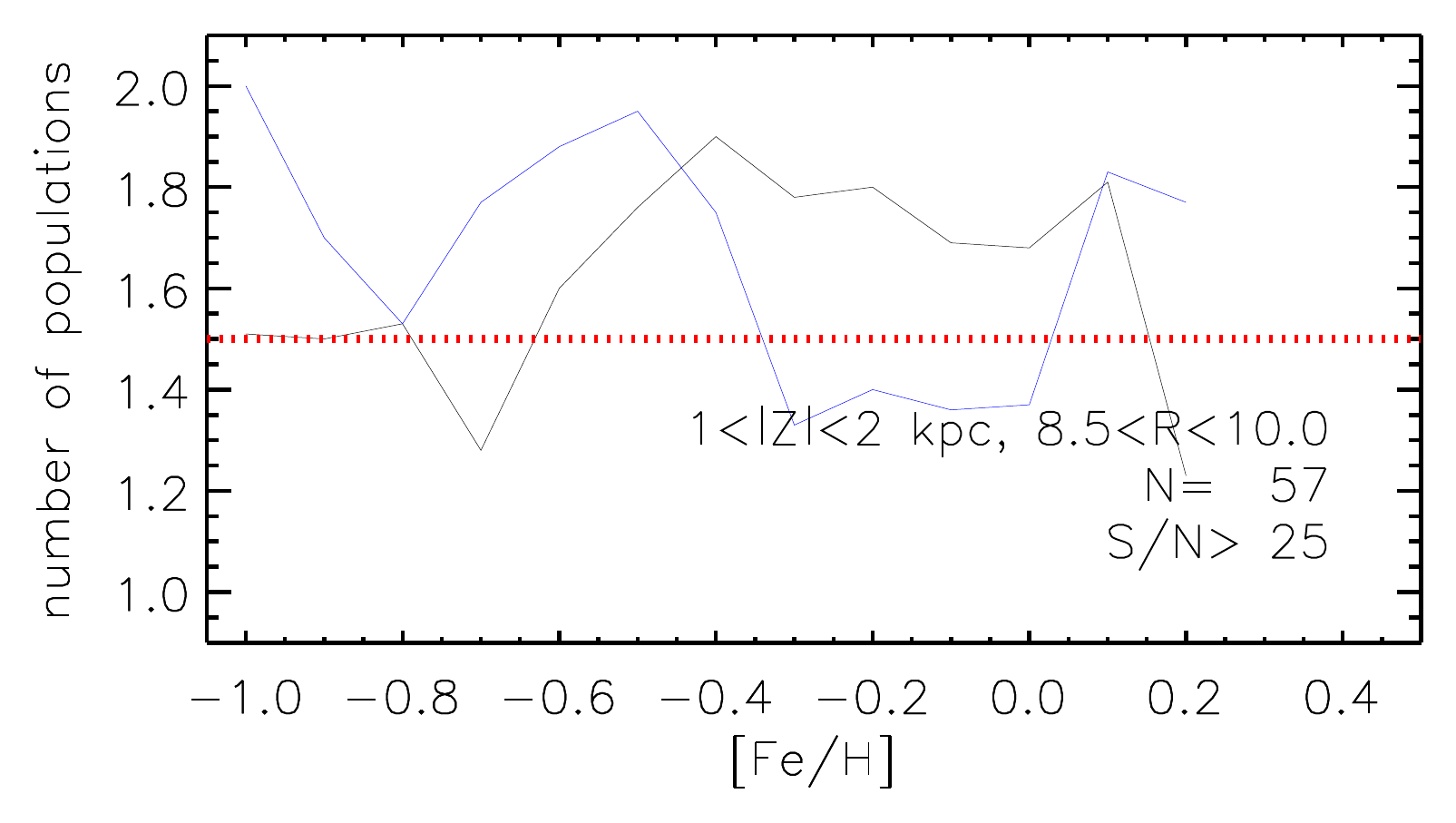}
\end{array}$
\caption{Number of required populations, averaged over 100 Monte-Carlo realisations, as a function of metallicity. For each realisation, the significance of the $\chi^2-$distribution is evaluated, and the likelihood of requiring one or two populations is evaluated. Blue (black) line represents the results for the low$-\alpha$ (high$-\alpha$) population. A value larger than 1.5 indicates that the low$-\alpha$ (high$-\alpha$) population cannot describe by itself the observed trends. }
\label{fig:chi2_significance}
\end{figure*}

 For the considered local sample, the relative proportion of the low$-\alpha$ over high$-\alpha$ stars varies from 30 per cent  to 80 per cent in the iron abundance range between $[-0.8,0.0]\dex$, reaching 50 per cent at [Fe/H]$\sim -0.4\dex$ (middle plots of Fig.~\ref{fig:proportions_MW}). 
 Below $\feh \sim -0.5\dex$, we find that the proportion of thin disc to thick disc stars  is slowly decreasing down to $\feh\approx-0.8\dex$. This result is in agreement  with the analysis of Sect.\,\ref{sect:vphi_analysis}, where we investigated the change in the relative proportions of the different stellar populations by analysing the variations in $\partial \vphi / \partial \mgfe$. 
  Indeed, recall that in Sect.\,\ref{sect:vphi_analysis} we found that below $-0.5\dex$, and at least down to $-0.8\dex$, the value of $\partial \vphi / \partial \mgfe$ varied only  slowly as a function of metallicity (albeit with large error bars) which is consistent with only a  small variation in the relative  number of stars of each disc within this metallicity range. This therefore also implies that the shape of the metallicity distribution functions of the discs are significantly non-Gaussian \citep[see also][]{Hayden15}.
  We note that these results are robust to S/N selections  and should not be dependent of the selection function, assuming there is no bias in $\afe$ at fixed $\feh$.
For the most metal-poor bin ($\feh=-0.9\dex$), we find that the proportion of the low$-\alpha$ stars increases up to 45 per cent. Combined with the significance of the fit, and the increase of the $\mgfe$ dispersion (see above), this indicates that this increase is not real, and is rather due to low number statistics.

 Finally, we notice that both sequences are seen at all the metallicities, from $-0.8\dex$ up to super-Solar values. However, the fact that the relative proportion  trend reverses for $\feh\gtrsim0.1$ (middle-top panel of Fig.~\ref{fig:proportions_MW})  suggests  that the number of stars in these bins is too small to perform a robust likelihood estimation and therefore separating the two populations when the sequences merge is challenging  (as already mentioned in Sect.~\ref{sect:SNR_cuts}).
 This is also indicated from the plot of Fig.~\ref{fig:chi2_significance}, where the significance of having only a thin disc rather than two populations (blue line), is noisy above solar metallicities.

  We now assess to what extent the derived proportions are in agreement with the expected metallicity and spatial distribution functions of the discs. To do so, we model the expected number, N,  of stars of a given disc as a function of Galactocentric radius, $R$, distance from the mid-plane, $Z$, and metallicity, $\feh$, as:
\begin{equation}
N(R, Z, {\feh}) = f \cdot \exp{\left[-\frac{Z}{h_Z} - \frac{R-R_0}{h_R}\right]} \cdot {\rm MDF},
\end{equation}
where $f$ is the local density of stars belonging to each of the discs, and ($h_Z$,$h_R$) are the scale-height and scale-length. The MDF is the adopted metallicity distribution function, defined as:

\begin{equation}
{\rm MDF} = \sum_{i=1}^3 a_i \cdot \\ \exp \left[- \frac{({\feh} - M_i)^2}{2 \sigma_i^2}  \right].
\end{equation}
It is obtained as the sum of three Gaussians, associated with the metal-poor, intermediate metallicity, and metal-rich regimes,  in order to reproduce  skewed distributions.   The factor $a_i$ is the weight  assigned to each of these metallicity regimes, while $M_i$ and $\sigma_i$ are the mean metallicity and dispersion of the associated  Gaussian.  We assume that the MDF has a fixed shape, but can shift to lower or higher mean metallicities as a function of $R$ and $Z$.
 We therefore define $M_i$, as follows:

\begin{equation}
M_i = \mu_i +Z\cdot\partial \feh / \partial Z + (R - R_0) \cdot \partial \feh / \partial R,
\end{equation}
where $R$ is expressed in kiloparsecs and $\mu_i$ is the mean of the metal-poor, intermediate and metal-rich regimes, as measured at the Solar neighbourhood. The input values for our simple expectation models are summarised in Tables~\ref{tab:MDF_discs} and \ref{tab:scales_discs}.

The adapted parameters are \emph{not} chosen in order to fit the Gaia-ESO data, but are a compilation derived from the literature, based on discs defined either by kinematics, chemical abundance or star-counts.  Therefore, the parameters of Table~\ref{tab:MDF_discs} roughly reproduce the skewness of the MDFs for the metal-weak regime of the thin and thick discs \citep{Wyse95,Kordopatis13c} and  the super-solar metallicity stellar distribution of RAVE \citep{Kordopatis15}. The vertical and radial metallicity gradients of the thin disc are adopted from \citet{Mikolaitis14} and \citet{Gazzano13}, respectively. Finally we adopt the scale-heights and scale-lengths of  \citet[][defined morphologically]{Juric08} and \citet[][defined chemically, see their Tables~1 and 2]{Bovy12b}, to illustrate how the star counts are expected to change when considering a thick disc that is more extended or less extended  than the thin disc. 

\begin{table}
\caption{Input values for the metallicity distribution functions of the discs.}
\begin{center}
\begin{tabular}{cccc}\hline \hline
			& thin disc & thick disc \\ \hline 
$f$ 			& 0.85 	& 0.14 \\
%$h_R (\kpc) $ 	&2.6; 3.8; 3.6  		& 3.6; 2.0; 2.0\\
%$h_Z (\kpc)$ 	& 0.3; 0.27; 0.25 	& 0.9; 0.7; 0.9\\
$a_1 (\dex)$ 		& 0.90 	& 0.80\\
$\mu_1 (\dex)$ 		& 0.0 	& $-0.5$ \\
$\sigma_1 (\dex)$ 	& 0.08 	& 0.35 \\
$a_2 (\dex)$  		& 0.1 	& 0.15 \\
$\mu_2 (\dex)$ 		& $-0.3$ 	& $-0.8$ \\
$\sigma_2 (\dex)$ 	& 0.3 	& 0.5 \\
$a_3 (\dex)$	 	& 0.02 	& 0.05 \\
$\mu_3 (\dex)$ 		& $+0.2$ 	& $+0.2$ \\
$\sigma_3 (\dex)$ 	& 0.5 	& 0.5 \\
$\partial \feh / \partial R~(\dex\kpc^{-1})$ & $-0.05$  & 0.00 \\
$\partial \feh / \partial Z~(\dex\kpc^{-1})$ & $-0.06$ & 0.00 \\
\hline
\end{tabular}
\end{center}
\label{tab:MDF_discs}
\end{table}%

\begin{table}
\caption{Adopted scale-lengths and scale-heights of the discs.}
\begin{center}
\begin{tabular}{ccc|cc}\hline \hline
& \multicolumn{2}{c|}{Thin disc} & \multicolumn{2}{c}{Thick disc}  \\
& $h_R$ & $h_Z$ & $h_R$ & $h_Z$ \\ 
& $(\kpc)$ & $(\kpc)$ & $(\kpc)$ & $(\kpc)$ \\ \hline
\citet{Juric08} & 2.6 & 0.30 & 3.6 & 0.9 \\
\citet{Bovy12b} & 3.8 & 0.27 & 2.0 & 0.7 \\
%K15       & 3.6 & 0.25 & 2.0 & 0.9 \\
\hline
\end{tabular}
\end{center}
\label{tab:scales_discs}
\end{table}%

The expected proportions of thin and thick discs are plotted in Fig.~\ref{fig:proportions_MW}.  One can see that  for the sample in the range $7.5 < R < 8.5\kpc$, close to the plane (middle top panel), the expected trends are not matching correctly the observed ones: the model of \citet{Juric08}  over- (under-) estimates the thick (thin) disc relative proportions, whereas the  inverse is noticed for the \citet{Bovy12b} model. We note, however,  that modifying  the \citet{Bovy12b} model, with a thin disc slightly more compact ($h_R=3.6\kpc, h_Z=0.25\kpc$) and a thick disc with a larger scale-height ($h_Z=0.9\kpc$), gives a trend in a better  agreement with the observations.

%%%%%%%%%%%%%%%%%%%%%%%%%%%%%%%%%
\subsubsection{Far from the plane}

For the stars being between 1 and $2\kpc$ from the Galactic plane (bottom plots of  Figs.~\ref{fig:scatter_and_trends} to \ref{fig:chi2_significance}), we find that the slopes of the mean $\alpha-$abundances are slightly flatter, compared to the derived values closer to the plane ($\mu_{\rm thin}=(-0.17 \pm 0.05) \times \feh + 0.04$, $\mu_{\rm thick}=(-0.21 \pm 0.05) \times \feh + 0.19$). 
In addition, the mean $\alpha-$enhancement of the discs at a given metallicity is  increased by $\sim0.04\dex$ for the thin disc and $\sim 0.07\dex$ for the thick disc (Fig.\,\ref{fig:trends_all_heights}), in agreement with the vertical $\afe$ gradients in the discs found by \citet{Mikolaitis14}.  

In addition, despite the fact that the selection function of Gaia-ESO is not taken into account (see Sect.~\ref{sect:dataset}),  we find that the relative weight of the two populations changes, with the relative importance of the thick disc being higher, as expected \citep[e.g.:][]{Minchev14}. At Solar metallicities, at these heights above the plane, the thick disc represents 50 per cent of the stars. 
Compared to our star-count models, the agreement with the \citet{Bovy12b} model is very good, in particular for the metal-poor part ($\feh\lesssim -0.4$), however with again an over-estimation of the thin disc relative proportion at the higher metallicities. %, with just a simple overestimation of the relative number of thick disc stars at the metal-poor end. 

Finally, the $\afe$ dispersions show mild trends (Fig.~\ref{fig:dispersions_and_KS}):  the thick disc's one decreases from $0.07\dex$ at low metallicities to $0.05\dex$ at super-solar values, as does the thin disc's, with however an important increase for metallicities below $-0.6\dex$. Nevertheless, this increase is only due to the low number of stars available  to fit for the low$-\alpha$ sequence, as shown by the relative proportion in Fig.~\ref{fig:proportions_MW} and the significance of the fit for metallicities below $\feh \sim -0.6\dex$ (Fig.~\ref{fig:chi2_significance}).

%%%%%%%%%%%%%%%%%%%%%%%%%%%%%%%%%%%%%%%%

\begin{figure}
\centering
\includegraphics[width=\linewidth, angle=0]{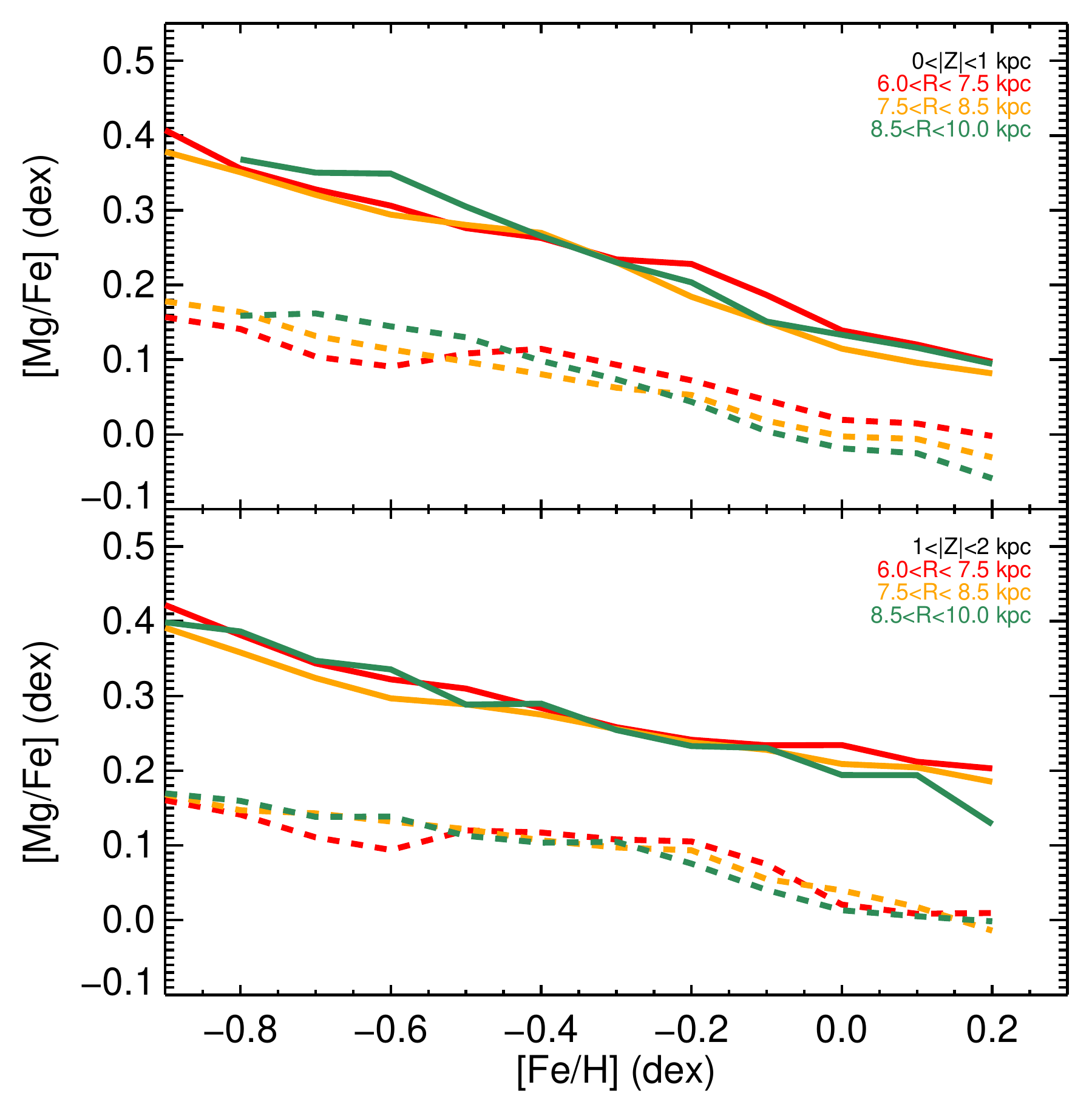}
\caption{Trends at different Galactocentric radii for given bins of distances from the Galactic plane. Plain and dashed lines are associated to the high$-\alpha$ and low$-\alpha$ populations, respectively. }
\label{fig:trends_all_radii}
\end{figure}

\subsection{Investigation beyond $7.5<R<8.5\kpc$}
\label{sect:other_radii}
The same fitting procedure as for the Solar suburbs  has been repeated for radii intervals between $6.0<R<7.5\kpc$ and $8.5<R<10.0\kpc$. The bin sizes are purposely wider than the ones at the Solar suburb, in order to have enough stars to analyse, and in order to minimise the impact of having an inhomogeneous sky coverage (see Fig.~\ref{fig:R_Z_plane}). 
Figure~\ref{fig:proportions_MW} shows how the relative proportion of each population changes as a function of the region in the Galaxy. One can see that the proportion of the thick disc decreases with $R$, both close and far from the plane, suggesting that the thick disc is more radially concentrated than the thin disc, in agreement with previous surveys \citep[e.g.:][]{Bensby11,Cheng12,Bovy12b,Anders14,Recio-Blanco14, Mikolaitis14},  and that the metal-weak tail of the low$-\alpha$ population does not extend below $-0.6$ in the $6.5<R<7.5\kpc$ ring. 
Our star-count models achieve a fitting of the relative proportions in different ways:  for the range $6.0<R<7.5\kpc$, the model of \citet{Juric08}  (thick disc longer than the thin disc) seems more appropriate, however, the \citet{Bovy12b} model, which is assuming a shorter scale-length for the thick disc, is in  better agreement for the stars  located at the largest radii.  This disagreement could indicate a varying scale-length or scale-height with $R$ for either  or both of the discs, as recently suggested by \citet{Minchev15}.

The $\mgfe - \feh$ slopes as a function of $\feh$  at the three Galactic regions are shown in Fig.~\ref{fig:trends_all_radii}, and summarised in Table\,\ref{tab:slopes_disc}.
Overall we find that as a function of $R$, the thick disc has similar trends, whereas the thin disc slopes become marginally steeper  (from $-0.14\pm 0.07$ at $R\leq 7.5\kpc$ to $-0.23\pm 0.03$ at $R\geq 8.5\kpc$). In addition the slopes become, on average, flatter as a function of $Z$, by $0.03\pm0.05$ for the thin disc and by $0.07\pm0.04$ for the thick disc.

The  result from the Gaia-ESO survey obtained here, that the high-$\alpha$ thick disc sequence varies little spatially, is 
 consistent with the conclusions from the APOGEE data for red clump stars published
recently  by \citet{Nidever14}. Those authors  found that the overall spatial variations of the
$\afe-\feh$ sequence of the high$-\alpha$ stars do not exceed 10 per
cent. 

Finally, we find that the mean $\alpha-$enhancement for both discs are increased higher from the plane, at all Galactocentric radii. 
This is indicative of a vertical gradient in $\alpha-$elements in the thin disc, in agreement with the value of $+0.04\dex\kpc^{-1}$ measured in \citet{Mikolaitis14}.

\begin{figure}
\centering
\includegraphics[width=\linewidth, angle=0]{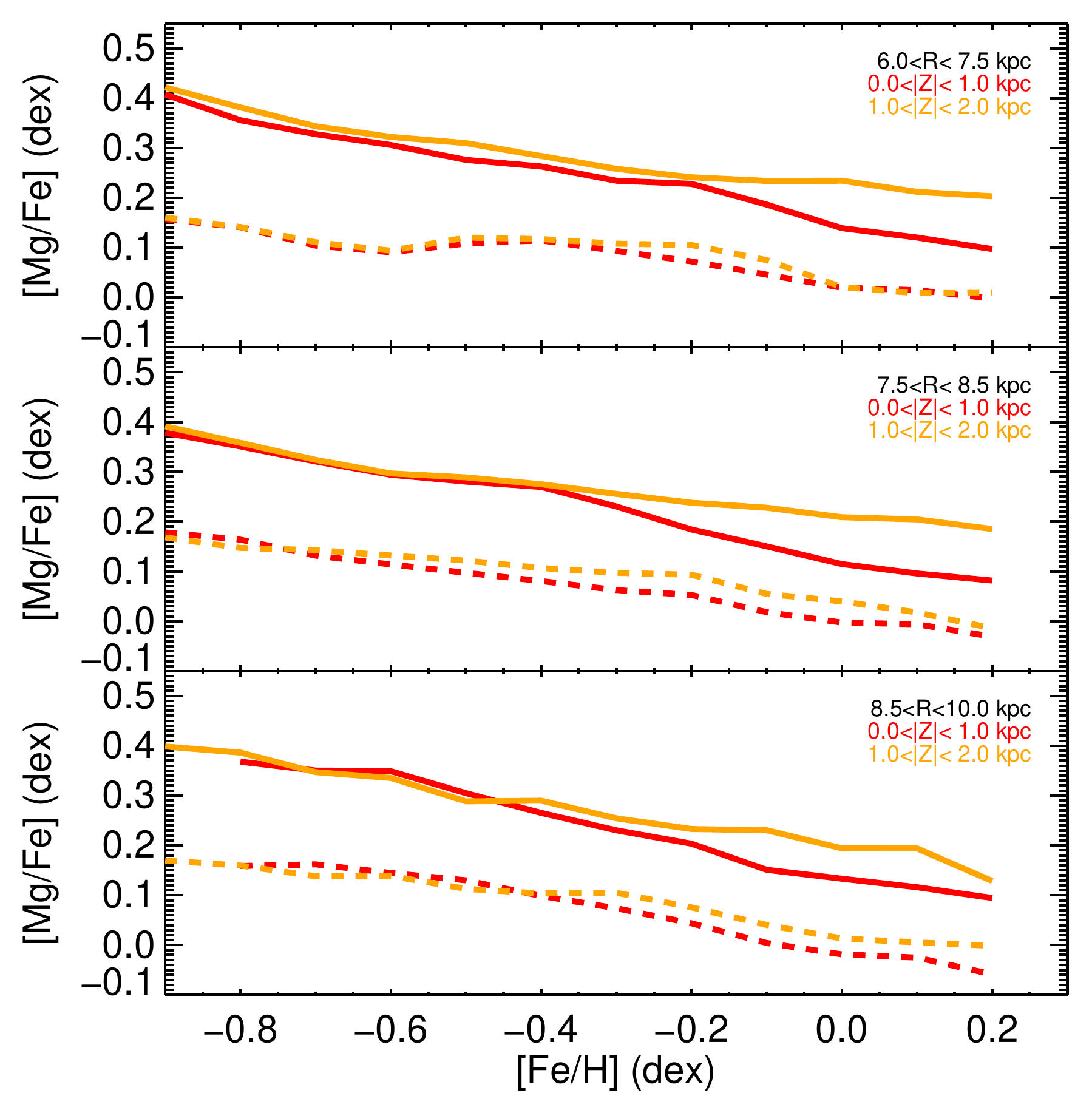}
\caption{Trends at different distances from the Galactic plane:  $6 < R < 7.5\kpc$ (top), $7.5 < R < 8.5\kpc$ (middle) and $8.5 < R < 10\kpc$  (lower panel).   }
\label{fig:trends_all_heights}
\end{figure}

%%%%%%%%%%%%%%%%%%%%%%%%%%%%%%%%%%%%%%%%

%%%%%%%%%%%%%%%%%%%%%%%%%%%%%
%%%%%%%%%%%%%%%%%%%%%%%%%%%%%
\section{Discussion: Implications for how the discs evolved}
 \label{sect:interpretation}
The steepness of the $\afe-\feh$  slopes contains information about how  discs have evolved,  encoding mainly  information on the time evolution of  their Star Formation Rates (SFR) and gas in/out flows, assumed to regulate the SFR.  
Typically, the $\afe$ decline with increasing $\feh$  is caused by the time delay distribution that the first supernovae of type Ia (mostly iron producers) have after the core-collapse supernovae (mainly producers of $\alpha-$elements).  
  After the adoption of an IMF, binary fraction and delay time distribution, the relative present to past SFR can be translated into a relative rate of core-collapse to Type Ia supernovae. In turn,  this defines a slope of $\Trend$ for the ISM and hence for newly formed stars, modulo possible elemental-dependent gas in/out flows.
 
 In the case that  most of the stars are formed during a starburst event, followed by a negligible star formation, then the slope of the decline -- tracing the abundance of  the gas -- is maximal:  the $\Trend$ slope connects the last star formed before the end of the starburst and the first one to form after the gas has been enriched by SNIa ejecta. 
On the other hand, flatter slopes, {\it i.e.} slower declines, imply more important contributions from  massive stars through core-collapse supernovae and thus more extended star formation.  
A typical slope for a closed-box star-forming system with a slowly decreasing SFR  as inferred for the solar neighbourhood and standard Type Ia delay time distribution is around $-0.3$ \citep[see e.g. Fig.\,1 of][]{Gilmore91}.  The flattening of the slope as the star-formation efficiency (defined as the reciprocal of the duration of star formation) decreases is illustrated in the top panel of Fig.~15 of \citet{Nidever14}. The middle panel of that figure in \citet{Nidever14} illustrates how changing the  rate of gas outflow can  also change the slope.

 Our work finds steeper slopes for the thick disc ($\approx -0.28\times\feh$) compared to the thin disc, with negligible spatial variations as a function of $R$ and $Z$. On the other hand, the thin disc slopes show a variation as a function of $R$, with the flatter slope being in the inner Galaxy ($-0.14\times\feh$) and the steeper in the outer Galaxy ($-0.23\times\feh$).

\citet{Edvardsson93}  undertook a pioneering study of kinematics and elemental abundances, for a sample of 189 F and G
dwarfs observed at the solar neighbourhood. They investigated the manner in which
the $\afe$\ abundance of a star in a given bin of $\feh$
depends on the star's mean orbital radius, using this as a proxy
for birth radius. 
These authors found that stars from the outer
Galaxy (mean orbital radius $R_m \geq 9\kpc$) follow an
$\afe-\feh$ relation that falls slightly below that followed by
stars with mean orbital radius close to the Solar Circle.  Further,
they found that for stars from the inner Galaxy ($R_m
\leq 7\kpc$), the $\afe$ values split into two groups: one at $\afe=0.2\dex$ for $\feh <-0.4\dex$, 
and one at $\afe=0.03\dex$ for $\feh \geq -0.4\dex$.
The conclusion of their analysis was that the variation  of the $\afe-\feh$ sequences is evidence of inside-out formation of the disc, as the inner disc region had a higher past to present SFR, and the outer parts form more slowly. 

 On the other hand, the two infall model of \citet{Chiappini97}, that also forms the thin disc in an inside-out fashion however with contribution of accreted pristine extra-galactic gas once the thick disc is formed, predicts opposite trends: the inner Galaxy has flatter $\afe-\feh$ trends, as a consequence of more important quantity of infalling gas and therefore more extended star formation history. Nevertheless, the difference in the trends that is predicted by this model is rather small and is only noticeable over a large radial range \citep[from $4\kpc$ up to $16\kpc$, see][their Fig.~14]{Chiappini97}.  It is clear that the results are quite sensitive to the details of the adopted model.

Our results, based upon an analysis using  the observed positions of the stars\footnote{As opposed to an estimate of the mean orbital radius used by \citet{Edvardsson93}. The radial excursions associated with epicyclic motions for thin disc stars at the Solar neighbourhood with $\sigma_V\sim30\kms$ are of the order $1\kpc$ \citep{BinneyTremaine08}. } find that the  slopes of the $\afe-\feh$ sequences are flatter for thin-disc stars in the inner Galaxy, compared to the outer parts. 
However, the radial range that is spanned by the Gaia-ESO targets, combined with the fact that the spatial variations are below $2\sigma$, cannot allow us to discard either of the models. 

On the other hand, we find almost no variations (differences are below $1\sigma$) of the trends for the thick disc, in agreement with recent results from \citet{Nidever14} using APOGEE data.
 To explain the invariance of the thick disc sequence, \citet{Nidever14} suggested the existence of a large-scale mixing in the early ``turbulent and compact'' disc. We note, however, that in such a  disc, the velocity dispersion of the gas should be high, in order to mix efficiently the metals from the inner Galaxy with regions at $R\approx 5\kpc$. This could potentially be incompatible with the observed velocity dispersion of the thick disc stars that formed from that gas, unless they were formed during dissipational collapse. Such a collapse would imply vertical gradients in metallicity, $\alpha-$elements and azimuthal velocity, all being still a matter of debate in the literature \citep[see][for an overview of the published values]{Mikolaitis14}, though being compatible with our data.   More theoretical work is needed to quantify the expected amplitude of any vertical gradient.

Indeed, we found that the $\alpha-$enhancement of the populations increase as a function of the distance from the mid-plane. For the thin disc, this vertical gradient is a signature of the age-velocity relation: older stars (the more $\alpha-$enhanced) have larger random velocities than newly born stars, plausibly  due to  interactions with gravitational perturbations such as spiral arms and Giant Molecular Clouds. On the other hand, the saturation of the thin disc age-velocity dispersion at $\sigma_Z \sim 20\kms$ \citep{Nordstrom04,Seabroke07}, well below the dispersion of the thick disc, rules out such an explanation for the vertical gradient in the thick disc.  In addition,  
such a vertical gradient is difficult to  explain by a single accretion event that formed the thick disc from the accreted stars, as for example suggested by \citet{Abadi03}. A gradient as observed would, however, naturally arise in the scenarios where the thick disc 
is formed {\it (i)} from the stars formed during dissipational settling \citep[e.g.][]{Brook04,Minchev13,Bird13}  {\it (ii) } 
by the dynamical heating of a pre-existent thin disc (with pre-existent vertical $\afe$ gradient resulting from an age-velocity dispersion relation) due to minor mergers \citep[e.g.:][]{Wyse95,Villalobos08, Minchev14}, or {\it (iii)}  by a thick disc formed by radial migration \citep{Schonrich09b}. We note, however, that the simulations of thick disc formation through stellar radial migration \citep[e.g.:][]{Loebman11}  have  difficulties reproducing many of the observed thick disc properties \citep{Minchev12b}.    
 
%%%%%%%%%%%%%%%%%%%%%%%%%%%%%%%%%%%%%%%%%%%%
\subsection{The low-metallicity extent of the thin disc}
\label{sect:MWThinDisc}

The extent of the thin disc sequence to low metallicities and its
overlap in iron abundance with that of the thick disc provides an
important constraint on the evolutionary sequence of the discs. The
low-metallicity limit is of particular interest in models in which
there is a hiatus in star formation between thick and thin discs, with
inflow of metal-poor gas, and good mixing, to reduce the mean
enrichment of the proto-thin-disc material \citep[e.g.][see also
discussion in \citealt{Wyse95}]{Haywood13, Snaith14b}.  Obviously the
lower the metal-poor limit of the thin disc, the larger the mass of
metal-poor gas that must be accreted. Our work has shown that, close
to the plane, the thin disc extends at least down to
$\feh\sim-0.8\dex$ in the  Galactocentric radial range $7.5 < R < 10\kpc$ and at least down to $\feh\sim-0.6\dex$ in the  range $6.5 < R<7.5\kpc$. A lower limit of $-0.6\dex$ also seems to be reached far from the plane ($1<|Z|<2\kpc$), at all the investigated radii.
The presence of thin-disc stars
below these metallicities is less secure. Indeed, given the
uncertainties on the relative proportion of the populations  the uncertainties on the 
$\afe$ dispersions  as well as the uncertainties on the results of the log-likehood ratio tests,
we cannot  state with certainty that 
the thin disc, as defined by the low$-\alpha$
elemental abundance sequence, extends down to $\feh\sim -1\dex$.  A
more robust determination of the proportion of thick- to thin-disc
stars requires that the selection function of the Gaia-ESO Survey be
corrected for, and this is beyond the scope of the present paper. We
note that when the thick and thin discs are defined through their
vertical kinematics or vertical extent from the mid-plane, very
similar results to those of this paper are obtained for the metal-poor
extent of the thin disc -- down to $\sim -0.8\dex$ in iron abundance
-- and for the relative proportions of thick- and thin-disc stars
below $-0.4\dex$ \citep{Wyse95}.

The metal-poor low$-\alpha$ stars could have formed after significant
metal-poor gas infall into the disc, or, assuming inefficient mixing,
in a relatively metal-poor region.   Alternatively these 
stars could have formed in the outer Galaxy, where the gas was
less-enriched, and migrated radially inwards.  In the absence of good
estimates of the individual stellar ages, it is difficult to
distinguish these possibilities.

\begin{figure*}
\centering
$\begin{array}{ccc}
\includegraphics[width=0.3\linewidth, angle=0]{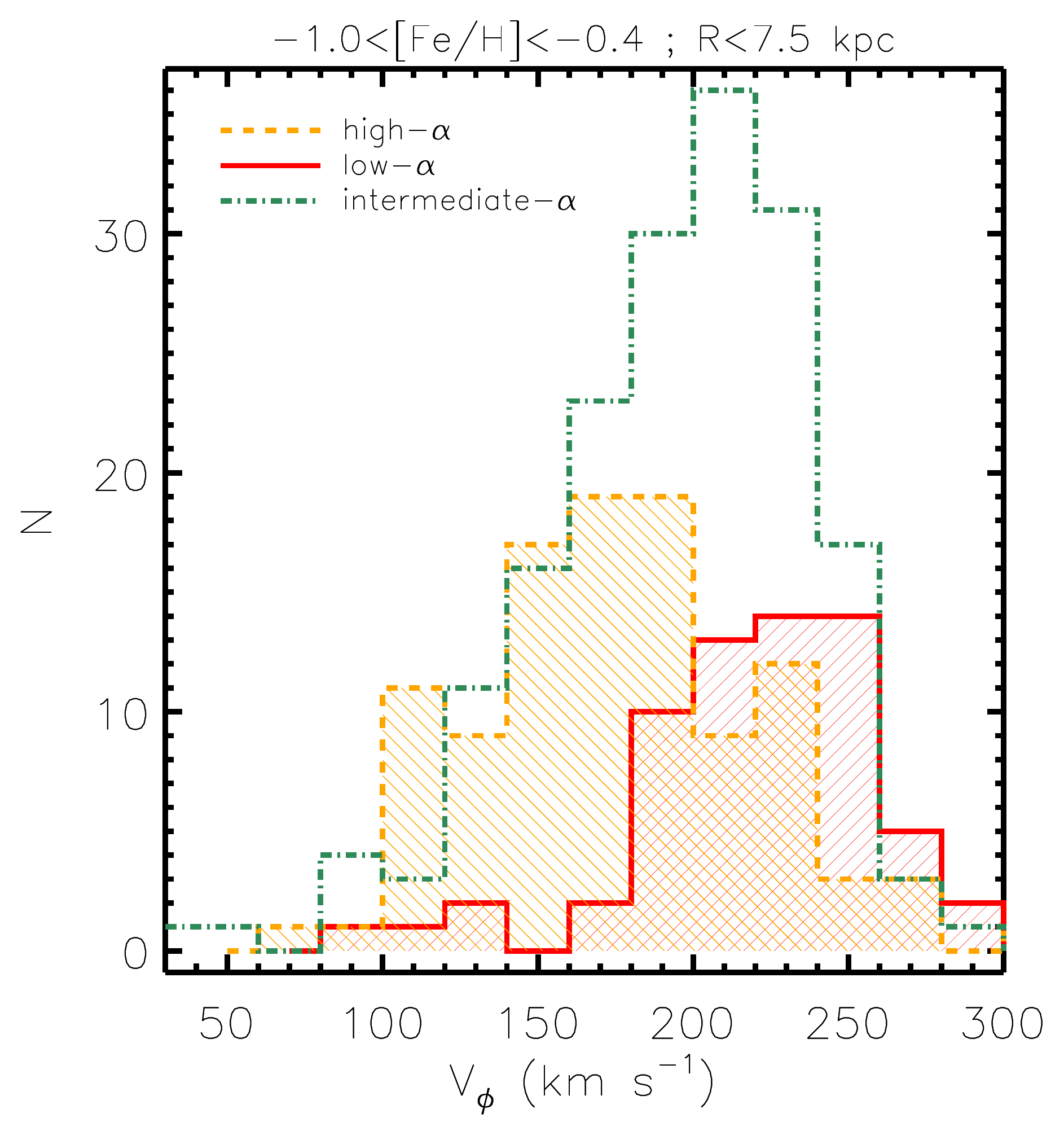}& 
\includegraphics[width=0.3\linewidth, angle=0]{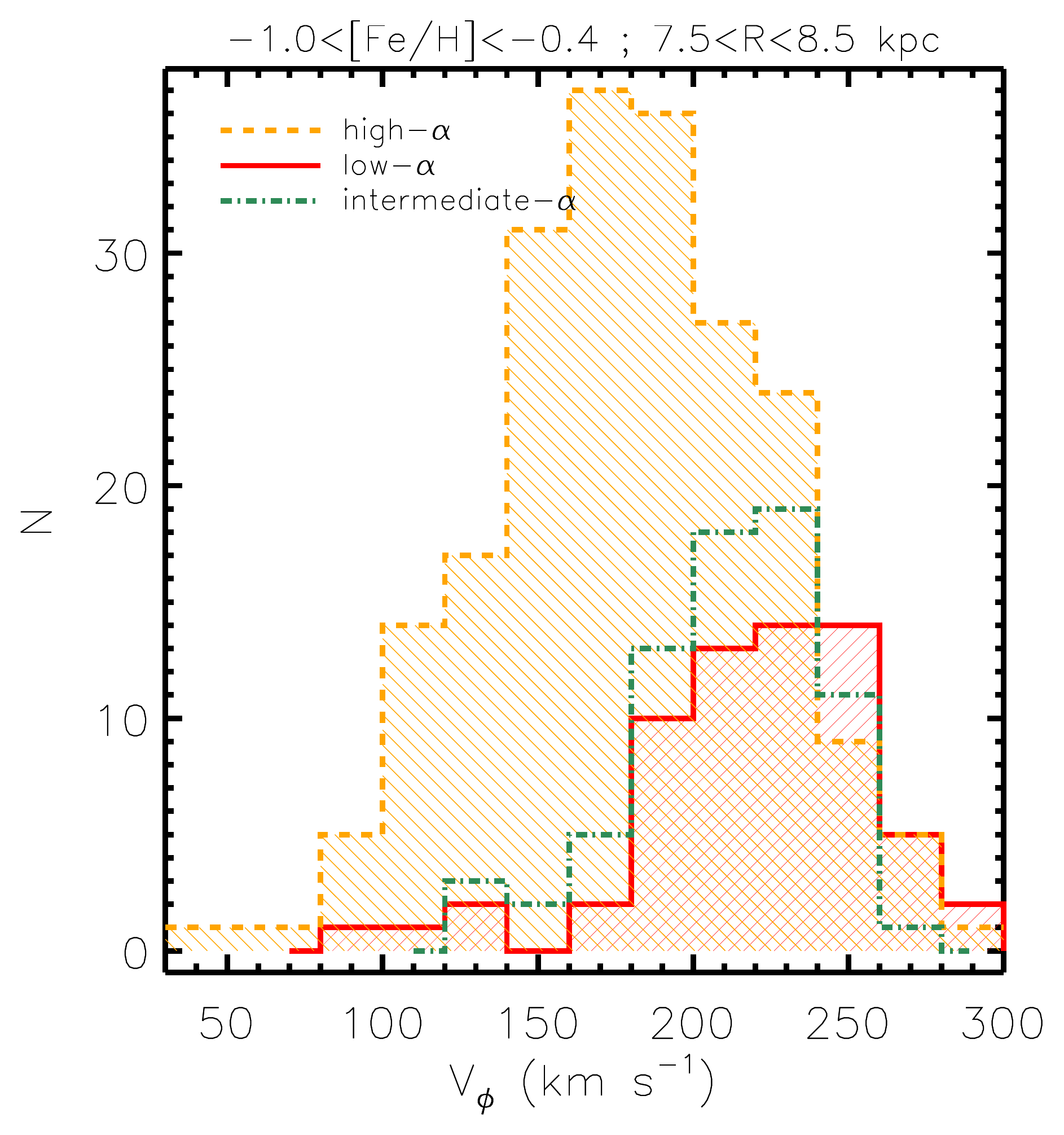}  &
\includegraphics[width=0.3\linewidth, angle=0]{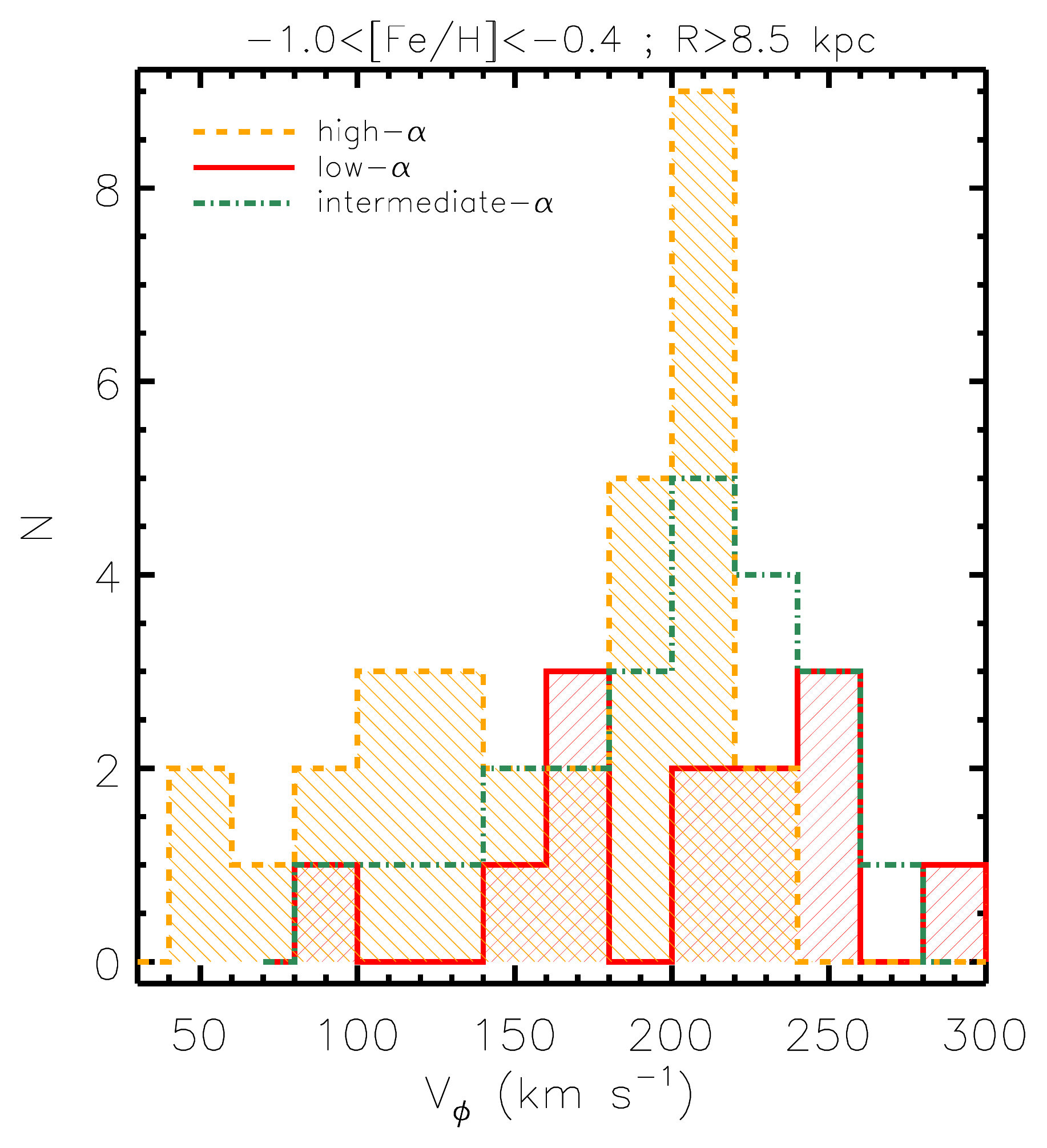}\\
\includegraphics[width=0.3\linewidth, angle=0]{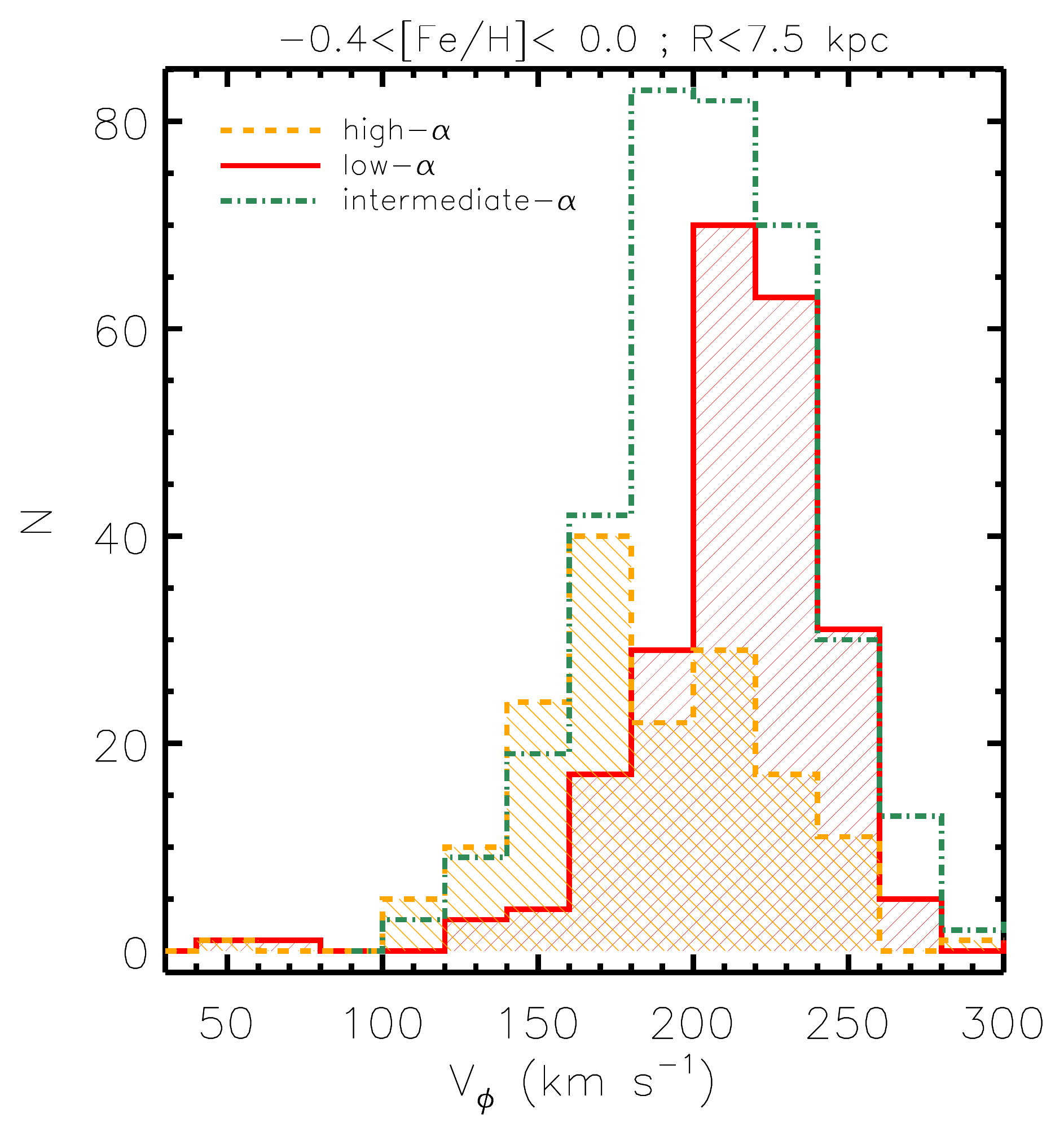}& 
\includegraphics[width=0.3\linewidth, angle=0]{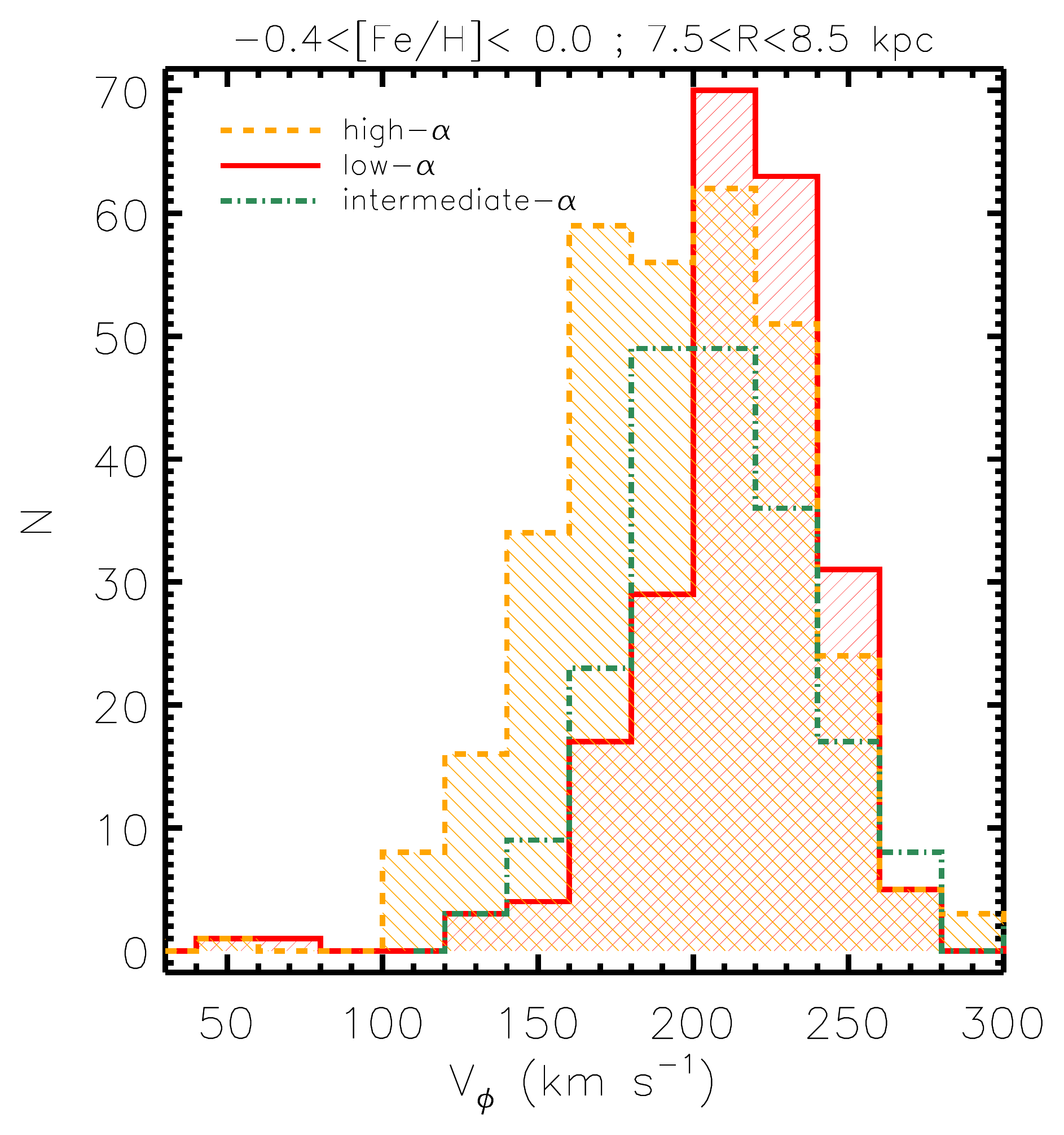}  &
\includegraphics[width=0.3\linewidth, angle=0]{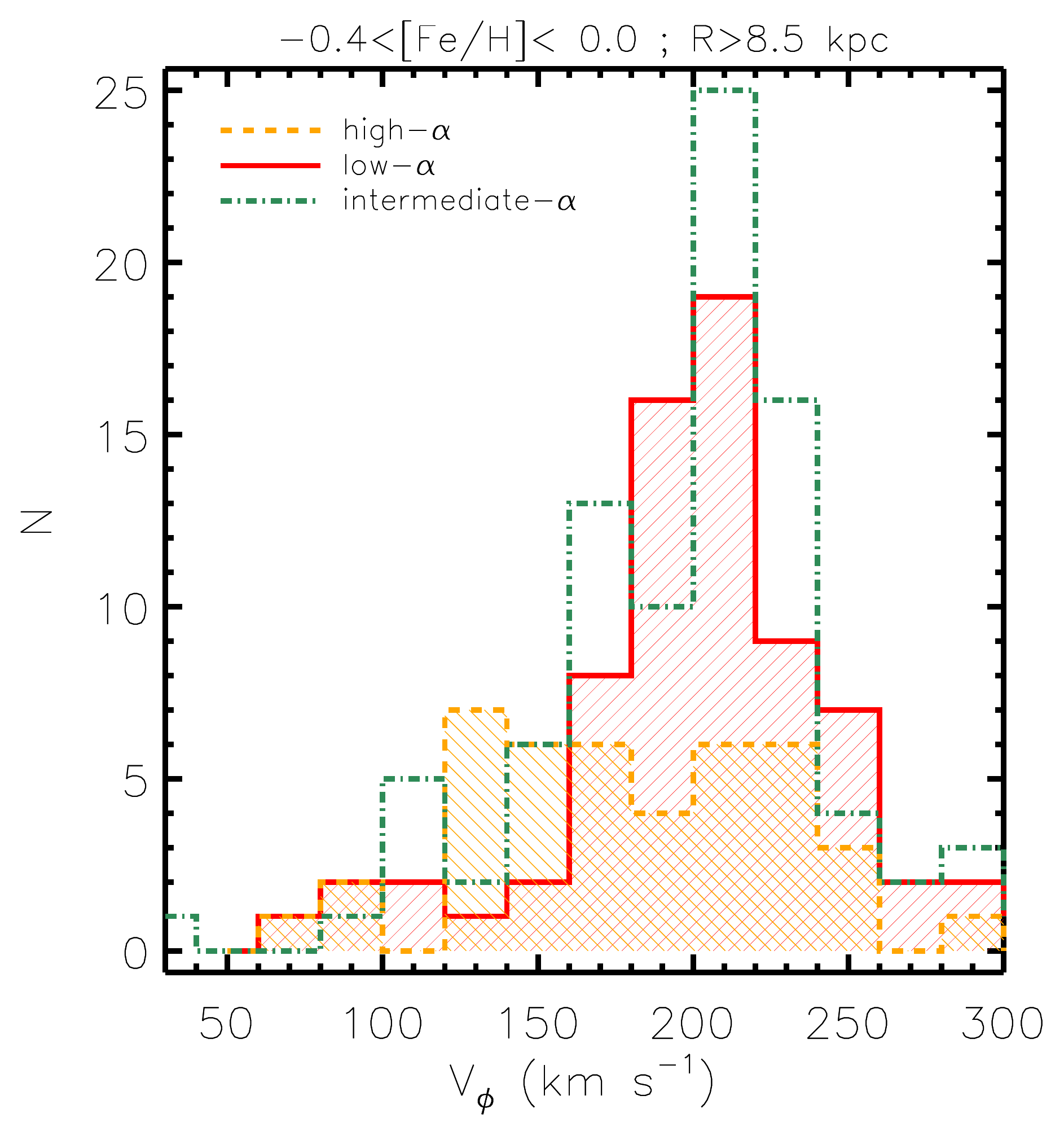}\\
\end{array}$
\caption{$\vphi-$histograms of the high$-\alpha$ (orange), low$-\alpha$ (red) and intermediate$-\alpha$ (green) stars at two different metallicity regimes: metal-poor (top) and intermediate (bottom) metallicity.  The stars are selected to be close to the Galactic plane ($|Z|<1\kpc$) at three Galactocentric radial ranges (left panels: $R\leq 7.5\kpc$, middle panels: $7.5 < R \leq 8.5\kpc$, right panels: $R> 8.5\kpc$). The Kolmogorov-Smirnov probabilities that the intermediate$-\alpha$ histograms are drawn from the distributions as the low$-$ or high$-\alpha$ histograms are shown in Table~\ref{tab:KS-tests}.}
\label{fig:velocity_histograms}
\end{figure*}

We nevertheless  searched within our Solar suburb sample for contamination by stars formed at  other radii. This contamination  could include stars indistinguishable by their kinematics \citep[{\it i.e.} radial migrators from corotation resonances with the spiral arms, see][]{Sellwood02} 
and stars on apo- and pericenters currently visiting the solar region \citep[which could also be radial migrators that changed their angular momentum at the Lindblad resonances with the spiral arms  and thus are on more elliptical orbits, see][]{Sellwood02}. 
 Since velocity can give us a partial hint on the origin of the stars,  we used the slopes defined in Table~\ref{tab:slopes_disc}, to select stars that have a $\mgfe$ ratio below these trends for the thin disc, and above these trends for the thick disc. We then investigated how their azimuthal velocity distributions compared. The stars residing between these cuts were then labelled as ``intermediate$-\alpha$'' stars ({\it i.e.} potentially being inside the gap).

%====================================================
The histograms of Fig.~\ref{fig:velocity_histograms}, obtained for the
 sample having $S/N \geq 10$, show that the low$-\alpha$
population has a broader velocity distribution at the metal-poor
regime  than at the metal-rich regime, for all the three investigated Galactic volumes. This is indicating
that most (but not all) of the these metal-poor low$-\alpha$ stars have been kinematically heated by interactions at Lindblad resonances and that they are very likely close to their pericentre ({\it i.e.} their guiding centre is at larger radius than their observed one). Indeed,   
 assuming a negative metallicity gradient for the interstellar medium \citep[e.g.][]{Genovali14}, it follows that many of these low metallicity stars would have been formed in the outer Galaxy. Furthermore, 
noticing that the velocity distribution of the low metallicity low$-\alpha$ stars is also shifted towards higher $\vphi$, this again implies that these stars are at their perigalacticon, increasing their velocity as their Galactocentric radius decreases. 
 We note that on the other hand, the
high$-\alpha$ population exhibits the typical correlation between the
azimuthal velocity and the metallicity of the thick disc 
\citep[measured for a thick disc defined by metallicity only,
e.g.,][]{Kordopatis11b,Kordopatis13c,Kordopatis13a}.   From that, we
conclude that it does not show any particularly odd kinematic
behaviour for its low metallicity stars.  

Finally, in order to assess whether there is a true gap between the high$-\alpha$ and low$-\alpha$ chemical paths, we have computed the Kolmogorov-Smirnov (KS) probabilities,  between the velocity distribution of the intermediate$-\alpha$ population and the ones of the  thin and thick discs (green, red and orange histograms in Fig.~\ref{fig:velocity_histograms}, respectively).  These probabilities quantify the significance of two datasets being drawn from the same distributions. The results shown in Table~\ref{tab:KS-tests} indicate that the intermediate$-\alpha$ velocity distribution is always more resembling to the  low$-\alpha$ velocity distribution than the high$-\alpha$ one, therefore suggesting that 
there is a true gap between the sequences, rather than
a trough. 

\begin{table}
\caption{Kolmogorov-Smirnov probabilities between the azimuthal velocity distribution of the intermediate-$\alpha$ population and the ones of the low$-$ and high$-\alpha$ populations.}
\begin{center}
\begin{tabular}{cc|cc}\hline \hline
&		& low$-\alpha$ & high$-\alpha$ \\ \hline 
$R\leq7.5\kpc$&$\meta <-0.4$& $2.1\cdot10^{-3}$ & $3.8\cdot10^{-4}$\\
&$\meta >-0.4$& $1.4\cdot10^{-5}$ & $5.1\cdot10^{-9}$\\ %\hline
$7.5 < R\leq8.5\kpc$&$\meta <-0.4$& $0.14$ & $8.7\cdot10^{-4}$\\
&$\meta >-0.4$& $3.2\cdot10^{-4}$ & $1.0\cdot10^{-4}$\\ %\hline
$R\geq7.5\kpc$&$\meta <-0.4$& $0.42$ & $0.30$\\
&$\meta >-0.4$& $0.51$ & $0.02$\\ %\hline
\hline
\end{tabular}
\end{center}
\label{tab:KS-tests}
\end{table}%

%%%%%%%%%%%%%%%%%%%%%%%%%%%%%%
\subsection{The high-metallicity extent of the thick disc}

Similar to the considerations of the low metallicity tail of the thin disc, the
high-metallicity extent of the thick disc holds information about the
degree of interdependence in the evolution paths of the
discs.
Indeed, \citet{Haywood13} and \citet{Bensby14} suggested that
at least part of the metal-poor end of the thin disc could have
started forming from either accreted gas at the end of the star
formation epoch of the thick disc and/or from gas
expelled from the thick disc \citep[see also][for an example of the latter type of 
model, or \citealt{Chiappini97} for the two infall model]{Gilmore86}.

We have demonstrated the presence of  high$-\alpha$ stars up to Solar metallicities, and probably up to $+0.2\dex$. 
This result is in agreement with previous studies, \citep[e.g.:][most recently]{Adibekyan13,Bensby14,Nidever14}; see also \citet{Bensby07, Kordopatis13c, Kordopatis15} for further evidence of metal-rich thick disc stars, defined kinematically.  
We find that the dispersion in $\mgfe$ at fixed iron abundance for thick-disc stars 
 is marginally smaller than the values derived for the thin disc. 
 Assuming that this dispersion reflects initial conditions, the smaller dispersion is indicative of a  better-mixed ISM for the thick disc compared to the thin disc.  Possible reasons for that could either be accreted gas that would change the ISM composition, or stellar radial migration that would mix stars formed in different regions of the Galaxy. 
However,   even if our results seem to indicate a separation between the sequences over the entire $\feh$ range, this might not be the case for the most metal-rich end, since our procedure always tries to fit two components to the data.  

 Finally, we note that \citet{Martig15} and \citet{Chiappini15} have found by combining APOGEE abundances and astero-seismic ages, high$-\alpha$ stars with  ages younger than $6\Gyr$, including in the super-Solar metallicity regime. 
Such young ages would imply that these stars were formed within the Galaxy, reflecting the complex star formation history and ISM mixing of the Milky Way in the inner most regions. The quality of our data does not allow us to derive reliable ages by isochrone fitting for our stars. We therefore cannot exclude the existence of such high$-\alpha$ metal-rich stars in our sample that would not belong to the thick disc but rather in the thin disc. 

%%%%%%%%%%%%%%%%%%%%%%
\section{Summary}
\label{sect:conclusions}
We have investigated the stellar abundances and kinematics coming from the second internal data release of the Gaia-ESO Survey in order to characterise,  in a statistical way, the high$-\alpha$ and low$-\alpha$ disc populations, commonly identified with  the thin and thick discs. 

By studying the way the azimuthal velocity correlates with the $\alpha-$abundances at a given metallicity, we evaluated the extent of the tails of the metallicity distribution function of the discs, without making any assumption on their shape.  We found that the thick disc extends up to super-Solar metallicities ($\feh \approx + 0.2$) and the thin disc  down to $\feh \approx -0.8$.  The kinematic data shows, in agreement with a decomposition of the discs according to their $\alpha-$abundances, that at the extended Solar neighbourhood, there is an almost constant relative fraction of thin disc and thick disc stars over the metallicity range  $-0.8 \lesssim \feh \lesssim -0.5$. 

In a complementary approach, we also investigated the distribution of stars in chemical space for different locations in the Galaxy. 
 We found a mild steepening (with a significance below $2\sigma$) of the decline of the $\Trend$ trend for the thin disc towards the larger Galactic radii. This could potentially suggest either a more extended star formation history in the inner Galaxy, or strong effects from radial migration, flattening the slope of the $\Trend$ relation over time.  However, the relatively short radial range ($6 \leq R \leq 10\kpc$) covered with the Gaia-ESO targets does not allow us to put strong constraints or discard any model.
  For the thick disc, no variations are detected across the Galaxy, suggesting that the processes that formed or mixed the thick disc stars were very efficient.

We found that the way the relative proportion of the disc stars varies
 across the Galaxy confirms the recently suggested hypotheses that the
 thick disc defined by $\alpha-$abundance has a shorter scale-length
 than the thin disc.  However, when comparing with toy models of thick
 and thin discs relative fractions having shorter or longer
 scale-lengths, we find evidence of varying scale-lengths and/or
 scale-heights with $R$.  In a recent analysis,
 \citet{Minchev15} suggested that in inside-out disc formation
 models, the thick disc (defined morphologically) is composed of the
 nested flares of mono-age populations. It was shown that flaring
 increases for older stellar samples, due to the conservation of
 vertical action when stars undergo radial migration. Since older populations
 have generally lower metallicities, such a  flaring thin disc would populate
 mostly the regions at $\feh<0.4\dex$ above the plane
 (Fig.~\ref{fig:proportions_MW}, bottom), thus decreasing the fraction
 of the chemically defined thick disc. 

Finally, 
by selecting the stars residing between the two chemical sequences, we have investigated their kinematics and found that they had typical thin disc velocities, suggesting that the thin and thick discs are following chemical paths separated by a true gap. However,  the quality of our data does not allow us to establish firmly whether the two chemical sequences merge at super-solar metallicities.
In future work, we plan to investigate part of the Gaia-ESO UVES sample \citep{Smiljanic14}, with stellar parameters obtained from spectra at a resolution of $R\sim 47\,000$,  in order to investigate more robustly the  metal-rich end of the discs.

%The goal of the description of this gap is to constrain the formation mechanisms of the Galactic thick disc. Indeed, various formation mechanisms have been suggested for the thick disc: accretion of extra-Galactic stars and gas, puffing up of the pre-existent thin disc, radial migration of the stars from the inner Galaxy to the Solar neighbourhood. Recently, \citet{Minchev14}, using the fourth data-release of RAVE \citep{Kordopatis13b} have suggested having identified signatures of a violent origin to the formation of the thick disc, together with strong signatures of radial migration for the most metal-poor and high$-\alpha$ stars. The presence of these signatures in the Gaia-ESO database is presented in Guiglion et al. (in prep).  

%It would be of particular interest to see wether his model can reproduce the properties that have been presented in this paper.

% Our results do not allow to discard the possibility of a link between the metal-rich end of the thick disc and the metal-poor tail of the thin disc. 
 
 %High resolution, high S/N spectra in order to have heavy elements etc. But even like that it will be very model dependent. 

\begin{acknowledgements}
The anonymous referee is thanked for useful comments that greatly improved the quality of this paper. 
GK acknowledges C.\,Gonzalez-Fernandez for fruitful comments and suggestions.% 
RFGW acknowledges support of NSF Grant OIA-1124403 and thanks the Aspen Center for Physics and NSF Grant \#1066293 for hospitality during the writing of this paper. T.B. was funded partly by grant No. 621-2009-3911 from The Swedish 
Research Council (VR) and partly by the project grant "The New Milky Way"
from the Knut and Alice Wallenberg foundation.
U.H. acknowledges support from the Swedish National Space Board (SNSB). 
V.A.  acknowledges the support from the Funda\c{c}\~ao para a Ci\^encia e a Tecnologia, FCT (Portugal) in the form of the fellowship SFRH/BPD/70574/2010. 
This research was partly supported by the Munich Institute for Astro- and Particle Physics (MIAPP) of the DFG cluster of excellence "Origin and Structure of the Universe". 
 Based on data products from observations made with ESO Telescopes at the La Silla Paranal Observatory under programme ID 188.B-3002. These data products have been processed by the Cambridge Astronomy Survey Unit (CASU) at the Institute of Astronomy, University of Cambridge, and by the FLAMES/UVES reduction team at INAF/Osservatorio Astrofisico di Arcetri. These data have been obtained from the Gaia-ESO Survey Data Archive, prepared and hosted by the Wide Field Astronomy Unit, Institute for Astronomy, University of Edinburgh, which is funded by the UK Science and Technology Facilities Council.
   2 This work was partly supported by the European Union FP7 programme through ERC grant number 320360 and by the Leverhulme Trust through grant RPG-2012-541. We acknowledge the support from INAF and Ministero dell' Istruzione, dell' Universit\`a' e della Ricerca (MIUR) in the form of the grant "Premiale VLT 2012". The results presented here benefit from discussions held during the Gaia-ESO workshops and conferences supported by the ESF (European Science Foundation) through the GREAT Research Network Programme.
\end{acknowledgements}

% \bsp % ``This paper has been produced using the ...''
\bibliographystyle{aa}
\bibliography{\alphabib}

\begin{thebibliography}{99}
\expandafter\ifx\csname natexlab\endcsname\relax\def\natexlab#1{#1}\fi

\bibitem[{{Abadi} {et~al.}(2003){Abadi}, {Navarro}, {Steinmetz}, \&
  {Eke}}]{Abadi03}
{Abadi}, M.~G., {Navarro}, J.~F., {Steinmetz}, M., \& {Eke}, V.~R. 2003, \apj,
  597, 21

\bibitem[{{Adibekyan} {et~al.}(2013){Adibekyan}, {Figueira}, {Santos},
  {Hakobyan}, {Sousa}, {Pace}, {Delgado Mena}, {Robin}, {Israelian}, \&
  {Gonz{\'a}lez Hern{\'a}ndez}}]{Adibekyan13}
{Adibekyan}, V.~Z., {Figueira}, P., {Santos}, N.~C., {et~al.} 2013, \aap, 554,
  A44

\bibitem[{{Adibekyan} {et~al.}(2012){Adibekyan}, {Sousa}, {Santos}, {Delgado
  Mena}, {Gonz{\'a}lez Hern{\'a}ndez}, {Israelian}, {Mayor}, \&
  {Khachatryan}}]{Adibekyan12}
{Adibekyan}, V.~Z., {Sousa}, S.~G., {Santos}, N.~C., {et~al.} 2012, \aap, 545,
  A32

\bibitem[{{Aihara} {et~al.}(2011){Aihara}, {Allende Prieto}, {An}, {Anderson},
  {Aubourg}, {Balbinot}, {Beers}, {Berlind}, {Bickerton}, {Bizyaev}, {Blanton},
  {Bochanski}, {Bolton}, {Bovy}, {Brandt}, {Brinkmann}, {Brown}, {Brownstein},
  {Busca}, {Campbell}, {Carr}, {Chen}, {Chiappini}, {Comparat}, {Connolly},
  {Cortes}, {Croft}, {Cuesta}, {da Costa}, {Davenport}, {Dawson}, {Dhital},
  {Ealet}, {Ebelke}, {Edmondson}, {Eisenstein}, {Escoffier}, {Esposito},
  {Evans}, {Fan}, {Femen{\'{\i}}a Castell{\'a}}, {Font-Ribera}, {Frinchaboy},
  {Ge}, {Gillespie}, {Gilmore}, {Gonz{\'a}lez Hern{\'a}ndez}, {Gott}, {Gould},
  {Grebel}, {Gunn}, {Hamilton}, {Harding}, {Harris}, {Hawley}, {Hearty}, {Ho},
  {Hogg}, {Holtzman}, {Honscheid}, {Inada}, {Ivans}, {Jiang}, {Johnson},
  {Jordan}, {Jordan}, {Kazin}, {Kirkby}, {Klaene}, {Knapp}, {Kneib},
  {Kochanek}, {Koesterke}, {Kollmeier}, {Kron}, {Lampeitl}, {Lang}, {Le Goff},
  {Lee}, {Lin}, {Long}, {Loomis}, {Lucatello}, {Lundgren}, {Lupton}, {Ma},
  {MacDonald}, {Mahadevan}, {Maia}, {Makler}, {Malanushenko}, {Malanushenko},
  {Mandelbaum}, {Maraston}, {Margala}, {Masters}, {McBride}, {McGehee},
  {McGreer}, {M{\'e}nard}, {Miralda-Escud{\'e}}, {Morrison}, {Mullally},
  {Muna}, {Munn}, {Murayama}, {Myers}, {Naugle}, {Neto}, {Nguyen}, {Nichol},
  {O'Connell}, {Ogando}, {Olmstead}, {Oravetz}, {Padmanabhan},
  {Palanque-Delabrouille}, {Pan}, {Pandey}, {P{\^a}ris}, {Percival},
  {Petitjean}, {Pfaffenberger}, {Pforr}, {Phleps}, {Pichon}, {Pieri}, {Prada},
  {Price-Whelan}, {Raddick}, {Ramos}, {Reyl{\'e}}, {Rich}, {Richards}, {Rix},
  {Robin}, {Rocha-Pinto}, {Rockosi}, {Roe}, {Rollinde}, {Ross}, {Ross},
  {Rossetto}, {S{\'a}nchez}, {Sayres}, {Schlegel}, {Schlesinger}, {Schmidt},
  {Schneider}, {Sheldon}, {Shu}, {Simmerer}, {Simmons}, {Sivarani}, {Snedden},
  {Sobeck}, {Steinmetz}, {Strauss}, {Szalay}, {Tanaka}, {Thakar}, {Thomas},
  {Tinker}, {Tofflemire}, {Tojeiro}, {Tremonti}, {Vandenberg}, {Vargas
  Maga{\~n}a}, {Verde}, {Vogt}, {Wake}, {Wang}, {Weaver}, {Weinberg}, {White},
  {White}, {Yanny}, {Yasuda}, {Yeche}, \& {Zehavi}}]{Aihara11}
{Aihara}, H., {Allende Prieto}, C., {An}, D., {et~al.} 2011, \apjs, 193, 29

\bibitem[{{Allende Prieto} {et~al.}(2006){Allende Prieto}, {Beers}, {Wilhelm},
  {Newberg}, {Rockosi}, {Yanny}, \& {Lee}}]{Allende-Prieto06}
{Allende Prieto}, C., {Beers}, T.~C., {Wilhelm}, R., {et~al.} 2006, \apj, 636,
  804

\bibitem[{{Anders} {et~al.}(2014){Anders}, {Chiappini}, {Santiago},
  {Rocha-Pinto}, {Girardi}, {da Costa}, {Maia}, {Steinmetz}, {Minchev},
  {Schultheis}, {Boeche}, {Miglio}, {Montalb{\'a}n}, {Schneider}, {Beers},
  {Cunha}, {Allende Prieto}, {Balbinot}, {Bizyaev}, {Brauer}, {Brinkmann},
  {Frinchaboy}, {Garc{\'{\i}}a P{\'e}rez}, {Hayden}, {Hearty}, {Holtzman},
  {Johnson}, {Kinemuchi}, {Majewski}, {Malanushenko}, {Malanushenko},
  {Nidever}, {O'Connell}, {Pan}, {Robin}, {Schiavon}, {Shetrone}, {Skrutskie},
  {Smith}, {Stassun}, \& {Zasowski}}]{Anders14}
{Anders}, F., {Chiappini}, C., {Santiago}, B.~X., {et~al.} 2014, \aap, 564,
  A115

\bibitem[{{Bensby} {et~al.}(2011){Bensby}, {Alves-Brito}, {Oey}, {Yong}, \&
  {Mel{\'e}ndez}}]{Bensby11}
{Bensby}, T., {Alves-Brito}, A., {Oey}, M.~S., {Yong}, D., \& {Mel{\'e}ndez},
  J. 2011, \apjl, 735, L46

\bibitem[{{Bensby} {et~al.}(2005){Bensby}, {Feltzing}, {Lundstr{\"o}m}, \&
  {Ilyin}}]{Bensby05}
{Bensby}, T., {Feltzing}, S., {Lundstr{\"o}m}, I., \& {Ilyin}, I. 2005, \aap,
  433, 185

\bibitem[{{Bensby} {et~al.}(2014){Bensby}, {Feltzing}, \& {Oey}}]{Bensby14}
{Bensby}, T., {Feltzing}, S., \& {Oey}, M.~S. 2014, \aap, 562, A71

\bibitem[{{Bensby} {et~al.}(2007){Bensby}, {Zenn}, {Oey}, \&
  {Feltzing}}]{Bensby07}
{Bensby}, T., {Zenn}, A.~R., {Oey}, M.~S., \& {Feltzing}, S. 2007, \apjl, 663,
  L13

\bibitem[{{Bergemann} {et~al.}(2014){Bergemann}, {Ruchti}, {Serenelli},
  {Feltzing}, {Alves-Brito}, {Asplund}, {Bensby}, {Gruyters}, {Heiter},
  {Hourihane}, {Korn}, {Lind}, {Marino}, {Jofre}, {Nordlander}, {Ryde},
  {Worley}, {Gilmore}, {Randich}, {Ferguson}, {Jeffries}, {Micela},
  {Negueruela}, {Prusti}, {Rix}, {Vallenari}, {Alfaro}, {Allende Prieto},
  {Bragaglia}, {Koposov}, {Lanzafame}, {Pancino}, {Recio-Blanco}, {Smiljanic},
  {Walton}, {Costado}, {Franciosini}, {Hill}, {Lardo}, {de Laverny}, {Magrini},
  {Maiorca}, {Masseron}, {Morbidelli}, {Sacco}, {Kordopatis}, \& {Tautvai{\v
  s}ien{\.e}}}]{Bergemann14}
{Bergemann}, M., {Ruchti}, G.~R., {Serenelli}, A., {et~al.} 2014, \aap, 565,
  A89

\bibitem[{{Binney}(2012)}]{Binney12b}
{Binney}, J. 2012, \mnras, 426, 1328

\bibitem[{{Binney} {et~al.}(2014){Binney}, {Burnett}, {Kordopatis},
  {Steinmetz}, {Gilmore}, {Bienayme}, {Bland-Hawthorn}, {Famaey}, {Grebel},
  {Helmi}, {Navarro}, {Parker}, {Reid}, {Seabroke}, {Siebert}, {Watson},
  {Williams}, {Wyse}, \& {Zwitter}}]{Binney14b}
{Binney}, J., {Burnett}, B., {Kordopatis}, G., {et~al.} 2014, \mnras, 439, 1231

\bibitem[{{Binney} \& {Tremaine}(2008)}]{BinneyTremaine08}
{Binney}, J. \& {Tremaine}, S. 2008, {Galactic Dynamics: Second Edition}
  (Princeton University Press)

\bibitem[{{Bird} {et~al.}(2013){Bird}, {Kazantzidis}, {Weinberg}, {Guedes},
  {Callegari}, {Mayer}, \& {Madau}}]{Bird13}
{Bird}, J.~C., {Kazantzidis}, S., {Weinberg}, D.~H., {et~al.} 2013, \apj, 773,
  43

\bibitem[{{Boeche} {et~al.}(2013){Boeche}, {Chiappini}, {Minchev}, {Williams},
  {Steinmetz}, {Sharma}, {Kordopatis}, {Bland-Hawthorn}, {Bienaym{\'e}},
  {Gibson}, {Gilmore}, {Grebel}, {Helmi}, {Munari}, {Navarro}, {Parker},
  {Reid}, {Seabroke}, {Siebert}, {Siviero}, {Watson}, {Wyse}, \&
  {Zwitter}}]{Boeche13}
{Boeche}, C., {Chiappini}, C., {Minchev}, I., {et~al.} 2013, \aap, 553, A19

\bibitem[{{Boeche} {et~al.}(2014){Boeche}, {Siebert}, {Piffl}, {Just},
  {Steinmetz}, {Grebel}, {Sharma}, {Kordopatis}, {Gilmore}, {Chiappini},
  {Freeman}, {Gibson}, {Munari}, {Siviero}, {Bienaym{\'e}}, {Navarro},
  {Parker}, {Reid}, {Seabroke}, {Watson}, {Wyse}, \& {Zwitter}}]{Boeche14}
{Boeche}, C., {Siebert}, A., {Piffl}, T., {et~al.} 2014, \aap, 568, A71

\bibitem[{{Bournaud} {et~al.}(2009){Bournaud}, {Elmegreen}, \&
  {Martig}}]{Bournaud09}
{Bournaud}, F., {Elmegreen}, B.~G., \& {Martig}, M. 2009, \apjl, 707, L1

\bibitem[{{Bovy} {et~al.}(2012{\natexlab{a}}){Bovy}, {Rix}, \& {Hogg}}]{Bovy12}
{Bovy}, J., {Rix}, H.-W., \& {Hogg}, D.~W. 2012{\natexlab{a}}, \apj, 751, 131

\bibitem[{{Bovy} {et~al.}(2012{\natexlab{b}}){Bovy}, {Rix}, {Liu}, \&
  {et~al.}}]{Bovy12b}
{Bovy}, J., {Rix}, H.-W., {Liu}, C., \& {et~al.} 2012{\natexlab{b}}, \apj, 753,
  148

\bibitem[{{Brook} {et~al.}(2004){Brook}, {Kawata}, {Gibson}, \&
  {Freeman}}]{Brook04}
{Brook}, C.~B., {Kawata}, D., {Gibson}, B.~K., \& {Freeman}, K.~C. 2004, \apj,
  612, 894

\bibitem[{{Carney} {et~al.}(1989){Carney}, {Latham}, \& {Laird}}]{Carney89}
{Carney}, B.~W., {Latham}, D.~W., \& {Laird}, J.~B. 1989, \aj, 97, 423

\bibitem[{{Carollo} {et~al.}(2010){Carollo}, {Beers}, {Chiba}, {Norris},
  {Freeman}, {Lee}, {Ivezi{\'c}}, {Rockosi}, \& {Yanny}}]{Carollo10}
{Carollo}, D., {Beers}, T.~C., {Chiba}, M., {et~al.} 2010, \apj, 712, 692

\bibitem[{{Cheng} {et~al.}(2012){Cheng}, {Rockosi}, {Morrison},
  {Sch{\"o}nrich}, {Lee}, {Beers}, {Bizyaev}, {Pan}, \& {Schneider}}]{Cheng12}
{Cheng}, J.~Y., {Rockosi}, C.~M., {Morrison}, H.~L., {et~al.} 2012, \apj, 746,
  149

\bibitem[{{Chiappini} {et~al.}(2015){Chiappini}, {Anders}, {Rodrigues},
  {Miglio}, {Montalb{\'a}n}, {Mosser}, {Girardi}, {Valentini}, {Noels},
  {Morel}, {Minchev}, {Steinmetz}, {Santiago}, {Schultheis}, {Martig}, {da
  Costa}, {Maia}, {Allende Prieto}, {de Assis Peralta}, {Hekker},
  {Theme{\ss}l}, {Kallinger}, {Garc{\'{\i}}a}, {Mathur}, {Baudin}, {Beers},
  {Cunha}, {Harding}, {Holtzman}, {Majewski}, {M{\'e}sz{\'a}ros}, {Nidever},
  {Pan}, {Schiavon}, {Shetrone}, {Schneider}, \& {Stassun}}]{Chiappini15}
{Chiappini}, C., {Anders}, F., {Rodrigues}, T.~S., {et~al.} 2015, \aap, 576,
  L12

\bibitem[{{Chiappini} {et~al.}(1997){Chiappini}, {Matteucci}, \&
  {Gratton}}]{Chiappini97}
{Chiappini}, C., {Matteucci}, F., \& {Gratton}, R. 1997, \apj, 477, 765

\bibitem[{{Chiba} \& {Beers}(2000)}]{Chiba00}
{Chiba}, M. \& {Beers}, T.~C. 2000, \aj, 119, 2843

\bibitem[{{Demarque} {et~al.}(2004){Demarque}, {Woo}, {Kim}, \&
  {Yi}}]{Demarque04}
{Demarque}, P., {Woo}, J.-H., {Kim}, Y.-C., \& {Yi}, S.~K. 2004, \apjs, 155,
  667

\bibitem[{{Dembo} \& {Steihaug}(1983)}]{Dembo83}
{Dembo}, R.~S. \& {Steihaug}, T. 1983, Math. Prog, 26, 190

\bibitem[{{Edvardsson} {et~al.}(1993){Edvardsson}, {Andersen}, {Gustafsson},
  {Lambert}, {Nissen}, \& {Tomkin}}]{Edvardsson93}
{Edvardsson}, B., {Andersen}, J., {Gustafsson}, B., {et~al.} 1993, \aap, 275,
  101

\bibitem[{{Eggen} {et~al.}(1962){Eggen}, {Lynden-Bell}, \& {Sandage}}]{Eggen62}
{Eggen}, O.~J., {Lynden-Bell}, D., \& {Sandage}, A.~R. 1962, \apj, 136, 748

\bibitem[{{Faure} {et~al.}(2014){Faure}, {Siebert}, \& {Famaey}}]{Faure14}
{Faure}, C., {Siebert}, A., \& {Famaey}, B. 2014, \mnras, 440, 2564

\bibitem[{{Flynn} {et~al.}(2006){Flynn}, {Holmberg}, {Portinari}, {Fuchs}, \&
  {Jahrei{\ss}}}]{Flynn06}
{Flynn}, C., {Holmberg}, J., {Portinari}, L., {Fuchs}, B., \& {Jahrei{\ss}}, H.
  2006, \mnras, 372, 1149

\bibitem[{{Freeman} \& {Bland-Hawthorn}(2002)}]{Freeman02}
{Freeman}, K. \& {Bland-Hawthorn}, J. 2002, \araa, 40, 487

\bibitem[{{Fuhrmann}(1998)}]{Fuhrmann98}
{Fuhrmann}, K. 1998, \aap, 338, 161

\bibitem[{{Fuhrmann}(2008)}]{Fuhrmann08}
{Fuhrmann}, K. 2008, \mnras, 384, 173

\bibitem[{{Fuhrmann}(2011)}]{Fuhrmann11}
{Fuhrmann}, K. 2011, \mnras, 414, 2893

\bibitem[{{Gazzano} {et~al.}(2013){Gazzano}, {Kordopatis}, {Deleuil}, \&
  {et~al}}]{Gazzano13}
{Gazzano}, J.-C., {Kordopatis}, G., {Deleuil}, M., \& {et~al}. 2013, \aap, 550,
  A125

\bibitem[{{Genovali} {et~al.}(2014){Genovali}, {Lemasle}, {Bono}, {Romaniello},
  {Fabrizio}, {Ferraro}, {Iannicola}, {Laney}, {Nonino}, {Bergemann},
  {Buonanno}, {Fran{\c c}ois}, {Inno}, {Kudritzki}, {Matsunaga}, {Pedicelli},
  {Primas}, \& {Th{\'e}venin}}]{Genovali14}
{Genovali}, K., {Lemasle}, B., {Bono}, G., {et~al.} 2014, \aap, 566, A37

\bibitem[{{Gilmore} {et~al.}(2012){Gilmore}, {Randich}, {Asplund}, {Binney},
  {Bonifacio}, {Drew}, {Feltzing}, {Ferguson}, {Jeffries}, {Micela},
  {Negueruela}, {Prusti}, {Rix}, {Vallenari}, {Alfaro}, {Allende-Prieto},
  {Babusiaux}, {Bensby}, {Blomme}, {Bragaglia}, {Flaccomio}, {Fran{\c c}ois},
  {Irwin}, {Koposov}, {Korn}, {Lanzafame}, {Pancino}, {Paunzen},
  {Recio-Blanco}, {Sacco}, {Smiljanic}, {Van Eck}, \& {Walton}}]{Gilmore12}
{Gilmore}, G., {Randich}, S., {Asplund}, M., {et~al.} 2012, The Messenger, 147,
  25

\bibitem[{{Gilmore} \& {Reid}(1983)}]{Gilmore83}
{Gilmore}, G. \& {Reid}, N. 1983, \mnras, 202, 1025

\bibitem[{{Gilmore} \& {Wyse}(1986)}]{Gilmore86}
{Gilmore}, G. \& {Wyse}, R.~F.~G. 1986, \nat, 322, 806

\bibitem[{{Gilmore} \& {Wyse}(1991)}]{Gilmore91}
{Gilmore}, G. \& {Wyse}, R.~F.~G. 1991, \apjl, 367, L55

\bibitem[{{Gilmore} {et~al.}(1989){Gilmore}, {Wyse}, \& {Kuijken}}]{Gilmore89}
{Gilmore}, G., {Wyse}, R.~F.~G., \& {Kuijken}, K. 1989, \araa, 27, 555

\bibitem[{{Grevesse} {et~al.}(2010){Grevesse}, {Asplund}, {Sauval}, \&
  {Scott}}]{Grevesse10}
{Grevesse}, N., {Asplund}, M., {Sauval}, A.~J., \& {Scott}, P. 2010, \apss,
  328, 179

\bibitem[{{Guiglion} {et~al.}(2014){Guiglion}, {Recio-Blanco}, \& {de
  Laverny}}]{Guiglion13}
{Guiglion}, G., {Recio-Blanco}, A., \& {de Laverny}, P. 2014, in IAU Symposium,
  Vol. 298, IAU Symposium, ed. S.~{Feltzing}, G.~{Zhao}, N.~A. {Walton}, \&
  P.~{Whitelock}, 408--408

\bibitem[{{Gustafsson} {et~al.}(2008){Gustafsson}, {Edvardsson}, {Eriksson},
  {J{\o}rgensen}, {Nordlund}, \& {Plez}}]{Gustafsson08}
{Gustafsson}, B., {Edvardsson}, B., {Eriksson}, K., {et~al.} 2008, \aap, 486,
  951

\bibitem[{{Hayden} {et~al.}(2015){Hayden}, {Bovy}, {Holtzman}, {Nidever},
  {Bird}, {Weinberg}, {Andrews}, {Allende Prieto}, {Anders}, {Beers},
  {Bizyaev}, {Chiappini}, {Cunha}, {Frinchaboy}, {Garc{\'{\i}}a-Her{\'n}andez},
  {Garc{\'{\i}}a P{\'e}rez}, {Girardi}, {Harding}, {Hearty}, {Johnson},
  {Majewski}, {M{\'e}sz{\'a}ros}, {Minchev}, {O'Connell}, {Pan}, {Robin},
  {Schiavon}, {Schneider}, {Schultheis}, {Shetrone}, {Skrutskie}, {Steinmetz},
  {Smith}, {Zamora}, \& {Zasowski}}]{Hayden15}
{Hayden}, M.~R., {Bovy}, J., {Holtzman}, J.~A., {et~al.} 2015, ArXiv e-prints

\bibitem[{{Haywood} {et~al.}(2013){Haywood}, {Di Matteo}, {Lehnert}, {Katz}, \&
  {G{\'o}mez}}]{Haywood13}
{Haywood}, M., {Di Matteo}, P., {Lehnert}, M.~D., {Katz}, D., \& {G{\'o}mez},
  A. 2013, \aap, 560, A109

\bibitem[{{Jofr{\'e}} {et~al.}(2014){Jofr{\'e}}, {Heiter}, {Soubiran},
  {Blanco-Cuaresma}, {Worley}, {Pancino}, {Cantat-Gaudin}, {Magrini},
  {Bergemann}, {Gonz{\'a}lez Hern{\'a}ndez}, {Hill}, {Lardo}, {de Laverny},
  {Lind}, {Masseron}, {Montes}, {Mucciarelli}, {Nordlander}, {Recio Blanco},
  {Sobeck}, {Sordo}, {Sousa}, {Tabernero}, {Vallenari}, \& {Van Eck}}]{Jofre14}
{Jofr{\'e}}, P., {Heiter}, U., {Soubiran}, C., {et~al.} 2014, \aap, 564, A133

\bibitem[{{Juri{\'c}} {et~al.}(2008){Juri{\'c}}, {Ivezi{\'c}}, {Brooks},
  {Lupton}, {Schlegel}, {Finkbeiner}, {Padmanabhan}, {Bond}, {Sesar},
  {Rockosi}, {Knapp}, {Gunn}, {Sumi}, {Schneider}, {Barentine}, {Brewington},
  {Brinkmann}, {Fukugita}, {Harvanek}, {Kleinman}, {Krzesinski}, {Long},
  {Neilsen}, {Nitta}, {Snedden}, \& {York}}]{Juric08}
{Juri{\'c}}, M., {Ivezi{\'c}}, {\v Z}., {Brooks}, A., {et~al.} 2008, \apj, 673,
  864

\bibitem[{{Kordopatis} {et~al.}(2015){Kordopatis}, {Binney}, {Gilmore}, {Wyse},
  {Belokurov}, {McMillan}, {Hatfield}, {Grebel}, {Steinmetz}, {Navarro},
  {Seabroke}, {Minchev}, {Chiappini}, {Bienaym{\'e}}, {Bland-Hawthorn},
  {Freeman}, {Gibson}, {Helmi}, {Munari}, {Parker}, {Reid}, {Siebert},
  {Siviero}, \& {Zwitter}}]{Kordopatis15}
{Kordopatis}, G., {Binney}, J., {Gilmore}, G., {et~al.} 2015, \mnras, 447, 3526

\bibitem[{{Kordopatis} {et~al.}(2013{\natexlab{a}}){Kordopatis}, {Gilmore},
  {Steinmetz}, {Boeche}, {Seabroke}, {Siebert}, {Zwitter}, {Binney}, {de
  Laverny}, {Recio-Blanco}, {Williams}, {Piffl}, {Enke}, {Roeser}, {Bijaoui},
  {Wyse}, {Freeman}, {Munari}, {Carrillo}, {Anguiano}, {Burton}, {Campbell},
  {Cass}, {Fiegert}, {Hartley}, {Parker}, {Reid}, {Ritter}, {Russell},
  {Stupar}, {Watson}, {Bienaym{\'e}}, {Bland-Hawthorn}, {Gerhard}, {Gibson},
  {Grebel}, {Helmi}, {Navarro}, {Conrad}, {Famaey}, {Faure}, {Just}, {Kos},
  {Matijevi{\v c}}, {McMillan}, {Minchev}, {Scholz}, {Sharma}, {Siviero}, {de
  Boer}, \& {{\v Z}erjal}}]{Kordopatis13b}
{Kordopatis}, G., {Gilmore}, G., {Steinmetz}, M., {et~al.} 2013{\natexlab{a}},
  \aj, 146, 134

\bibitem[{{Kordopatis} {et~al.}(2013{\natexlab{b}}){Kordopatis}, {Gilmore},
  {Wyse}, {Steinmetz}, {Siebert}, {Bienaym{\'e}}, {McMillan}, {Minchev},
  {Zwitter}, {Gibson}, {Seabroke}, {Grebel}, {Bland-Hawthorn}, {Boeche},
  {Freeman}, {Munari}, {Navarro}, {Parker}, {Reid}, \&
  {Siviero}}]{Kordopatis13c}
{Kordopatis}, G., {Gilmore}, G., {Wyse}, R.~F.~G., {et~al.} 2013{\natexlab{b}},
  \mnras, 436, 3231

\bibitem[{{Kordopatis} {et~al.}(2013{\natexlab{c}}){Kordopatis}, {Hill},
  {Irwin}, {Gilmore}, {Wyse}, {Tolstoy}, {de Laverny}, {Recio-Blanco},
  {Battaglia}, \& {Starkenburg}}]{Kordopatis13a}
{Kordopatis}, G., {Hill}, V., {Irwin}, M., {et~al.} 2013{\natexlab{c}}, \aap,
  555, A12

\bibitem[{{Kordopatis} {et~al.}(2011){Kordopatis}, {Recio-Blanco}, {de
  Laverny}, {Gilmore}, {Hill}, {Wyse}, {Helmi}, {Bijaoui}, {Zoccali}, \&
  {Bienaym{\'e}}}]{Kordopatis11b}
{Kordopatis}, G., {Recio-Blanco}, A., {de Laverny}, P., {et~al.} 2011, \aap,
  535, A107

\bibitem[{{Lee} {et~al.}(2011){Lee}, {Beers}, {An}, {Ivezi{\'c}}, {Just},
  {Rockosi}, {Morrison}, {Johnson}, {Sch{\"o}nrich}, {Bird}, {Yanny},
  {Harding}, \& {Rocha-Pinto}}]{Lee11}
{Lee}, Y.~S., {Beers}, T.~C., {An}, D., {et~al.} 2011, \apj, 738, 187

\bibitem[{{Loebman} {et~al.}(2011){Loebman}, {Ro{\v s}kar}, {Debattista}, \&
  {et~al.}}]{Loebman11}
{Loebman}, S.~R., {Ro{\v s}kar}, R., {Debattista}, V.~P., \& {et~al.} 2011,
  \apj, 737, 8

\bibitem[{{Martig} {et~al.}(2014){Martig}, {Rix}, {Silva Aguirre}, {Hekker},
  {Mosser}, {Elsworth}, {Bovy}, {Stello}, {Anders}, {Garc{\'{\i}}a}, {Tayar},
  {Rodrigues}, {Basu}, {Carrera}, {Ceillier}, {Chaplin}, {Chiappini},
  {Frinchaboy}, {Garc{\'{\i}}a-Hern{\'a}ndez}, {Hearty}, {Holtzman}, {Johnson},
  {Mathur}, {M{\'e}sz{\'a}ros}, {Miglio}, {Nidever}, {Pinsonneault},
  {Schiavon}, {Schneider}, {Serenelli}, {Shetrone}, \& {Zamora}}]{Martig15}
{Martig}, M., {Rix}, H.-W., {Silva Aguirre}, V., {et~al.} 2014, ArXiv e-prints

\bibitem[{{Mikolaitis} {et~al.}(2014){Mikolaitis}, {Hill}, {Recio-Blanco}, {de
  Laverny}, {Allende Prieto}, {Kordopatis}, {Tautvai{\v s}iene}, {Romano},
  {Gilmore}, {Randich}, {Feltzing}, {Micela}, {Vallenari}, {Alfaro}, {Bensby},
  {Bragaglia}, {Flaccomio}, {Lanzafame}, {Pancino}, {Smiljanic}, {Bergemann},
  {Carraro}, {Costado}, {Damiani}, {Hourihane}, {Jofr{\'e}}, {Lardo},
  {Magrini}, {Maiorca}, {Morbidelli}, {Sbordone}, {Sousa}, {Worley}, \&
  {Zaggia}}]{Mikolaitis14}
{Mikolaitis}, {\v S}., {Hill}, V., {Recio-Blanco}, A., {et~al.} 2014, \aap,
  572, A33

\bibitem[{{Minchev} {et~al.}(2013){Minchev}, {Chiappini}, \&
  {Martig}}]{Minchev13}
{Minchev}, I., {Chiappini}, C., \& {Martig}, M. 2013, \aap, 558, A9

\bibitem[{{Minchev} {et~al.}(2014){Minchev}, {Chiappini}, {Martig},
  {Steinmetz}, {de Jong}, {Boeche}, {Scannapieco}, {Zwitter}, {Wyse}, {Binney},
  {Bland-Hawthorn}, {Bienaym{\'e}}, {Famaey}, {Freeman}, {Gibson}, {Grebel},
  {Gilmore}, {Helmi}, {Kordopatis}, {Lee}, {Munari}, {Navarro}, {Parker},
  {Quillen}, {Reid}, {Siebert}, {Siviero}, {Seabroke}, {Watson}, \&
  {Williams}}]{Minchev14}
{Minchev}, I., {Chiappini}, C., {Martig}, M., {et~al.} 2014, \apjl, 781, L20

\bibitem[{{Minchev} {et~al.}(2012){Minchev}, {Famaey}, {Quillen}, {Dehnen},
  {Martig}, \& {Siebert}}]{Minchev12b}
{Minchev}, I., {Famaey}, B., {Quillen}, A.~C., {et~al.} 2012, \aap, 548, A127

\bibitem[{{Minchev} {et~al.}(2015){Minchev}, {Martig}, {Streich},
  {Scannapieco}, {de Jong}, \& {Steinmetz}}]{Minchev15}
{Minchev}, I., {Martig}, M., {Streich}, D., {et~al.} 2015, ArXiv e-prints

\bibitem[{{Nidever} {et~al.}(2014){Nidever}, {Bovy}, {Bird}, {Andrews},
  {Hayden}, {Holtzman}, {Majewski}, {Smith}, {Robin}, {Garc{\'{\i}}a
  P{\'e}rez}, {Cunha}, {Allende Prieto}, {Zasowski}, {Schiavon}, {Johnson},
  {Weinberg}, {Feuillet}, {Schneider}, {Shetrone}, {Sobeck},
  {Garc{\'{\i}}a-Hern{\'a}ndez}, {Zamora}, {Rix}, {Beers}, {Wilson},
  {O'Connell}, {Minchev}, {Chiappini}, {Anders}, {Bizyaev}, {Brewington},
  {Ebelke}, {Frinchaboy}, {Ge}, {Kinemuchi}, {Malanushenko}, {Malanushenko},
  {Marchante}, {M{\'e}sz{\'a}ros}, {Oravetz}, {Pan}, {Simmons}, \&
  {Skrutskie}}]{Nidever14}
{Nidever}, D.~L., {Bovy}, J., {Bird}, J.~C., {et~al.} 2014, \apj, 796, 38

\bibitem[{{Nissen}(2015)}]{Nissen15}
{Nissen}, P.~E. 2015, ArXiv e-prints

\bibitem[{{Nordstr{\"o}m} {et~al.}(2004){Nordstr{\"o}m}, {Mayor}, {Andersen},
  {Holmberg}, {Pont}, {J{\o}rgensen}, {Olsen}, {Udry}, \&
  {Mowlavi}}]{Nordstrom04}
{Nordstr{\"o}m}, B., {Mayor}, M., {Andersen}, J., {et~al.} 2004, \aap, 418, 989

\bibitem[{{Norris} {et~al.}(1985){Norris}, {Bessell}, \& {Pickles}}]{Norris85}
{Norris}, J., {Bessell}, M.~S., \& {Pickles}, A.~J. 1985, \apjs, 58, 463

\bibitem[{{Recio-Blanco} {et~al.}(2006){Recio-Blanco}, {Bijaoui}, \& {de
  Laverny}}]{Recio-Blanco06}
{Recio-Blanco}, A., {Bijaoui}, A., \& {de Laverny}, P. 2006, \mnras, 370, 141

\bibitem[{{Recio-Blanco} {et~al.}(2014){Recio-Blanco}, {de Laverny},
  {Kordopatis}, {Helmi}, {Hill}, {Gilmore}, {Wyse}, {Adibekyan}, {Randich},
  {Asplund}, {Feltzing}, {Jeffries}, {Micela}, {Vallenari}, {Alfaro}, {Allende
  Prieto}, {Bensby}, {Bragaglia}, {Flaccomio}, {Koposov}, {Korn}, {Lanzafame},
  {Pancino}, {Smiljanic}, {Jackson}, {Lewis}, {Magrini}, {Morbidelli},
  {Prisinzano}, {Sacco}, {Worley}, {Hourihane}, {Bergemann}, {Costado},
  {Heiter}, {Joffre}, {Lardo}, {Lind}, \& {Maiorca}}]{Recio-Blanco14}
{Recio-Blanco}, A., {de Laverny}, P., {Kordopatis}, G., {et~al.} 2014, \aap,
  567, A5

\bibitem[{{Reddy} \& {Lambert}(2008)}]{Reddy08}
{Reddy}, B.~E. \& {Lambert}, D.~L. 2008, \mnras, 391, 95

\bibitem[{{Reddy} {et~al.}(2006){Reddy}, {Lambert}, \& {Allende
  Prieto}}]{Reddy06}
{Reddy}, B.~E., {Lambert}, D.~L., \& {Allende Prieto}, C. 2006, \mnras, 367,
  1329

\bibitem[{{Rix} \& {Bovy}(2013)}]{Rix13}
{Rix}, H.-W. \& {Bovy}, J. 2013, \aapr, 21, 61

\bibitem[{{Roeser} {et~al.}(2010){Roeser}, {Demleitner}, \&
  {Schilbach}}]{Roeser10}
{Roeser}, S., {Demleitner}, M., \& {Schilbach}, E. 2010, \aj, 139, 2440

\bibitem[{{Ruchti} {et~al.}(2010){Ruchti}, {Fulbright}, {Wyse}, {Gilmore},
  {Bienaym{\'e}}, {Binney}, {Bland-Hawthorn}, {Campbell}, {Freeman}, {Gibson},
  {Grebel}, {Helmi}, {Munari}, {Navarro}, {Parker}, {Reid}, {Seabroke},
  {Siebert}, {Siviero}, {Steinmetz}, {Watson}, {Williams}, \&
  {Zwitter}}]{Ruchti10}
{Ruchti}, G.~R., {Fulbright}, J.~P., {Wyse}, R.~F.~G., {et~al.} 2010, \apjl,
  721, L92

\bibitem[{{Ruchti} {et~al.}(2011){Ruchti}, {Fulbright}, {Wyse}, {Gilmore},
  {Bienaym{\'e}}, {Bland-Hawthorn}, {Gibson}, {Grebel}, {Helmi}, {Munari},
  {Navarro}, {Parker}, {Reid}, {Seabroke}, {Siebert}, {Siviero}, {Steinmetz},
  {Watson}, {Williams}, \& {Zwitter}}]{Ruchti11}
{Ruchti}, G.~R., {Fulbright}, J.~P., {Wyse}, R.~F.~G., {et~al.} 2011, \apj,
  737, 9

\bibitem[{{Schlegel} {et~al.}(1998){Schlegel}, {Finkbeiner}, \&
  {Davis}}]{Schlegel98}
{Schlegel}, D.~J., {Finkbeiner}, D.~P., \& {Davis}, M. 1998, \apj, 500, 525

\bibitem[{{Sch{\"o}nrich} \& {Binney}(2009)}]{Schonrich09b}
{Sch{\"o}nrich}, R. \& {Binney}, J. 2009, \mnras, 399, 1145

\bibitem[{{Sch{\"o}nrich} {et~al.}(2010){Sch{\"o}nrich}, {Binney}, \&
  {Dehnen}}]{Schonrich10}
{Sch{\"o}nrich}, R., {Binney}, J., \& {Dehnen}, W. 2010, \mnras, 403, 1829

\bibitem[{{Schultheis} {et~al.}(2015){Schultheis}, {Kordopatis},
  {Recio-Blanco}, {de Laverny}, {Hill}, {Gilmore}, {Alfaro}, {Costado},
  {Bensby}, {Damiani}, {Feltzing}, {Flaccomio}, {Lardo}, {Jofre}, {Prisinzano},
  {Zaggia}, {Jimenez-Esteban}, {Morbidelli}, {Lanzafame}, {Hourihane},
  {Worley}, \& {Francois}}]{Schultheis15}
{Schultheis}, M., {Kordopatis}, G., {Recio-Blanco}, A., {et~al.} 2015, \aap,
  577, A77

\bibitem[{{Seabroke} \& {Gilmore}(2007)}]{Seabroke07}
{Seabroke}, G.~M. \& {Gilmore}, G. 2007, \mnras, 380, 1348

\bibitem[{{Sellwood} \& {Binney}(2002)}]{Sellwood02}
{Sellwood}, J.~A. \& {Binney}, J.~J. 2002, \mnras, 336, 785

\bibitem[{{Sharma} {et~al.}(2014){Sharma}, {Bland-Hawthorn}, {Binney},
  {Freeman}, {Steinmetz}, {Boeche}, {Bienaym{\'e}}, {Gibson}, {Gilmore},
  {Grebel}, {Helmi}, {Kordopatis}, {Munari}, {Navarro}, {Parker}, {Reid},
  {Seabroke}, {Siebert}, {Watson}, {Williams}, {Wyse}, \& {Zwitter}}]{Sharma14}
{Sharma}, S., {Bland-Hawthorn}, J., {Binney}, J., {et~al.} 2014, \apj, 793, 51

\bibitem[{{Siebert} {et~al.}(2011{\natexlab{a}}){Siebert}, {Famaey}, {Minchev},
  {Seabroke}, {Binney}, {Burnett}, {Freeman}, {Williams}, {Bienaym{\'e}},
  {Bland-Hawthorn}, {Campbell}, {Fulbright}, {Gibson}, {Gilmore}, {Grebel},
  {Helmi}, {Munari}, {Navarro}, {Parker}, {Reid}, {Siviero}, {Steinmetz},
  {Watson}, {Wyse}, \& {Zwitter}}]{Siebert11a}
{Siebert}, A., {Famaey}, B., {Minchev}, I., {et~al.} 2011{\natexlab{a}},
  \mnras, 412, 2026

\bibitem[{{Siebert} {et~al.}(2011{\natexlab{b}}){Siebert}, {Williams},
  {Siviero}, {Reid}, {Boeche}, {Steinmetz}, {Fulbright}, {Munari}, {Zwitter},
  {Watson}, {Wyse}, {de Jong}, {Enke}, {Anguiano}, {Burton}, {Cass}, {Fiegert},
  {Hartley}, {Ritter}, {Russel}, {Stupar}, {Bienaym{\'e}}, {Freeman},
  {Gilmore}, {Grebel}, {Helmi}, {Navarro}, {Binney}, {Bland-Hawthorn},
  {Campbell}, {Famaey}, {Gerhard}, {Gibson}, {Matijevi{\v c}}, {Parker},
  {Seabroke}, {Sharma}, {Smith}, \& {Wylie-de Boer}}]{Siebert11}
{Siebert}, A., {Williams}, M.~E.~K., {Siviero}, A., {et~al.}
  2011{\natexlab{b}}, \aj, 141, 187

\bibitem[{{Smiljanic} {et~al.}(2014){Smiljanic}, {Korn}, {Bergemann}, {Frasca},
  {Magrini}, {Masseron}, {Pancino}, {Ruchti}, {San Roman}, {Sbordone}, {Sousa},
  {Tabernero}, {Tautvai{\v s}ien{\.e}}, {Valentini}, {Weber}, {Worley},
  {Adibekyan}, {Allende Prieto}, {Barisevi{\v c}ius}, {Biazzo},
  {Blanco-Cuaresma}, {Bonifacio}, {Bragaglia}, {Caffau}, {Cantat-Gaudin},
  {Chorniy}, {de Laverny}, {Delgado-Mena}, {Donati}, {Duffau}, {Franciosini},
  {Friel}, {Geisler}, {Gonz{\'a}lez Hern{\'a}ndez}, {Gruyters}, {Guiglion},
  {Hansen}, {Heiter}, {Hill}, {Jacobson}, {Jofre}, {J{\"o}nsson}, {Lanzafame},
  {Lardo}, {Ludwig}, {Maiorca}, {Mikolaitis}, {Montes}, {Morel}, {Mucciarelli},
  {Mu{\~n}oz}, {Nordlander}, {Pasquini}, {Puzeras}, {Recio-Blanco}, {Ryde},
  {Sacco}, {Santos}, {Serenelli}, {Sordo}, {Soubiran}, {Spina}, {Steffen},
  {Vallenari}, {Van Eck}, {Villanova}, {Gilmore}, {Randich}, {Asplund},
  {Binney}, {Drew}, {Feltzing}, {Ferguson}, {Jeffries}, {Micela}, {Negueruela},
  {Prusti}, {Rix}, {Alfaro}, {Babusiaux}, {Bensby}, {Blomme}, {Flaccomio},
  {Fran{\c c}ois}, {Irwin}, {Koposov}, {Walton}, {Bayo}, {Carraro}, {Costado},
  {Damiani}, {Edvardsson}, {Hourihane}, {Jackson}, {Lewis}, {Lind}, {Marconi},
  {Martayan}, {Monaco}, {Morbidelli}, {Prisinzano}, \& {Zaggia}}]{Smiljanic14}
{Smiljanic}, R., {Korn}, A.~J., {Bergemann}, M., {et~al.} 2014, \aap, 570, A122

\bibitem[{{Snaith} {et~al.}(2014){Snaith}, {Haywood}, {Di Matteo}, {Lehnert},
  {Combes}, {Katz}, \& {G{\'o}mez}}]{Snaith14b}
{Snaith}, O., {Haywood}, M., {Di Matteo}, P., {et~al.} 2014, ArXiv e-prints

\bibitem[{{Soubiran} {et~al.}(2003){Soubiran}, {Bienaym{\'e}}, \&
  {Siebert}}]{Soubiran03}
{Soubiran}, C., {Bienaym{\'e}}, O., \& {Siebert}, A. 2003, \aap, 398, 141

\bibitem[{{Steinmetz} {et~al.}(2006){Steinmetz}, {Zwitter}, {Siebert},
  {Watson}, {Freeman}, {Munari}, {Campbell}, {Williams}, {Seabroke}, {Wyse},
  {Parker}, {Bienaym{\'e}}, {Roeser}, {Gibson}, {Gilmore}, {Grebel}, {Helmi},
  {Navarro}, {Burton}, {Cass}, {Dawe}, {Fiegert}, {Hartley}, {Russell},
  {Saunders}, {Enke}, {Bailin}, {Binney}, {Bland-Hawthorn}, {Boeche}, {Dehnen},
  {Eisenstein}, {Evans}, {Fiorucci}, {Fulbright}, {Gerhard}, {Jauregi}, {Kelz},
  {Mijovi{\'c}}, {Minchev}, {Parmentier}, {Pe{\~n}arrubia}, {Quillen}, {Read},
  {Ruchti}, {Scholz}, {Siviero}, {Smith}, {Sordo}, {Veltz}, {Vidrih}, {von
  Berlepsch}, {Boyle}, \& {Schilbach}}]{Steinmetz06}
{Steinmetz}, M., {Zwitter}, T., {Siebert}, A., {et~al.} 2006, \aj, 132, 1645

\bibitem[{{Tacconi} {et~al.}(2013){Tacconi}, {Neri}, {Genzel}, {Combes},
  {Bolatto}, {Cooper}, {Wuyts}, {Bournaud}, {Burkert}, {Comerford}, {Cox},
  {Davis}, {F{\"o}rster Schreiber}, {Garc{\'{\i}}a-Burillo}, {Gracia-Carpio},
  {Lutz}, {Naab}, {Newman}, {Omont}, {Saintonge}, {Shapiro Griffin}, {Shapley},
  {Sternberg}, \& {Weiner}}]{Tacconi13}
{Tacconi}, L.~J., {Neri}, R., {Genzel}, R., {et~al.} 2013, \apj, 768, 74

\bibitem[{{Valenti} \& {Piskunov}(1996)}]{Valenti96}
{Valenti}, J.~A. \& {Piskunov}, N. 1996, \aaps, 118, 595

\bibitem[{{Villalobos} \& {Helmi}(2008)}]{Villalobos08}
{Villalobos}, {\'A}. \& {Helmi}, A. 2008, \mnras, 391, 1806

\bibitem[{{Widrow} {et~al.}(2014){Widrow}, {Barber}, {Chequers}, \&
  {Cheng}}]{Widrow14}
{Widrow}, L.~M., {Barber}, J., {Chequers}, M.~H., \& {Cheng}, E. 2014, \mnras,
  440, 1971

\bibitem[{{Widrow} {et~al.}(2012){Widrow}, {Gardner}, {Yanny}, {Dodelson}, \&
  {Chen}}]{Widrow12}
{Widrow}, L.~M., {Gardner}, S., {Yanny}, B., {Dodelson}, S., \& {Chen}, H.-Y.
  2012, \apjl, 750, L41

\bibitem[{{Williams} {et~al.}(2013){Williams}, {Steinmetz}, {Binney},
  {Siebert}, {Enke}, {Famaey}, {Minchev}, {de Jong}, {Boeche}, {Freeman},
  {Bienaym{\'e}}, {Bland-Hawthorn}, {Gibson}, {Gilmore}, {Grebel}, {Helmi},
  {Kordopatis}, {Munari}, {Navarro}, {Parker}, {Reid}, {Seabroke}, {Sharma},
  {Siviero}, {Watson}, {Wyse}, \& {Zwitter}}]{Williams13}
{Williams}, M.~E.~K., {Steinmetz}, M., {Binney}, J., {et~al.} 2013, \mnras,
  436, 101

\bibitem[{{Wyse}(2001)}]{Wyse01}
{Wyse}, R.~F.~G. 2001, in Astronomical Society of the Pacific Conference
  Series, Vol. 230, Galaxy Disks and Disk Galaxies, ed. J.~G. {Funes} \& E.~M.
  {Corsini}, 71--80

\bibitem[{{Wyse} \& {Gilmore}(1988)}]{Wyse88}
{Wyse}, R.~F.~G. \& {Gilmore}, G. 1988, \aj, 95, 1404

\bibitem[{{Wyse} \& {Gilmore}(1995)}]{Wyse95}
{Wyse}, R.~F.~G. \& {Gilmore}, G. 1995, \aj, 110, 2771

\bibitem[{{Yanny} {et~al.}(2009){Yanny}, {Rockosi}, {Newberg}, {Knapp},
  {Adelman-McCarthy}, {Alcorn}, {Allam}, {Allende Prieto}, {An}, {Anderson},
  {Anderson}, {Bailer-Jones}, {Bastian}, {Beers}, {Bell}, {Belokurov},
  {Bizyaev}, {Blythe}, {Bochanski}, {Boroski}, {Brinchmann}, {Brinkmann},
  {Brewington}, {Carey}, {Cudworth}, {Evans}, {Evans}, {Gates}, {G{\"a}nsicke},
  {Gillespie}, {Gilmore}, {Nebot Gomez-Moran}, {Grebel}, {Greenwell}, {Gunn},
  {Jordan}, {Jordan}, {Harding}, {Harris}, {Hendry}, {Holder}, {Ivans},
  {Ivezi{\v c}}, {Jester}, {Johnson}, {Kent}, {Kleinman}, {Kniazev},
  {Krzesinski}, {Kron}, {Kuropatkin}, {Lebedeva}, {Lee}, {French Leger},
  {L{\'e}pine}, {Levine}, {Lin}, {Long}, {Loomis}, {Lupton}, {Malanushenko},
  {Malanushenko}, {Margon}, {Martinez-Delgado}, {McGehee}, {Monet}, {Morrison},
  {Munn}, {Neilsen}, {Nitta}, {Norris}, {Oravetz}, {Owen}, {Padmanabhan},
  {Pan}, {Peterson}, {Pier}, {Platson}, {Re Fiorentin}, {Richards}, {Rix},
  {Schlegel}, {Schneider}, {Schreiber}, {Schwope}, {Sibley}, {Simmons},
  {Snedden}, {Allyn Smith}, {Stark}, {Stauffer}, {Steinmetz}, {Stoughton},
  {SubbaRao}, {Szalay}, {Szkody}, {Thakar}, {Sivarani}, {Tucker}, {Uomoto},
  {Vanden Berk}, {Vidrih}, {Wadadekar}, {Watters}, {Wilhelm}, {Wyse}, {Yarger},
  \& {Zucker}}]{Yanny09}
{Yanny}, B., {Rockosi}, C., {Newberg}, H.~J., {et~al.} 2009, \aj, 137, 4377

\end{thebibliography}

\label{lastpage}

\end{document}